\documentclass[aps,pra,twocolumn,floatfix,superscriptaddress,showpacs,longbibliography]{revtex4-1}
\usepackage{xfrac}
\usepackage{amsfonts}
\usepackage{dsfont}
\usepackage{amsmath}
\usepackage{amssymb}
\usepackage{amsthm}
\usepackage{mathtools}
\usepackage{graphicx}
\usepackage[caption=false]{subfig}
\usepackage{floatrow}
\usepackage{color}
\usepackage{bm}
\usepackage[normalem]{ulem}
\usepackage[hidelinks]{hyperref}
\hypersetup{
     colorlinks   = true,
     citecolor    = blue,
     linkcolor    = blue
}
\usepackage{mathptmx}
\usepackage{times}
\usepackage{enumerate}

\newcommand{\ignore}[1]{}
\newcommand{\nobibentry}[1]{{\let\nocite\ignore\bibentry{#1}}}

\newcommand{\bibfnamefont}[1]{#1}
\newcommand{\bibnamefont}[1]{#1}
\newcommand{\qav}[1]{\left<#1\right>}

\newcommand{\expect}[1]{\langle\hspace{-2pt}\langle #1 \rangle \hspace{-2pt}\rangle}
\newcommand{\ket}[1]{\left\vert#1\right\rangle}

\def\ketbra#1#2{{\vert#1\rangle\!\langle#2\vert}}
\newcommand{\andreu}[1]{{\color{black}#1}}

\DeclareMathAlphabet{\mathcal}{OMS}{cmsy}{m}{n}

\theoremstyle{definition}

\theoremstyle{plain}

\theoremstyle{plain}

\begin{document}

\title{Open quantum systems coupled to finite baths: A hierarchy of master equations }

\author{Andreu Riera-Campeny}
\affiliation{F\'{\i}sica Te\`{o}rica: Informaci\'{o} i Fen\`{o}mens Qu\`{a}ntics. Departament de F\'{\i}sica, Universitat Aut\`{o}noma de Barcelona, 08193 Bellaterra, Spain}
\author{Anna Sanpera}
\affiliation{F\'{\i}sica Te\`{o}rica: Informaci\'{o} i Fen\`{o}mens Qu\`{a}ntics. Departament de F\'{\i}sica, Universitat Aut\`{o}noma de Barcelona, 08193 Bellaterra, Spain}
\affiliation{ICREA, Psg. Llu\' is Companys 23, 08001 Barcelona, Spain.}
\author{Philipp Strasberg}
\affiliation{F\'{\i}sica Te\`{o}rica: Informaci\'{o} i Fen\`{o}mens Qu\`{a}ntics. Departament de F\'{\i}sica, Universitat Aut\`{o}noma de Barcelona, 08193 Bellaterra, Spain}

\begin{abstract}
An open quantum system in contact with an infinite bath approaches equilibrium, while the state of the bath remains unchanged. If the bath is finite, the open system still relaxes to equilibrium, but it induces a dynamical evolution of the bath state. In this work, we study the dynamics of open quantum systems in contact with finite baths. We obtain a hierarchy of master equations that improve their accuracy by including more dynamical information of the bath. For instance, as the least accurate but simplest description in the hierarchy we obtain the conventional Born-Markov-secular master equation. Remarkably, our framework works even if the measurements of the bath energy are imperfect, which, not only is more realistic, but also unifies the theoretical description. Also, we discuss this formalism in detail for a particular non-interacting environment where the Boltzmann temperature and the Kubo-Martin-Schwinger relation naturally arise. Finally, we apply our hierarchy of master equations to study the central spin model.
\end{abstract}		

\maketitle

\section{Introduction}

Large environments, as compared to the size of the open system, often cause the latter to relax to equilibrium while keeping their macroscopic properties unchanged. Since the influence of the open system on the environment is imperceptible, those environments act as \textit{infinite} baths. However, not all baths are infinite. In some situations, the interaction with the system during the equilibration process induces a dynamical evolution of the environment. In turn, the evolution of the environment has a feedback effect on the open system dynamics. Such environments are the central object of study of this work. Since a finite influence from the system can produce an appreciable change in their state, we refer to them as \textit{finite} baths.

The miniaturization of quantum experiments towards the microscopic scale is leading also to a more detailed description of their surroundings \cite{Brantut2012, Brantut2013, Muller2015, Pekola2016, Halbertal2016, Muller2019, Karimi2020, Hausler2021}. As a consequence, there is often more information available about the bath and, in the case of finite baths, this information evolves dynamically. To date, see for instance \cite{deVega2017,Weimer2021}, most theoretical and numerical tools to describe the evolution of open quantum systems rely on completely tracing out the environment. However, this \textit{modus operandi} ignores the dynamical nature of the bath, which could be potentially used to obtain more accurate predictions about the open system dynamics. It is then timely to investigate novel theoretical techniques which profit from this dynamical information by including, to some extent, a dynamically evolving bath.

To this end, we extend the much used quantum master equation approach \cite{Gardiner2000, Breuer2002, Schaller2014, deVega2017}. While the conventional derivation is based on the assumption that the environment is infinite, memoryless and at a fixed temperature, we overcome these limitations by including additional information about the bath into the description. In particular, we focus on including the information about the (coarse-grained) energy of the bath. Master equations in this spirit were first derived in Ref.~\cite{Esposito2003a}, refined in Ref.~\cite{Esposito2007}, and formalized in Refs.~\cite{Breuer2006} and \cite{Breuer2007}. Lately, the authors extended the range of validity of this type of master equations and established connections with their nonequilibrium thermodynamic description in Ref.~\cite{Riera-Campeny2021}. We refer to this formalism as the Extended Microcanonical Master Equation (EMME). Recently, in Ref.~\cite{Donvil2021}, this formalism was applied to understand ultrasensitive calorimetric measurements reported in Refs. \cite{Karimi2020, Ronzani2018, Kokkoniemi2019, Senior2020}.

\andreu{It is important to distinguish our approach from conventional non-Markovian methods (see~\cite{deVega2017} and references therein), where the environment is still completely traced out such that no information about the state of the bath is available. While the effect of a dynamically evolving bath \emph{on the system dynamics} is (sometimes exactly, sometimes approximately) taken into account in these methods, the effect of the system \emph{on the bath} remains inaccessible. This is because one still traces out the bath completely and, moreover, most approaches today still treat the bath as infinite by using a continuum representation or thermodynamic limit for the bath modes. We remark that Markovian embedding strategies, such as reaction coordinate methods~\cite{Nazir2018}, treat parts of strongly coupled bath modes explicitly, but they nevertheless still rely on the idea of an infinitely large residual bath. Here, we are less interested in strong coupling effects than to describe environments which can change from a macroscopic point of view.}

\andreu{In this work, we derive several results that are of theoretical and practical relevance. In order of importance they are: (1) the derivation of a Born-Markov-secular (BMS) master equation that includes finite baths effects through a nonequilibrium time-dependent temperature, dubbed $\beta^\star(t)$; (2) the connection with previous works on nonequilibrium thermodynamics, showing that the same nonequilibrium temperature $\beta^\star(t)$ is \textit{also} thermodynamically meaningful (as we extensively discuss in the conclusion section); and (3) the extension of the formalism to generalized measurements. Finally, (4) we apply the formalism to the central spin system as an example. The outline for the rest of the paper is as follows. }

\floatsetup[figure]{style=plain,subcapbesideposition=top}
\begin{figure}
{\includegraphics[width=\textwidth]{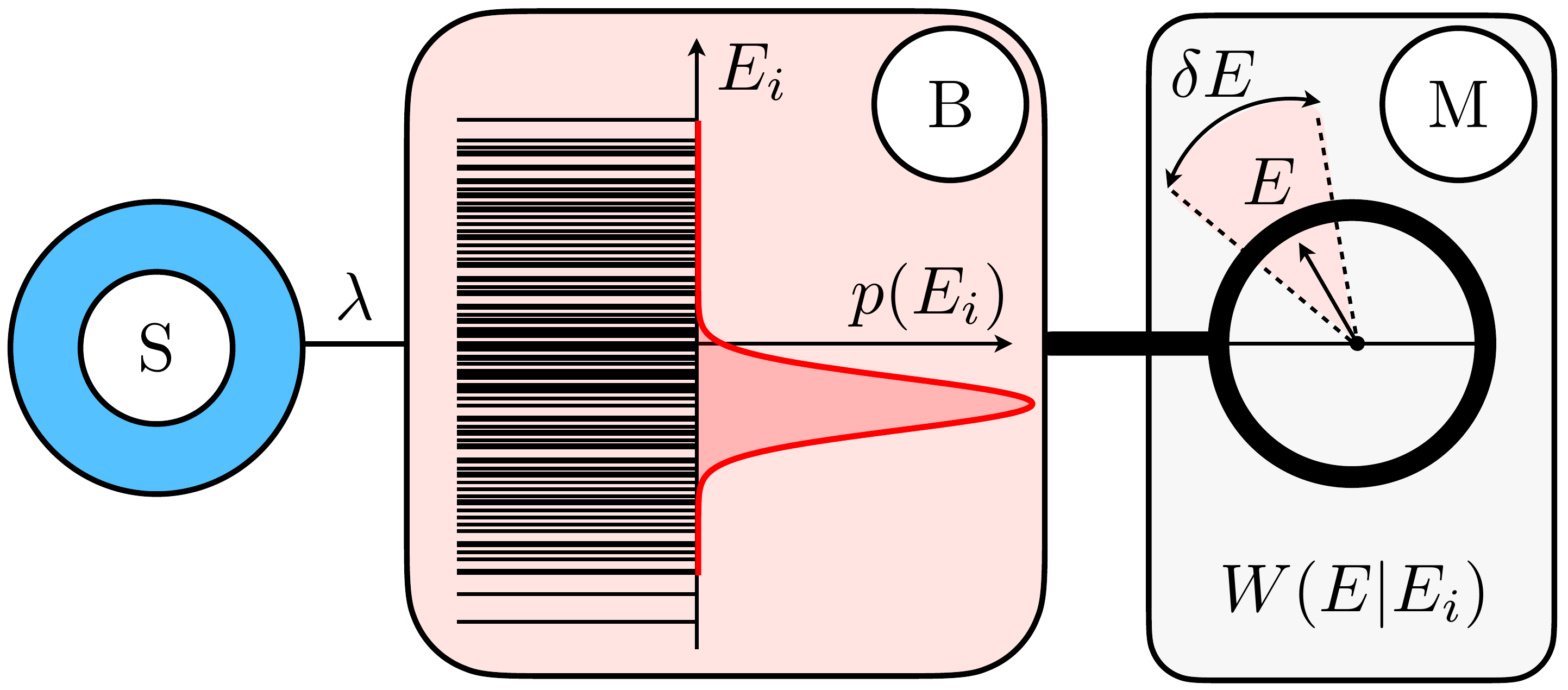}}
  \caption{Sketch of the system (labeled by S) interacting with the finite bath (labeled by B). The bath energy levels are depicted with horizontal black lines, and their corresponding energy distribution $p(E_i)$ is plotted in the red solid line. The measurement apparatus (labeled by M) gives an output $E$ given an input $E_i$ according to the weighting function $W(E|E_i)$, which introduces a finite precision of order $\delta E$. \label{fig:sketch}} 
\end{figure}

In Sec.~\ref{sec:preliminaries}, we introduce the framework of open systems coupled to finite baths, including imperfect measurements of the bath energy. This framework is used to derive the EMME, which is our general starting point for further investigations.

Section~\ref{sec:limits} is devoted to investigate under which conditions our equation reduces to the standard BMS master equation. As a result, we obtain a hierarchy of master equations for open quantum systems coupled to finite baths. In this hierarchy, including more dynamical information about the bath yields a more accurate, although more complicated master equation description. In particular, the EMME is best approximated by a BMS master equation with an effective time-dependent temperature $\beta^\star(t)$, which is determined self-consistently.

In Sec.~\ref{sec:noninteracting}, we use our results to discuss the general case of a non-interacting bath that couples locally to the system, which describes well many practical cases. Also, we derive the Kubo-Martin-Schwinger relation for such systems, where the temperature is set by the Boltzmann temperature.

In Sec.~\ref{sec:models}, we apply the results of our previous sections to the central spin system, which serves as a paradigmatic example of a finite bath.

Finally, we present our conclusions in Sec.~\ref{sec:conclusions}, conjecturing a parallelism between the hierarchy of master equations and their corresponding second law~\cite{Strasberg2021b}. A summary of this emerging picture is provided at the end in Fig.~\ref{fig:big_conclusions}.

\section{The EMME with imperfect measurements}\label{sec:preliminaries}

\andreu{In this section, we extend the derivation of the EMME to imperfect measurements. Our motivation is twofold. First, from a theoretical point of view imperfect measurements are simply more general. Second, from the practical point of view measurements of the bath energy will be always subject to an error. For instance, imperfect measurements can lead to an apparent heating~\cite{Donvil2021} in the ultrasensitive calorimetric measurements of Refs.~\cite{Karimi2020, Ronzani2018, Kokkoniemi2019, Senior2020}.} 

Consider an open quantum system (labeled by S) that interacts weakly with a finite bath (labeled by B), as we sketch in Fig.~\ref{fig:sketch}. Because the bath is finite, to describe its dynamics dynamical information about the system \textit{and} the finite bath has to be taken into account. We consider that, while in principle all the information about the open quantum system is accessible, only partial information about the bath is available. This partial bath information is obtained using a measurement apparatus, labeled by M, that has finite precision, and it is mathematically described by a positive operator valued measure or POVM.

\subsection{Setup}
 
Let $H_\text{S} = \sum_k \varepsilon_k \ketbra{k}{k}$ and $H_\text{B} = \sum_i E_i\ketbra{E_i}{E_i}$ be the system and bath Hamiltonian respectively. The system and the bath are coupled through the Hamiltonian $H_\text{int} = \lambda S\otimes B$, where $\lambda$ represents the interaction energy scale. Throughout the article, we assume that $B$ contains only off-diagonal terms; i.e. $\langle E_i|B|E_i\rangle = 0$ for all $E_i$, which simplifies the discussion. However, as we show in App.~\ref{app:offdiagonal_b}, if $B$ contains diagonal terms, it is possible to put forward the same arguments by first redefining the Lamb shift Hamiltonian (see below). Also, the extension to general couplings of the form $H_\text{int} = \lambda \sum_\alpha S^\alpha \otimes B^\alpha$ is possible and analogous procedures can be found elsewhere (see for instance \cite{Riera-Campeny2021}).

At the time $t=0$, the system-bath composite is found in the initial state $\rho(0)$. In the interaction picture with respect to $H_0 = H_\text{S}+H_\text{B}$, the evolution of the closed system is generated by the Liouville-von Neumann equation ($\hbar=1$)
\begin{align}
\partial_t \tilde{\rho}(t) = -i [\tilde{H}_\text{int}(t),\tilde{\rho}(t)] \coloneqq \mathcal{L}(t)[\tilde{\rho}(t)], \label{eq:vonNeumann}
\end{align}
where we use the tilde to denote operators in the interaction picture; for instance $\tilde{\rho}(t) = \exp(i H_0 t)\rho(0) \exp(-i H_0 t)$. 

\subsection{Imperfect measurements}


Partial knowledge of the bath state is obtained via a measurement apparatus described by the POVM $\{P(E)\}$, see for instance \cite{Nielsen2002}, with POVM elements 
\begin{align}
P(E) = \sum_i W(E|E_i) \ketbra{E_i}{E_i},
\end{align}
where the outputs $E$ can be discrete or continuous. Hereafter, we use the integral notation $\int dE$ to describe the sum or integral over all outputs of a POVM, regardless of whether they are of discrete or continuous nature. The weighting function $W(E|E_i)$ corresponds to the positive and normalized conditional probability of obtaining the outcome $E$ given that the bath energy was $E_i$. We denote by $\delta E$ the typical energy width over which the weighting function $W(E|E_i)$ spreads, and also introduce the volumes (or volume densities) $V(E) = \text{tr}[P(E)]$, which measure the number of bath states whose energy is compatible with the output $E$. Two natural choices for such weighting functions are the indicator weighting function $W_\text{I}(E|E_i)$ with discrete outputs $E \in \{m\delta E\}$ and $m\in\mathbb{Z}$, and the Gaussian weighting function $W_\text{G}(E|E_i)$ with a continuous output $E\in\mathbb{R}$. The first case is defined as
\begin{align}
W_\text{I}(E|E_i) &= \mathbf{1}_E(E_i) \coloneqq \left\{\begin{matrix}
1 &\text{if } |E-E_i|\leq \delta E/2 \\
0 &\text{else}
\end{matrix}\right.,\label{eq:indicator_measure}
\end{align}
whose POVM elements correspond to orthogonal projectors that divide the bath spectrum in a set of non-overlapping energy windows. We denote those projectors by $\Pi(E) = \sum_{E_i} \mathbf{1}_E(E_i)\ketbra{E_i}{E_i}$. This case was also investigated in Refs.~\cite{Esposito2003a, Esposito2007, Breuer2006, Breuer2007, Riera-Campeny2021}. For the second case, we introduce the notation $\mathcal{N}(x,\sigma) = (\sqrt{2\pi}\sigma)^{-1}\exp[- x^2/(2\sigma^2)]$ for the normal distribution. Then, the Gaussian weighting function is compactly defined as
\begin{align}
W_\text{G}(E|E_i) = \mathcal{N}(E-E_i,\delta E).
\end{align}

\subsection{Derivation of the master equation}

Our main goal is to derive a master equation for the unnormalized conditional state of the system 
\begin{align}
\rho_\text{S}(E)= \text{tr}_\text{B}[\rho P(E)],\label{eq:rhoSE}
\end{align}
which describes the state of the open quantum system S, provided that the bath B was measured with energy $E$. We note that both, the reduced state of the system $\rho_\text{S}\coloneqq \int dE \rho_\text{S}(E)$ and the probability density that the bath has energy $E$, that is, $p(E)= \text{tr}[\rho_\text{S}(E)]$, can be computed from the unnormalized state \eqref{eq:rhoSE}. 

To derive the master equation, we perform the ``generalized Born approximation''
\begin{align}
\rho(t) \approx \int dE \rho_\text{S}(E,t)\otimes\omega_\text{B}(E),\label{eq:born_approx}
\end{align}
where $\omega_\text{B}(E) = P(E)/V(E)$, for all times $t\geq 0$. Since for $W_\text{I}(E|E_i)$ the state $\omega_\text{B}(E)$ corresponds to the microcanonical state at energy $E$, we refer to $\omega_\text{B}(E)$ as the microcanonical state even when the POVM elements $P(E)$ are not orthogonal. Then, truncating the formal solution of Eq.~\eqref{eq:vonNeumann} to second order in $\lambda$, multiplying by $P(E)$ and tracing over the bath yields
\begin{align}
\partial_t \tilde{\rho}_\text{S}(E;t) = \text{tr}_\text{B}\left\{\int_0^t dt' P(E)\mathcal{L}(t)\left[\mathcal{L}(t')[\rho(t')]\right]\right\}.\label{eq:finite_time_redfield}
\end{align}
Using $\tilde{H}_\text{int}(t) = \lambda \tilde{S}(t)\otimes \tilde{B}(t)$, and performing the first Markov approximation $\rho_\text{S}(E,t') \mapsto \rho_\text{S}(E,t)$ we arrive at
\begin{widetext}
\begin{align}
\partial_t \tilde{\rho}_\text{S}(E;t) = \lambda^2 \int_0^t dt'\int dE' \left[\left\langle \tilde{B}(t) P(E) \tilde{B}(t')\right\rangle_{E'} \tilde{S}(t') \tilde{\rho}_\text{S}(E',t) \tilde{S}(t) - \left\langle P(E) \tilde{B}(t)\tilde{B}(t')\right\rangle_{E'} \tilde{S}(t) \tilde{S}(t') \tilde{\rho}_\text{S}(E',t) + \text{h.c.}\right],
\end{align}
where we have used the shorthand notation $\langle\cdots \rangle_{E'} = \text{tr}_\text{B}[\cdots \omega_\text{B}(E')]$. We introduce the decomposition $\tilde{S}(t) = \sum_\omega S_\omega \exp(-i\omega t)$ with $S_\omega = \sum_{kq} \langle k|S|q\rangle \delta_{\varepsilon_q-\varepsilon_k,\omega}$, perform the change of variables $\tau = t-t'$, and move back to the Schr\"odinger picture to obtain
\begin{align}
\partial_t \rho_\text{S}(E) =-i[H_\text{S},\rho_\text{S}(E)]+ \lambda^2 \sum_\omega \int_0^t dt'\int dE' e^{i\omega \tau} \left[\left\langle \tilde{B}(\tau) P(E) B\right\rangle_{E'} S_\omega \rho_\text{S}(E') S- \left\langle P(E) \tilde{B}(\tau)B\right\rangle_{E'} S S_\omega \rho_\text{S}(E') + \text{h.c.}\right].
\end{align}
To proceed further, we perform the standard Markov and secular approximations. The first one corresponds to sending the upper limit of the integral $t\to\infty$ and holds for a sufficiently rapidly decaying correlation function. The second corresponds to performing a time-average in the interaction picture (see for instance \cite{Schaller2014, Breuer2002}). Then, explicitly writing the Hermitian conjugate and collecting terms we arrive at
\begin{align}
\partial_t \rho_\text{S}(E) =& -i[H_\text{S},\rho_\text{S}(E)]-i \int dE' [H_\text{LS}(E,E'),\rho_\text{S}(E')] 
+\sum_\omega \int dE' \left[ \gamma_1(E,E';\omega) S_\omega \rho_\text{S}(E') S_\omega^\dagger \right.-\left. \frac{\gamma_2(E,E';\omega)}{2} \left\{S_\omega^\dagger S_\omega, \rho_\text{S}(E')\right\}\right],\label{eq:emme_povm}
\end{align}
\end{widetext}
where we have used $2 \theta(\tau) = 1+\text{sgn}(\tau)$, being $\text{sgn}(\tau)$ the sign function, to define
\begin{align}
&H_\text{LS}(E,E') = \frac{\lambda^2}{2i} \sum_\omega \int_\mathbb{R} d\tau  e^{i\omega \tau} \text{sgn}(\tau) \left\langle P(E) \tilde{B}(\tau)B\right\rangle_{E'} S_\omega^\dagger S_\omega,\nonumber \\
&\gamma_1(E,E';\omega) = \lambda^2 \int_\mathbb{R} d\tau e^{i\omega \tau} \left\langle \tilde{B}(\tau) P(E) B\right\rangle_{E'},\nonumber \\
&\gamma_2(E,E';\omega) = \lambda^2 \int_\mathbb{R} d\tau e^{i\omega \tau} \left\langle P(E) \tilde{B}(\tau)B\right\rangle_{E'}.\label{eq:definitions}
\end{align}

In App.~\ref{app:energy_conservation}, we proof that Eq.~\eqref{eq:emme_povm} preserves the average total energy 
\begin{align}
U = U_\text{S}+U_\text{B} \coloneqq \int dE \text{tr}_\text{S}[(H_\text{S}+E\mathbf{1}_\text{S}) \rho_\text{S}(E)],
\end{align}
for unbiased measurements, which satisfy $\int dE E W(E|E_i) = E_i$. For biased measurements, such as projective measurements, energy conservation holds up to the measurement uncertainty $\delta E$. Further identifying the heat flux as $\dot{Q} = -\partial_t \int dE E p(E)$, this conservation law establishes a connection with the first law of thermodynamics of an autonomous system; i.e. $\partial_t U_\text{S} = \dot{Q}$. It is worth mentioning that the present formalism could be extended to include slow driving or multiple heat baths as done, e.g., in Ref.~\cite{Riera-Campeny2021}. However, to remain focused, we consider a time-independent Hamiltonian and a single heat bath only. 

Finally, we note that for the projective measurements arising from the weighting functions $W_\text{I}(E|E_i)$ one has $\gamma_2(E,E'';\omega) = \delta_{EE''} \sum_{E'} \gamma_1(E',E;\omega)$ which leads to the original form of the EMME as presented in Ref.~\cite{Esposito2003a,Esposito2007,Breuer2006,Breuer2007,Riera-Campeny2021}. For completeness, we show in App.~\ref{app:projection_techniques} that the present derivation is equivalent to the one using projection operator techniques to second order in $\lambda$.

\section{Reduced system dynamics}\label{sec:limits}

\floatsetup[figure]{style=plain,subcapbesideposition=top}
\begin{figure}
{\includegraphics[width=\textwidth]{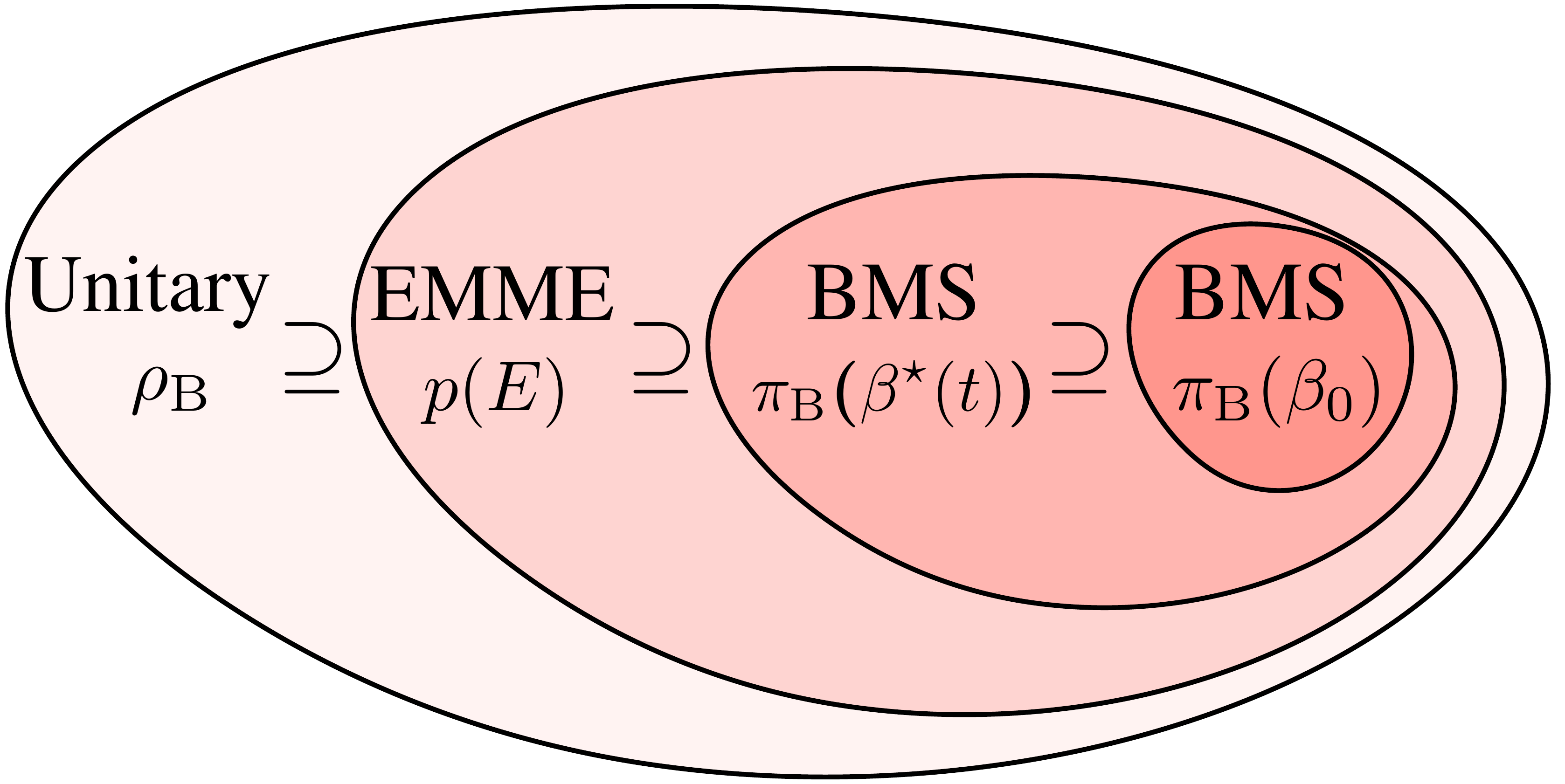}}
  \caption{Sketch of the hierarchy and validity range of the different discussed master equations.\label{fig:sketch2}} 
\end{figure}

In the conventional theory of open quantum systems, one is only interested in the evolution of the reduced state of the system $\rho_\text{S}$. This state encodes all information that can be extracted by measuring locally the open quantum system at a single time; that is, without considering multi-time statistics Ref.~\cite{Milz2021}. A dynamical equation for the reduced state $\rho_\text{S}$ can be found by taking advantage of the derivation presented in the last section. Namely, taking as a starting point Eq.~\eqref{eq:emme_povm}, one can marginalize over the energy $E$ to obtain an equation for $\rho_\text{S} = \int dE \rho_\text{S}(E)$. Clearly, this is different from the conventional approach, where the bath is traced out completely from the start \cite{deVega2017,Weimer2021,Gardiner2000,Breuer2002,Schaller2014}. In the conventional approach, the BMS master equation for the reduced state $\rho_\text{S}$ is obtained in the limit of an infinite, memoryless, thermal, and weakly-coupled bath~\cite{Breuer2002}. In the following, we consider both methods and investigate under which circumstances they become equivalent. This process gives rise to a hierarchy of master equations for the reduced state $\rho_\text{S}$ that can be summarized as follows. 

First, the most general master equation that takes into account all the environmental dynamical information is equivalent to unitary evolution. Second, an open quantum system that exchanges energy with its finite environment can be described using the EMME, which keeps track of the dynamically evolving bath energy distribution $p(E) \coloneqq \text{tr}[\rho P(E)]$. Third, in some cases it may suffice to keep track of the bath average energy, which is in one-to-one correspondence with a certain (time-dependent) effective nonequilibrium temperature $\beta^\star$. Then, the dynamics is generated by the BMS master equation at this inverse temperature $\beta^\star$. Finally, if one fully ignores the finiteness of the bath and assumes it is found in a constant thermal state at inverse temperature $\beta_0$, the dynamics are generated by the BMS master equation at this constant temperature. We sketch this hierarchy in Fig.~\ref{fig:sketch2}.

\subsection{The reduced EMME}

As discussed above, the first method takes as a starting point the EMME in Eq.~\eqref{eq:emme_povm}. Integrating over the energy $E$, one obtains the equation
\begin{align}
\partial_t \rho_\text{S} =& -i[H_\text{S},\rho_\text{S}]-i \int dE [H_\text{LS}(E),\rho_\text{S}(E)] \nonumber \\
&+\sum_\omega \int dE \kappa(E;\omega) {\bm (} S_\omega \rho_\text{S}(E) S_\omega^\dagger 
- \frac{1}{2} \left\{S_\omega^\dagger S_\omega, \rho_\text{S}(E)\right\}{\bm )}.\label{eq:emme_povm_reduced}
\end{align} 
where the quantities
\begin{align}
&H_\text{LS}(E) = \int dE' H_\text{LS}(E',E), \nonumber\\
&\kappa(E; \omega) = \int dE' \gamma_1(E',E;\omega) = \int dE' \gamma_2(E',E;\omega),
\end{align}
have been defined. More explicitly, the function $\kappa(E;\omega)$ can be written in terms of the bath correlation function as $\kappa(E;\omega) = \lambda^2 \int_\mathbb{R} d\tau e^{i\omega \tau} \langle \tilde{B}(\tau) B\rangle_{E}$. Importantly, Eq.~\eqref{eq:emme_povm_reduced} is not a closed equation for the reduced density matrix $\rho_\text{S}$, since it explicitly depends on the unnormalized conditional state $\rho_\text{S}(E)$.

\subsection{The BMS master equation}

We recall the BMS master equation, which is a dynamical equation for the reduced state of the system $\rho_\text{S}$. It is derived assuming that the bath is found in a thermal state $\pi_\text{B}(\beta)$ at the inverse temperature $\beta$. We define this thermal state as
\begin{align}
\pi_\text{B}(\beta) \coloneqq \int dE \frac{V(E) e^{-\beta E}}{Z_\text{B}(\beta)} \omega_\text{B}(E)\approx \frac{e^{-\beta H_\text{B}}}{\text{tr}(e^{-\beta H_\text{B}})}, \label{eq:thermal_state}
\end{align} 
where $Z_\text{B}(\beta) \coloneqq \int dE V(E) \exp(-\beta E)$ is the partition function. Note that, since $\omega_\text{B}(E)$ does not always correspond to the microcanonical state, $\pi_\text{B}(\beta)$ does not correspond to the ``usual'' thermal state either. However, $\pi_\text{B}(\beta)$ coincides with the thermal state in the limit of $\delta E\to 0$.

In order to establish a relation between the BMS master equation and the reduced EMME it is useful to regard the thermal state $\pi_\text{B}(\beta)$ as an average over the canonical distribution of the microcanonical state $\omega_\text{B}(E)$. To be precise, we introduce the thermal expectation value
\begin{align}
    \expect{f(E)}_\beta = \int dE \frac{V(E) e^{-\beta E}}{Z_\text{B}(\beta)} f(E),\label{eq:thermal_expectation}
\end{align}
with which we find the relation $\pi_\text{B}(\beta) = \expect{\omega_\text{B}(E)}_\beta$. With the initial condition $\rho(0) = \rho_\text{S}(0) \otimes \pi_\text{B}(\beta)$, one can derive the well-known BMS master equation \cite{Gardiner2000, Breuer2002, Schaller2014}
\begin{align}
\partial_t \rho_\text{S} =& -i[H_\text{S}+H_\text{LS}(\beta),\rho_\text{S}]\nonumber \\
&+\sum_\omega \kappa(\beta;\omega) \left( S_\omega \rho_\text{S} S_\omega^\dagger - \frac{1}{2} \left\{S_\omega^\dagger S_\omega, \rho_\text{S}\right\}\right),\label{eq:bms}
\end{align} 
where we have defined, using the expectation value in Eq.~\eqref{eq:thermal_expectation}, the quantities 
\begin{align}
&H_\text{LS}(\beta) = \expect{H_\text{LS}(E)}_\beta, \nonumber\\
&\kappa(\beta;\omega) = \expect{\kappa(E;\omega)}_\beta.
\end{align}
For conciseness of the notation, we hereafter abbreviate Eq.~\eqref{eq:bms} as $\partial_t\rho_\text{S} \eqqcolon \mathcal{L}_\text{S}(\beta)[\rho_\text{S}]$.

\subsection{Matching the two descriptions: Limiting cases}

The first limit (i) corresponds to the case where the state $\rho_\text{S}(E)$ remains approximately uncorrelated at all times; that is, $\rho_\text{S}(E) \approx \rho_\text{S} p(E)$. Then, one recovers a closed equation for the reduced state of the system in the form
\begin{align}
\partial_t \rho_\text{S} =& -i[H_\text{S}+\expect{H_\text{LS}(E)}_{p},\rho_\text{S}]\nonumber \\
&+\sum_\omega \expect{\kappa(E;\omega)}_{p} \left( S_\omega \rho_\text{S} S_\omega^\dagger - \frac{1}{2} \left\{S_\omega^\dagger S_\omega, \rho_\text{S}\right\}\right).\label{eq:emme_uncorrelated}
\end{align} 
where $\expect{f(E)}_{p} = \int dE f(E) p(E)$. If, moreover, the distribution $p(E)$ happens to be thermal, then $\expect{\cdots}_p = \expect{\cdots}_\beta$ and one recovers exactly Eq.~\eqref{eq:bms}. However, this limit is unsatisfactory, since dissipation and noise are often a consequence of building and destroying system-bath correlations. More importantly, $\rho_\text{S}(E) \approx \rho_\text{S}\otimes p(E)$ is an \emph{ad hoc} assumption where one has not used any physical properties of the bath. 

There exists a second limit (ii) that causes Eq.~\eqref{eq:emme_povm_reduced} to reduce to Eq.~\eqref{eq:bms}. Let $\Delta E$ be the uncertainty associated with the distribution $p(E)$ around the average energy $U_\text{B} = \expect{E}_p$. Then, if the functions $\kappa(E;\omega)$ and $H_\text{LS}(E)$ that govern the influence of the bath on the system are approximately constant over the range of energies $\Delta E$, one can marginalize the state $\rho_\text{S}(E)$ over the energy $E$ in Eq.~\eqref{eq:emme_povm_reduced} to find a closed equation for $\rho_\text{S}$ that is formally equal to the BMS. If, moreover, the equivalence of ensembles holds for the bath in the sense that $H_\text{LS}(U_\text{B}) \approx H_\text{LS}(\beta)$ and $\kappa(U_\text{B},\omega)\approx \kappa(\beta;\omega)$ for some inverse temperature $\beta$ one recovers exactly Eq.~\eqref{eq:bms}. A similar discussion is conducted in Ref.~\cite{Esposito2007}.

\subsection{A hierarchy of master equations}\label{subsec:hierarchy}

The philosophy behind the present approach is as follows. Incorporating dynamical information about the environment leads to more accurate predictions about the open system dynamics. In this spirit, we wonder whether it exists a hierarchy of such master equations, where neglecting certain dynamical information about the environment gives rise to simpler master equations, but with a more restricted range of validity. The answer is positive, and it requires interpolating between the discussed limiting cases (i) and (ii). 




Typically, one wants to describe the evolution of a system under conditions that correspond to neither (i) nor (ii), but that can be relatively close to both limits at the same time. To discuss this situation, it is convenient to introduce a perturbative parameter $\eta$ that keeps track of the degree of closeness to (i) and (ii). Namely, we introduce the differences $\delta\rho_\text{S}(E) \coloneqq \rho_\text{S}(E)-\rho_\text{S} p(E)$, $\delta H_\text{LS}(E) \coloneqq H_\text{LS}(E)-H_\text{LS}(\beta)$, and $\delta \kappa(E;\omega) \coloneqq \kappa(E;\omega)-\kappa(\beta;\omega)$, and assume that they are of the same order $\eta$. For the sake of the discussion, we also fix the initial state of the bath to be a thermal state at temperature $\beta_0$; that is,  $\rho_\text{B}(0) = \pi_\text{B}(\beta_0)$, and we allow the parameter $\beta=\beta(t)$ to depend on time with the initial condition $\beta(0) = \beta_0$. Reexpressing Eq.~\eqref{eq:emme_povm_reduced} in terms of the differences we obtain to first order in $\eta$
\begin{align}
\partial_t \rho_\text{S} \approx& \mathcal{L}_\text{S}(\beta)[\rho_\text{S}]  -i [\expect{\delta H_\text{LS}(E)}_{p},\rho_\text{S}]\label{eq:emme_eta_expansion}\\
&+ \sum_\omega \expect{\delta\kappa(E;\omega)}_{p} \left( S_\omega \rho_\text{S} S_\omega^\dagger -\frac{1}{2} \left\{S_\omega^\dagger S_\omega, \rho_\text{S}\right\} \right). \nonumber
\end{align}

Intriguingly, the EMME and the BMS can coincide to first order in $\eta$ provided that we can find a time-dependent inverse temperature $\beta$ such that $\expect{ \delta H_\text{LS}(E)}_{p}$ and $\expect{\delta\kappa(E;\omega)}_{p}$ vanish for all times. However, it is hard to even know if such an inverse temperature exists. Nonetheless, one can always find linear approximations in the energy $E$ around the average energy
\begin{align}
& H_\text{LS}(E)\approx H_\text{LS}(U_\text{B}) + [\partial_{E} H_\text{LS}(E)]_{E=U_\text{B}}(E-U_\text{B}),\nonumber \\
& \kappa(E;\omega) \approx \kappa(U_\text{B};\omega) + [\partial_{E} \kappa(E;\omega)]_{E=U_\text{B}}(E-U_\text{B}),    
\end{align}
for a sufficiently small $\Delta E$ and a sufficiently smooth bath spectrum. In that case, there always exists a time-dependent choice $\beta(t)$ for which the first order of Eq.~\eqref{eq:emme_eta_expansion} vanishes. This choice corresponds to the solution $\beta^\star$ of the equation
\begin{align}
\expect{E}_{\beta^\star} \overset{!}{\coloneqq} U_\text{B},\label{eq:noneq_temperature}
\end{align}
that is, the inverse temperature of a thermal state that has the same average energy as the actual state of the bath. Therefore, the role of the time-dependent solution of Eq.~\eqref{eq:noneq_temperature} is to update the temperature of the bath according to the current average bath energy. This relation can be made even more apparent by taking the derivative of Eq.~\eqref{eq:noneq_temperature}, which leads to
\begin{align}
\frac{d}{dt}\beta^\star = \frac{\beta^{\star 2} }{\mathcal{C}(\beta^\star)}\dot{Q}, \label{eq:beta_evolution}
\end{align}
where $\mathcal{C}(\beta) \coloneqq -\beta^2 d \expect{E}_\beta /d\beta$ is the canonical heat capacity with respect to the inverse temperature, and $\dot{Q} =- d U_\text{B}/dt$ the heat flux. Equation~\eqref{eq:beta_evolution} explicitly shows that in order to use the BMS master equations for a finite bath, one has to update the bath temperature due to the heat flux exchanged with the system. Only in the limit of an infinite bath, for which the extensive heat capacity $\mathcal{C}(\beta^\star)$ tends to infinity, one is allowed to set $\beta^\star(t) =\beta_0$ and be still correct to first order in $\eta$ at all times.

Interestingly, the same nonequilibrium temperature $\beta^\star$ has been proposed as a definition of nonequilibrium temperature in phenomenological nonequilibrium thermodynamics, see Refs.~\cite{Muschik1977,Muschik1977b}. Moreover, it has been recently shown to appear in a microscopic derivation of Clausius' inequality and the entropy production \cite{Riera-Campeny2021,Strasberg2021,Strasberg2021b}. We discuss again this connection at the end in Fig.~\ref{fig:big_conclusions}.
  

\section{Correlation functions and emergent canonical distribution}\label{sec:noninteracting}

In this section, we take a step back and consider the EMME~\eqref{eq:emme_povm} again. From a practical point of view, using Eq.~\eqref{eq:emme_povm} to describe the dynamics of an open quantum system requires computing the Lamb-shift Hamiltonian $H_\text{LS}(E)$ as well as the Fourier transformed correlation functions $\gamma_1(E,E';\omega)$ and $\gamma_2(E,E';\omega)$. This task is setup dependent, and it can be difficult to get an intuition for an arbitrarily general coupling operator $B$ or bath Hamiltonian $H_\text{B}$. However, there is a relatively large class of environments for which one can take advantage of arguments of statistical mechanics to proceed further in the calculations. This class corresponds to environments of non-interacting parts that couple locally to the system, and we investigate them in the following.

\subsection{Piecewise non-interacting bath}

\floatsetup[figure]{style=plain,subcapbesideposition=top}
\begin{figure}
{\includegraphics[width=\textwidth]{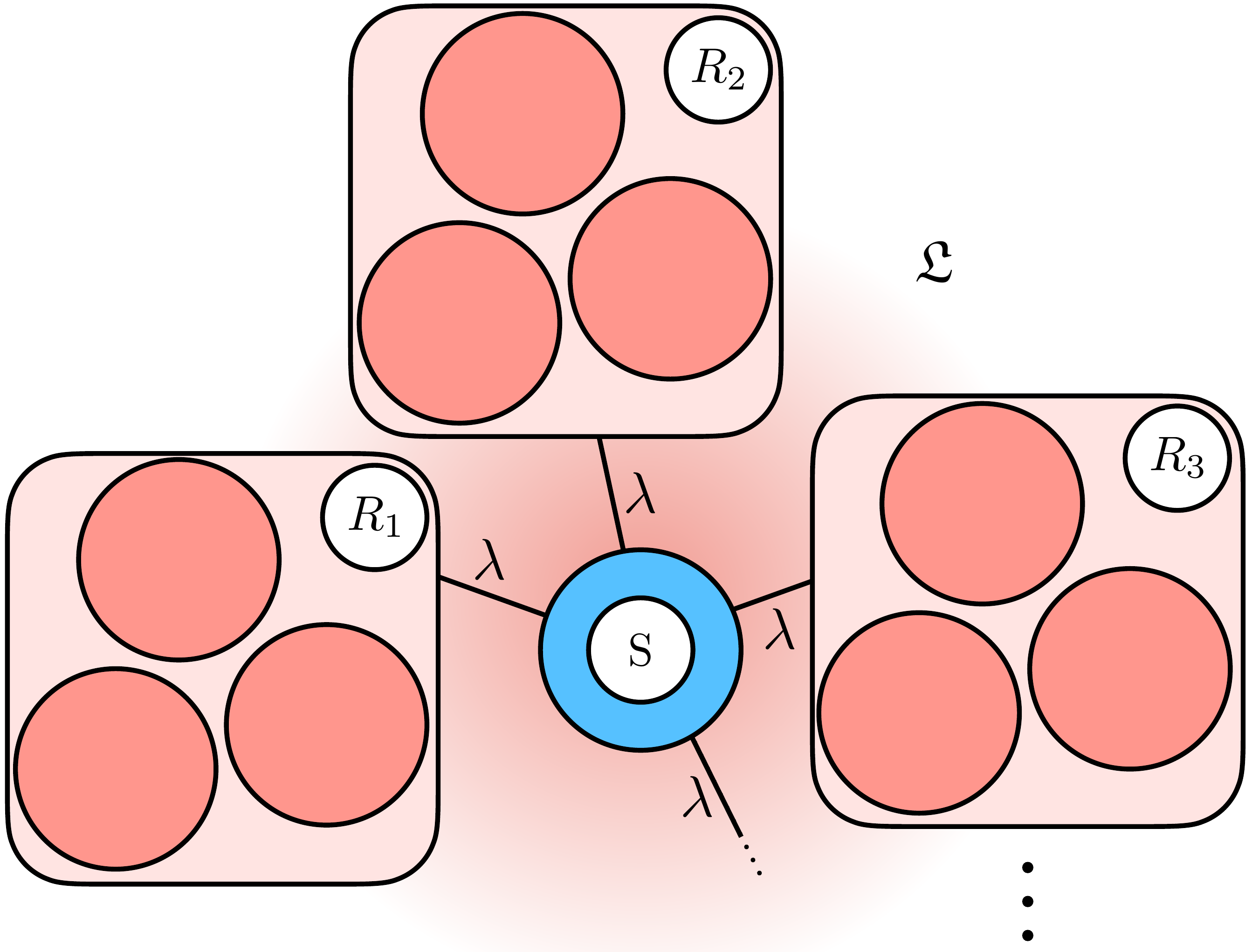}}
  \caption{Sketch of the piecewise non-interacting spin bath. The central system (in blue, label S) is weakly coupled to several regions (in red, labels $R_1, R_2, R_3,\cdots$) that conform a lattice $\mathfrak{L}$. \label{fig:sketch3}} 
\end{figure}

We consider the bath to be embedded in a finite lattice $\mathfrak{L}$. To each lattice site $r\in \mathfrak{L}$ we associate a local Hilbert space of dimension $d(r)$. Then, we partition $\mathfrak{L} = \cup_{R} R$ into regions $R$, and associate to each region $R$ a local Hamiltonian $H_R$ that only involves sites $r\in R$ and has dimension $d(R) = \prod_{r\in R} d(r)$. Then, the bath Hamiltonian
\begin{align}
H_\text{B} = \sum_{R} H_R,\label{eq:loc_ham}
\end{align}
is piecewise non-interacting; i.e., $[H_R,H_{R'}] = 0$. We sketch this scenario in Fig.~\ref{fig:sketch3}. Importantly, the notion of local is not restricted to spatially local. The structure in Eq.~\eqref{eq:loc_ham} could also be, for instance, with respect to momentum space. 

We introduce the notation $\ket{n(R)}$ for the $n^{th}$ excited state of the Hamiltonian $H_R$; that is, $H_R\ket{n(R)} = E_{n(R)}\ket{n(R)}$. Then, the eigenenergies of the bath Hamiltonian are given by $E_\textbf{n} \coloneqq \sum_R E_{n(R)}$, where $\textbf{n}$ is a vector of components $n(R)$. Moreover, we assume that the open quantum system interacts locally with each $H_R$, giving rise to an interaction of the form
\begin{align}
H_\text{int} =\lambda S\otimes  \sum_{R} B_R,\label{eq:loc_int}
\end{align}
where again $B_R$ only involves sites $r\in R$. Despite being restrictive, many well-known models of open quantum systems like the central spin~\cite{Gaudin1976} or the Caldeira-Leggett~\cite{Caldeira1983} model fall in this category.

With a Hamiltonian of the form \eqref{eq:loc_ham} and an interaction of the form \eqref{eq:loc_int}, the computation of the correlation functions simplifies and can be written as a sum of local correlation functions; for instance, $\langle \tilde{B}(\tau) B \rangle_{E'} = \sum_{R} \langle B_R(\tau) B_R \rangle_{E'}$. Moreover, the correlation function
\begin{align}
\langle B_R(\tau) B_R \rangle_{E'} = \text{tr}_R\{B_R(\tau) B_R  \text{tr}_{\bar{R}}[\omega_\text{B}(E')]\},
\end{align}
depends only on the reduced state $\text{tr}_{\bar{R}}[\omega_\text{B}(E')]$ of the region $R$, where $\bar{R}$ is the complementary set of $R$ such that $R\cup \bar{R}=\mathfrak{L}$. The trace over $\bar{R}$ can be now performed as follows. Define the conditional POVM elements
\begin{align}
P{\bm (}E|n(R){\bm )} &\coloneqq \ketbra{n(R)}{n(R)} P(E)\ketbra{n(R)}{n(R)} \nonumber\\
&= \sum_{\bar{\textbf{n}}} W(E|E_{n(R)} +E_{\bar{\textbf{n}}})\ketbra{n(R),\bar{\textbf{n}}}{n(R),\bar{\textbf{n}}},\label{eq:thermal_canonical_principle}
\end{align}
where $\bar{\textbf{n}}$ sums over the complementary components of $n(R)$ and also define the corresponding conditional volumes $V{\bm (}E|n(R){\bm )} = \text{tr}[P{\bm (}E|n(R){\bm )}]$. Then, we arrive at the exact formula
\begin{align}
\text{tr}_{\bar{R}}[\omega_\text{B}(E)] =& \sum_{n(R)} \frac{V{\bm (}E|n(R){\bm )}}{V(E)} \ketbra{n(R)}{n(R)},\label{eq:reduced_state}
\end{align}
that can be used to compute the exactly the function $\kappa(E;\omega)$ as shown in App.~\ref{app:correlation_functions}. 

\subsection{Emergent canonical reduced state}

Despite being formally exact, the expression of the reduced state~\eqref{eq:reduced_state} is not yet transparent. In this subsection we discuss how, under reasonable assumptions, the canonical distribution arises as the reduced description of a large system in a microcanonical state of fixed energy, a well-known argument in statistical mechanics~\cite{Landau2013}. 

To this end, we introduce the Boltzmann entropy ($k_\text{B}=1$)
\begin{align}
\mathcal{S}(E) \coloneqq \log {\bm (}V(E)\delta E {\bm )},\label{eq:boltzmann_entropy}
\end{align}
as well as the Boltzmann inverse temperature
\begin{align}
\beta(E) \coloneqq \frac{\partial}{\partial E} \mathcal{S}(E),\label{eq:boltzmann_temperature}
\end{align}
that corresponds to its derivative. Essentially, provided a sufficiently large bath, one can replace $V(E|n(R))$ by $\exp[\mathcal{S}(E) -\beta(E) E_{n(R)}]/\delta E$ in Eq.~\eqref{eq:reduced_state}, which yields the canonical distribution
\begin{align}
    \text{tr}_{\bar{R}}[\omega_\text{B}(E)] \approx  \sum_{n(R)} \frac{e^{-\beta(E) E_{n(R)}}}{Z_\text{R}{\bm (} \beta(E) {\bm )}}\ketbra{n(R)}{n(R)},\label{eq:canonical_from_mirco}
\end{align} 
where $Z_{R}(\beta) = \sum_{n(R)} \exp(-\beta E_{n(R)})$.

To find the relation~\eqref{eq:canonical_from_mirco} formally, we have to make further assumptions. Namely, (\textit{i}) we assume the weighting function to be a function of only the difference $W(E|E_i) = W(E-E_i)$; (\textit{ii}) we assume that $V(E) \approx V_{\bar{R}}(E) d(R)$, being $V_{\bar{R}}(E)$ the volume of the complementary region, and (\textit{iii}) we assume that the Boltzmann entropy $\mathcal{S}(E) = \log {\bm (}V(E)\delta E {\bm )}$ is a sufficiently smooth function of $E$ so that it can be Taylor expanded to first order in the local energy scale $E_n(R)$. 

Assumption (\textit{i}) is to be expected in many practical cases. For instance, both $W_\text{I}(E|E_i)$ and $W_\text{G}(E|E_i)$ fall in this category. Assumption (\textit{ii}), is expected for large baths where the eigenstate distribution of $\bar{R}$ has reached its limiting value, and attaching to it the extra region $R$ is equivalent to multiplying the limiting distribution by the local dimension $d(R)$. Finally, assumption (\textit{iii}) is also to be expected for large baths with many regions $R$, since the local energies $E_{n(R)}$ are a small contribution to the total energy. Then, with the help of the Boltzmann inverse temperature $\beta(E)$, one can expand $\mathcal{S}(E-E_{n(R)})\approx \mathcal{S}(E)-\beta(E)E_{n(R)}$. 

Putting assumptions (\textit{i})--(\textit{iii}) together, we arrive at the desired result 
\begin{align}
    \text{tr}_{\bar{R}}[\omega_\text{B}(E)] \approx & \sum_{n(R)} \frac{e^{-\beta(E) E_{n(R)}}}{Z_\text{R}{\bm (} \beta(E) {\bm )}}\ketbra{n(R)}{n(R)} =  \pi_R{\bm (}\beta(E){\bm )},\label{eq:canonical_from_mirco_2}
\end{align} 
where $\pi_R(\beta) \propto \exp(-\beta H_R)$ is the thermal state of the region $R$. Finally, we believe that the above assumptions (\textit{i})-(\textit{iii}) are not crucial to the derivation of Eq.~\eqref{eq:canonical_from_mirco_2}, since the thermal state has been shown to arise as the reduced state for the overwhelming majority of quantum states~\cite{Goldstein2006,Popescu2006}.

In the next subsection, we exploit the thermal character of the reduced state of the region $R$ to derive the well-known Kubo-Martin-Schwinger relation. 

\subsection{Kubo-Martin-Schwinger relation}

In quantum statistical mechanics, the Kubo-Martin-Schwinger relation~\cite{Kubo1957,Martin1959} is a property of the two-time correlation functions of a system in thermal equilibrium. Particularizing to the bath coupling operators of a bath in a thermal state, the Kubo-Martin-Schwinger relation yields
\begin{align}
    \text{tr}[\tilde{B}(-\tau-i\beta)^\dagger B \pi_\text{B}(\beta)] = \text{tr}[\tilde{B}(\tau)^\dagger B \pi_\text{B}(\beta)], 
\end{align}
where the left term is evaluated at a time with non-zero imaginary part. In the BMS master equation, the dissipation rate $\kappa(\beta,\omega)$ is obtained as the Fourier transform of the right-hand-side of the above equation, the Kubo-Martin-Schwinger relation implies the local detailed balance condition
\begin{align}
    \kappa(\beta;-\omega) =& \lambda^2 \int_\mathbb{R} d\tau \text{tr}[\tilde{B}(\tau)^\dagger B \pi_\text{B}(\beta)] e^{i\omega\tau} \nonumber\\
    =& e^{-\beta \omega}\kappa(\beta;\omega)~\label{eq:kms_original}.
\end{align}
which, in turn, gives rise to a thermal stationary distribution for the system S. 

In the context of the EMME, we expect a similar relation to hold. Using the approximation~\eqref{eq:canonical_from_mirco_2}, we note that 
\begin{align}
    \sum_R \text{tr}[\tilde{B}_R(\tau)^\dagger B_R \omega_\text{B}(E)] \approx \sum_R \text{tr}_\text{R}[\tilde{B}_R(\tau)^\dagger B_R \pi_R{\bm(}\beta(E){\bm )}]. 
\end{align}
Since each local term $R$ is now a two-point correlation function at equilibrium, the microcanonical analogue of the Kubo-Martin-Schwinger relation is 
\begin{align}
    \sum_R \text{tr}[\tilde{B}_R(-\tau-i\beta(E))^\dagger& B_R \pi_R{\bm(}\beta(E){\bm )}] \nonumber\\
    &= \sum_R \text{tr}[\tilde{B}_R(\tau)^\dagger B_R \pi_R{\bm(}\beta(E){\bm )}]. 
\end{align}
Similarly, the corresponding dissipation rates obtained via the Fourier transform fulfill
\begin{align}
\kappa(E;-\omega) \approx e^{-\beta(E) \omega}\kappa(E;\omega), \label{eq:kms_relation}
\end{align}
which has the form of the local detailed balance condition where the temperature is fixed by the Boltzmann inverse temperature $\beta(E)$.

\section{The central spin system}\label{sec:models}

\floatsetup[figure]{style=plain,subcapbesideposition=top}
\begin{figure}
{\includegraphics[width=0.8\textwidth]{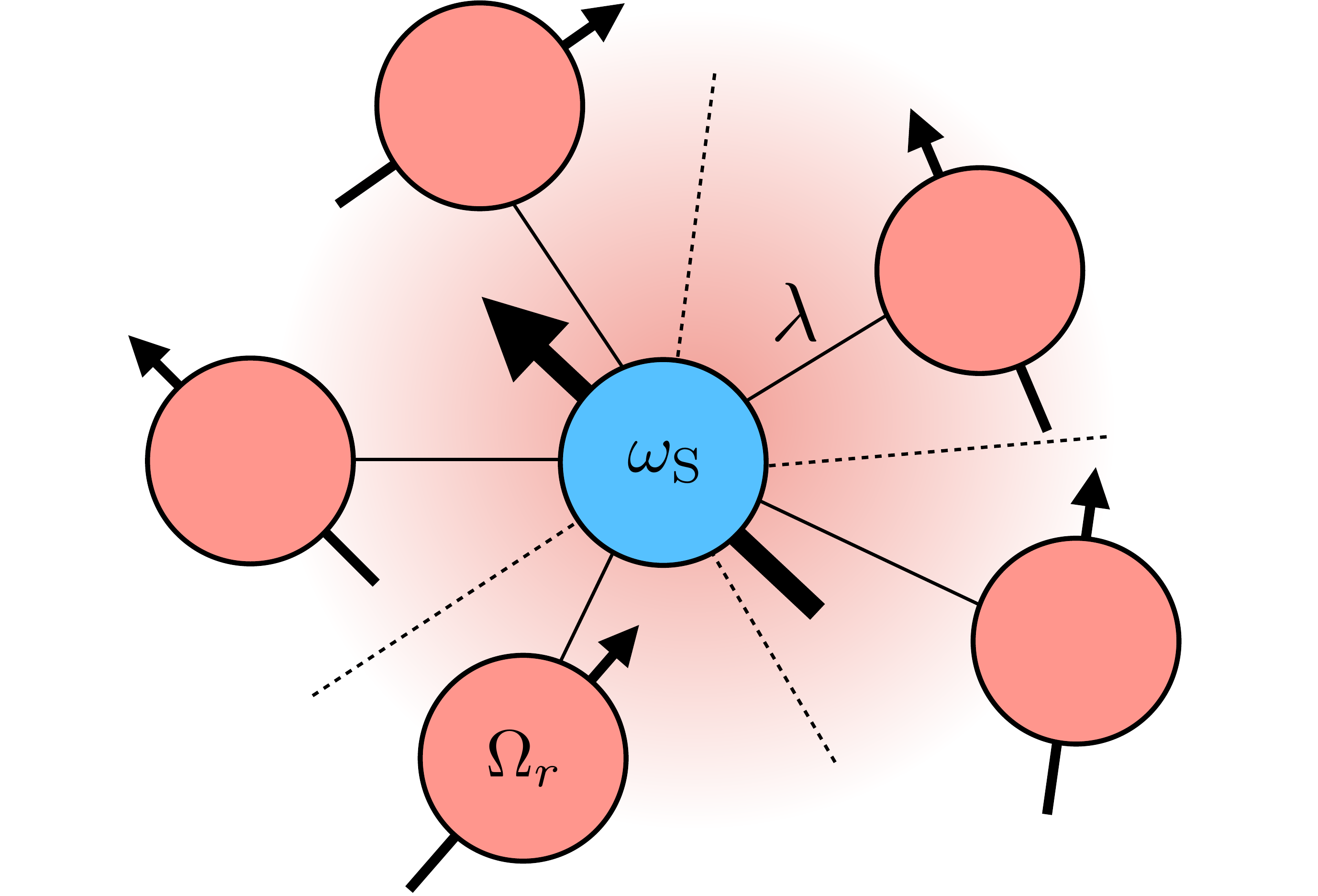}}
  \caption{Sketch of the central spin model, where the central spin has an energy scale $\omega_\text{S}$ and the spins in the bath have different energy splittings $\Omega_r$. The interaction strength between the system and the spins in the environment is of order $\lambda$. \label{fig:central_spin}} 
\end{figure}

In this section, we synthesize the previous two sections by numerically studying the dynamics of the central spin system model \cite{Prokofev2000} (see Fig.~\ref{fig:central_spin} for a sketch). This model represents a central spin-s particle that interacts locally with a collection of surrounding spin particles that act as the bath, as it is realized in platforms such as nitrogen-vacancy centers in diamond~\cite{London2013,Sushkov2014,Scwartz2018} or quantum dots~\cite{Hanson2007,Urbaszek2013}. More recently, this model has been also used to theoretically describe the behavior of the spin degrees of freedom of a polycrystalline solid made of an ensemble of Triphenylphosphine molecules \cite{Niknam2020,Niknam2021}. We focus on the two types of weighting functions corresponding to the indicator $W_\text{I}(E|E_i)$ and the Gaussian $W_\text{G}(E|E_i)$ cases.  

\begin{figure}
{\includegraphics[width=\textwidth]{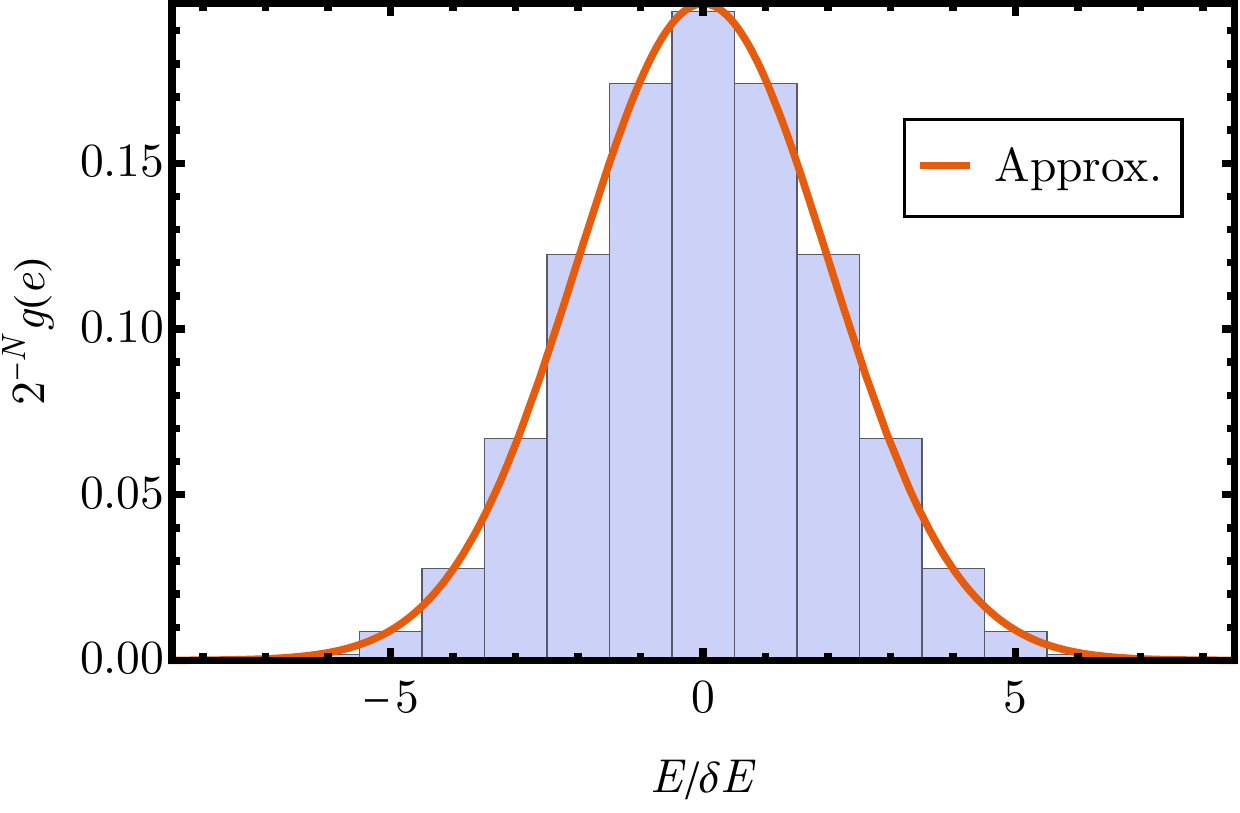}}
  \caption{Histogram of the energy levels of a non-interacting spin bath with random frequencies $\Omega_r$ extracted from a single realization of the Gaussian distribution with mean $\Omega_0$ and variance $\sigma_\Omega = 0.2\Omega_0$ for $N=14$ spins. In orange, the comparison to the normal distribution $\mathcal{N}(e,\sigma_N)$ (solid red line).\label{fig:spectrum}}
\end{figure}

The Hamiltonian of the non-interacting spin bath is microscopically described by
\begin{align}
H_\text{B} = \sum_{r=1}^N \frac{\Omega_r}{2} \sigma_r^z,\label{eq:non-interacting}
\end{align}
where $\sigma_r^{x,y,z}$ are the Pauli matrices and $\Omega_r$ is the Zeeman energy of the $r^{th}$ spin. We consider the frequencies $\Omega_r$ to be sampled from a given underlying probability distribution $p_Z(\Omega)$ of Zeeman energies, with average $\Omega_0$ and variance $\sigma_\Omega$. Its eigenenergies are given by
\begin{align}
E_{\textbf{n}} \coloneqq \sum_{r=1}^N n_r \frac{\Omega_r}{2},\label{eq:energies}
\end{align}
where $\textbf{n} = (n_1,\cdots,n_N)$ has components $n_r\in \{-1,1\}$. Interestingly, every energy $E_\textbf{n}$ can be regarded as the endpoint of a random walk of $N$ steps and irregular step sizes $\Omega_r/2$. In that scenario, the central limit theorem applies (see App.~\ref{app:lindeberg}) and the distribution of end points, and consequently of energies $E_\textbf{n}$, is given by a normal distribution of variance $\sigma_N^2$
\begin{align}
\sigma^2_N = \sum_{r=1}^N \frac{\Omega_r^2}{4}.
\end{align}
Then, it is possible to approximate the density of states of the bath 
\begin{align}
g(e) \coloneqq \sum_{\textbf{n}}\delta(e-E_\textbf{n}) \approx  \frac{2^N}{\sqrt{2\pi} \sigma_N} \exp\left(-\frac{e^2}{2 \sigma_N^2 }\right).\label{eq:gaussian_fit}
\end{align}
In Fig.~\ref{fig:spectrum}, we compare the histogram of the exact spectrum with the Gaussian fit in Eq.~\eqref{eq:gaussian_fit}, showing a very good agreement.

\floatsetup[figure]{style=plain,subcapbesideposition=top}

\subsection{Volume terms and correlation functions}\label{subsec:volumes}

\floatsetup[figure]{style=plain,subcapbesideposition=top}
\begin{figure}
{\includegraphics[width=\textwidth]{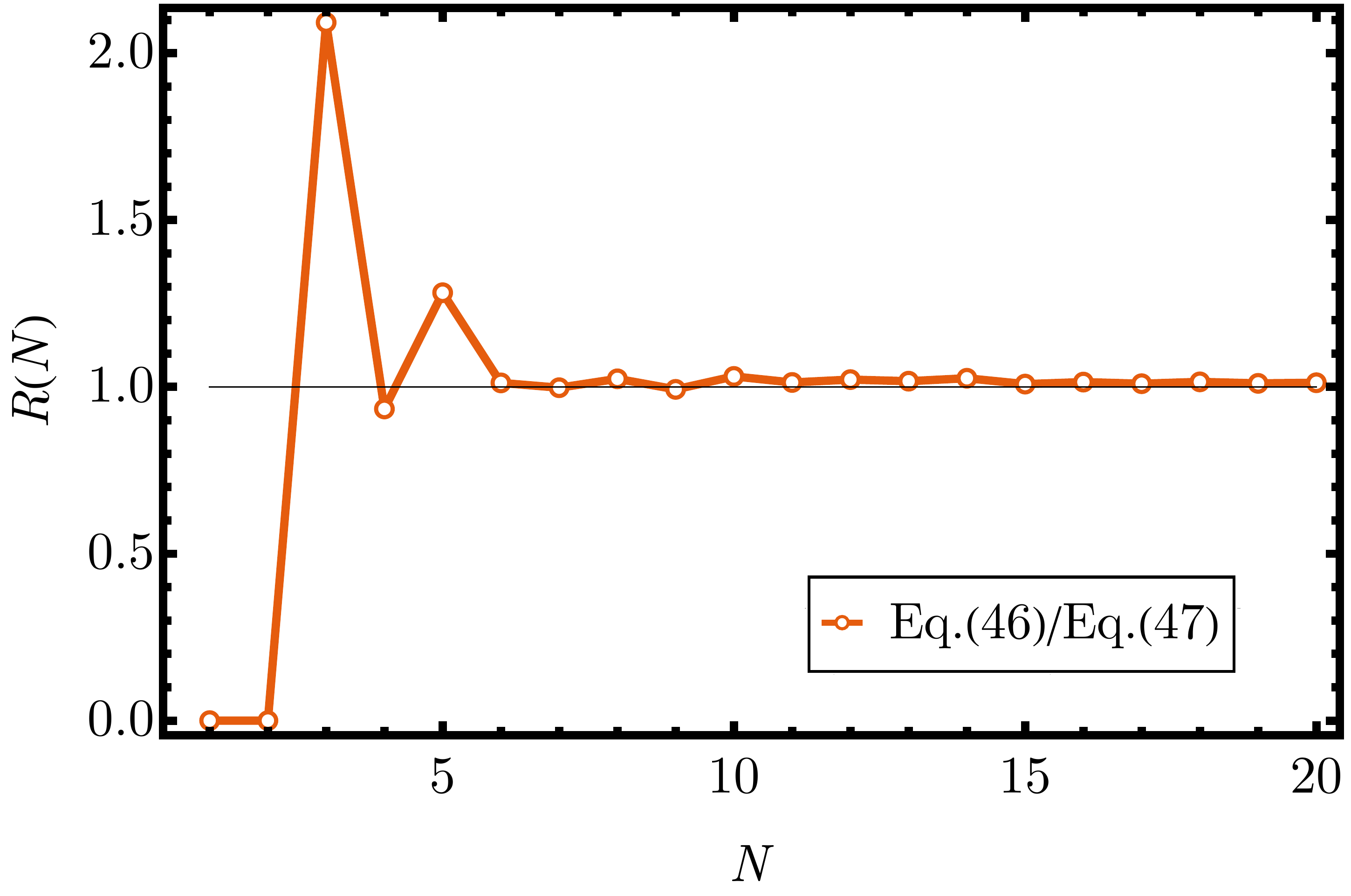}}
  \caption{Ratio $R(N)$ between the exact value of $f(E,E';\Omega)$ in Eq.~\eqref{eq:exact_f} and its approximated value as shown in Eq.~\eqref{eq:approx_f} as a function of the number of spins $N$, whose frequencies are extracted from a single realization of the distribution $p_Z(\Omega) = \mathcal{N}(\Omega-\Omega_0,\sigma_\Omega)$. The parameters are given in terms of the coarse-graining $\delta E$ and correspond to $\Omega_0=\delta E$, $\sigma_\Omega = 0.2\delta E$, $\Omega = \Omega_0$, $E = -\delta E$ and $E' = -2\delta E$. \label{fig:ratio}} 
\end{figure}

\floatsetup[figure]{style=plain,subcapbesideposition=top}
\begin{figure*}
{\includegraphics[width=\textwidth]{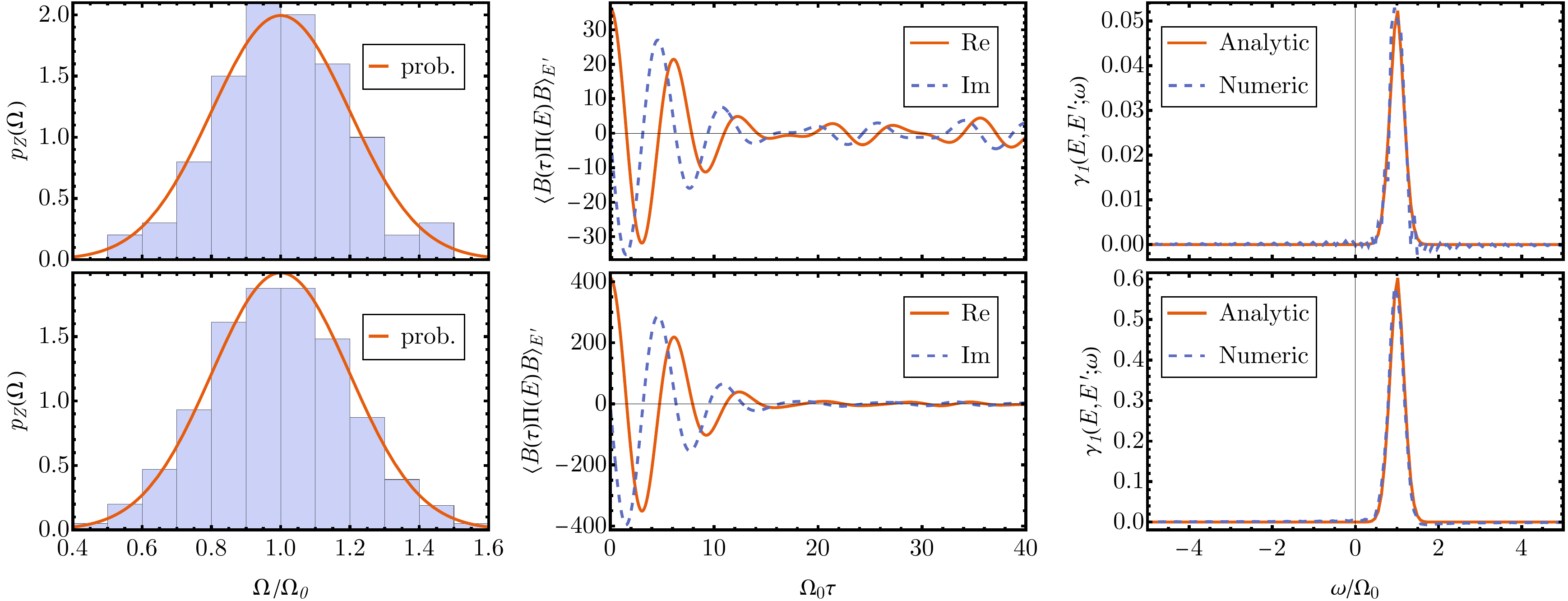}}\\
\caption{First column: Histogram of the Zeeman energies and fit to the underlying probability distribution $\mathcal{N}(\Omega-\Omega_0,\sigma_\Omega)$ (orange solid line). Second column: real (orange solid line) and imaginary (blue dashed line) parts of the correlation function $\langle B(\tau) \Pi(E) B\rangle_{E'}$ with the weighting function $W_\text{I}(E|E_i)$. Third column: analytic expression of the dissipation rates in Eq.~\eqref{eq:gamma_vs_f} (solid orange line) versus its numerical value computed using the numeric Fourier transform of the second column (blue dashed line). The particle numbers are $N=100$ (first row), and $N=1000$ (second row). In the three columns, each frequency $\Omega_r$ is extracted from a single-shot realization of $p_Z(\Omega) = \mathcal{N}(\Omega-\Omega_0,\sigma_\Omega)$ with $\sigma_\Omega = 0.2\Omega_0$. For the second and third column we have chosen $\delta E = \Omega_0$, the energies $E = -8\delta E$ and $E' = -9 \delta E$, and fixed $c_r = c_0 = 1$. In the third column we have set $\lambda = 0.01\delta E$. \label{fig:decay_correlations}}
\end{figure*}

A crucial ingredient to describe the evolution of a system using the EMME is having access to the volume terms $V(E)$. In terms of the density of states $g(e)$, the volumes can be written as
\begin{align}
V(E) = \int de W(E|e) g(e),\label{eq:volumes_integral}
\end{align}
We can find closed expressions for the volume terms in the cases of $W_\text{I}(E|E_i)$ and $W_\text{G}(E|E_i)$. They read respectively
\begin{align}
&V_\text{I}(E) \approx 2^{N-1}\left[\text{erf}\left(\frac{E+\delta E/2}{\sqrt{2}\sigma_N}\right)-\text{erf}\left(\frac{E-\delta E/2}{\sqrt{2}\sigma_N}\right)\right],\nonumber\\
&V_\text{G}(E)\approx \frac{2^N}{\sqrt{2\pi(\delta E^2 + \sigma_N^2)}}\exp\left[-\frac{E^2}{2 (\delta E^2 +\sigma_N^2)}\right],\label{eq:volumes_noninteracting}
\end{align}
where $\text{erf}$ is the error function, and which become equivalent in the limit $\delta E\to 0$.

Provided the analytical expressions for the volume $V_\text{I}(E)$ and volume density $V_\text{G}(E)$, we can also compute analytically the Boltzmann entropy $\mathcal{S}_\text{B}(E)$ by simply taking the logarithm. In particular, for the Gaussian volume $V_\text{G}(E)$, we find the linear relation $\beta(E) = -E/\sigma_{N}^2$, between the Boltzmann temperature and the energy $E$. The same relation also holds for $V_\text{I}(E)$, provided that $\delta E/(\sqrt{2}\sigma_{N})$ is small enough to Taylor expand the error function $\text{erf}(x+\delta x) \approx \text{erf}(x) + \exp(x^2) \delta x/\sqrt{\pi}$. Interestingly, the microcanonical heat capacity $\mathcal{C}(E) \coloneqq -\beta(E)^2 [\partial_E \beta(E)]^{-1}$ for this model turns out to be simply $\mathcal{C}(E) = \beta(E)^2\sigma_{N}^{2}$ which, as expected, is extensive with the number of spins $N$. 

Equipped with the expression of the volume terms, we proceed to compute the bath correlation functions for the Hamiltonian~\eqref{eq:non-interacting} and the interaction
\begin{align}
H_\text{int} = \lambda S\otimes \sum_{r=1}^N c_r \sigma^x_r.\label{eq:sb-interaction}
\end{align}
To this end, we note that $H_\text{B}$ and $H_\text{int}$ have the form given by~\eqref{eq:loc_ham} and~\eqref{eq:loc_int} respectively. In particular, the regions $R$ contain a single site $r$ and the corresponding Hamiltonians $H_r = \Omega_r \sigma_r^z/2$ have dimension $d(R) = d(r) = 2$. Therefore, following the discussion of Sec.~\ref{sec:noninteracting} and App.~\ref{app:correlation_functions}, we find
\begin{align}
\langle B(\tau) P(E) B\rangle_{E'} =&  \frac{1}{V(E')} \left( \sum_{r=1}^N |c_r|^2e^{-i\Omega_r \tau} f(E,E';\Omega_r)\right.\nonumber\\
&+ \left.\sum_{r=1}^N  |c_r|^2 e^{i\Omega_r t} f(E,E';-\Omega_r)\right).\label{eq:nonint_correlation} 
\end{align}
In Eq.~\eqref{eq:nonint_correlation} we have introduced the function
\begin{align}
f(E,E';\Omega_r) = \sum_{\bar{\textbf{n}}}W(E|E_{\bar{\textbf{n}}}+\Omega_r/2) W(E'|E_{\bar{\textbf{n}}}-\Omega_r/2),\label{eq:exact_f}
\end{align}
that has the symmetry $f(E,E';\Omega_r) = f(E',E;-\Omega_r)$. The exact computation of $f(E,E';\Omega_r)$ is in general complicated. However, we can make use of the density of states $g(e)$ in Eq.~\eqref{eq:gaussian_fit} to approximate
\begin{align}
f(E,E';\Omega_r) \approx \int de \frac{g(e)}{2} W\left(E\left|e+\frac{\Omega_r}{2}\right.\right) W\left(E'\left|e-\frac{\Omega_r}{2}\right.\right).\label{eq:approx_f}
\end{align}
where we have used that, for a sufficiently large $N$, removing a particle approximately scales down $g(e)$ by a factor of two. In Fig.~\ref{fig:ratio}, we show the ratio $R(N)$ of the exact value of $f(E,E';\Omega)$ in Eq.~\eqref{eq:exact_f} over its approximated value as computed with Eq.~\eqref{eq:approx_f} as a function of the number of spins $N$. We observe that, for a relatively small number of particles $N\sim 10$, it is justified to use Eq.~\eqref{eq:approx_f} to evaluate the function $f(E,E';\Omega_r)$. 


In the derivation of the EMME, we have used that the correlation functions decay rapidly in time. This approximation is crucial to obtain a time-local equation for the evolution of $\rho_\text{S}(E)$ and thus, its validity has to be assessed. With the help of $f(E,E';\Omega_r)$ it is possible to evaluate efficiently the correlation function $\langle B(\tau) P(E) B \rangle_{E'}$ for a large number of particles. In Fig.~\ref{fig:decay_correlations} we show the decay of the correlation functions for a non-interacting spin-bath using the weighting function $W_I(E|E_i)$. We explore the particle numbers $N=100$ and $N=1000$ corresponding to the first and second row, respectively. In the first column, we show the histogram of Zeeman energies $\{\Omega_r\}$ of a particular bath realization, which we compare with the underlying probability distribution $p_Z(\Omega) = \mathcal{N}(\Omega-\Omega_0,\sigma_\Omega)$ (solid orange line). As expected, increasing the particle number $N$ leads to a better agreement between the underlying distribution and the actual realization. In the second column of Fig.~\ref{fig:decay_correlations}, we show the decay of the correlation functions as a function of time for a particular choice of the energies $E$ and $E'$. We observe that the correlation functions rapidly decay in time compared to the system relaxation time, which is an indicator for the validity of the Markovian approximation.

Ultimately, we are interested in computing the dissipation rates; e.g. $\gamma_1(E,E';\Omega)$, that enter the EMME. If the number of spins of the bath is very large and their splittings $\Omega_r$ densely fill a spectral region, the dissipation rates are conveniently written in terms of the spectral density $J(\Omega) = 2\pi \lambda^2\sum_{r} c_r^2 \delta (\Omega-\Omega_r)$ defined for $\Omega \geq 0$. Continuing $J(\Omega)$ towards negative frequencies as $J(-\Omega) = J(\Omega)$ we find the relation
\begin{align}
\gamma_1(E,E';\Omega) = J(\Omega) f(E,E';\Omega)/V(E').\label{eq:gamma_vs_f}
\end{align}
We note that, for a spin-independent coupling $c_r = c_0$ for all $r$, the spectral density is linked to the distribution of Zeeman energies through
\begin{align}
J(\Omega) = 2\pi\lambda^2 c_0^2 N \left[\frac{1}{N}\sum_{r=1}^N \delta(\Omega-\Omega_r)\right],
\end{align}
where the term between brackets converges to $p_Z(\Omega)$ as $N$ tends to infinity. In the last column of Fig.~\ref{fig:decay_correlations}, we compare the numerical value of the dissipation rates $\gamma_1(E,E';\Omega)$, obtained via a numeric Fourier transform, versus its analytic value in Eq.~\eqref{eq:gamma_vs_f}. We find a good agreement between both expressions even for $N=100$, which improves with increasing $N$.

\subsection{Hierarchy of master equations for the central spin system}\label{subsec:comparison}

\floatsetup[figure]{style=plain,subcapbesideposition=top}
\begin{figure*}
{\includegraphics[width=0.66\textwidth]{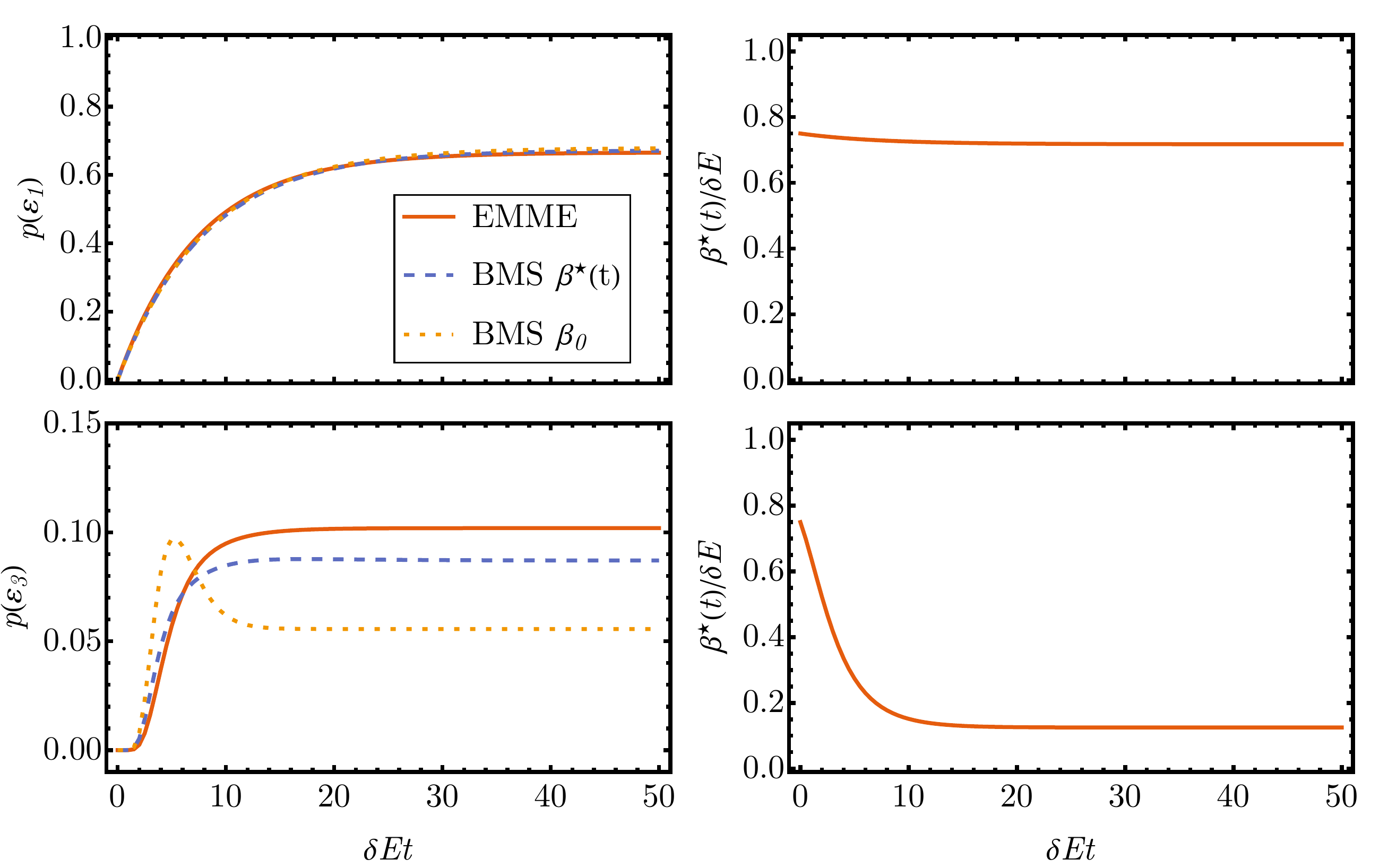}}\\
\caption{(left-column) Comparison between the EMME dynamics (solid orange line), the BMS with the effective nonequilibrium temperature $\beta^\star(t)$ (blue dashed line), and the BMS at fixed temperature $\beta_0$ (yellow dotted line) for: (top-row) the first excited state of a spin-1/2 particle and (bottom-row) the third excited state of a spin-10 particle; (right-column) corresponding evolution of the effective non-equilibrium temperature (solid orange line). Initially, the system-bath composite is found in the state $\rho(0) = \ketbra{2s}{2s}\otimes\pi_\text{B}(\beta_0)$ with $\beta_0 = 0.75 \delta E$. We take the parameters $N = 100$ spins, $\omega_\text{S} = \delta E$, $\lambda = 0.01\delta E$, $\Omega_0 = \delta E$, $\sigma_\Omega = 0.2\delta E$, and a spin independent coupling $c_r\to 1$.  \label{fig:evolution}}
\end{figure*}

\floatsetup[figure]{style=plain,subcapbesideposition=top}
\begin{figure*}
{\includegraphics[width=0.66\textwidth]{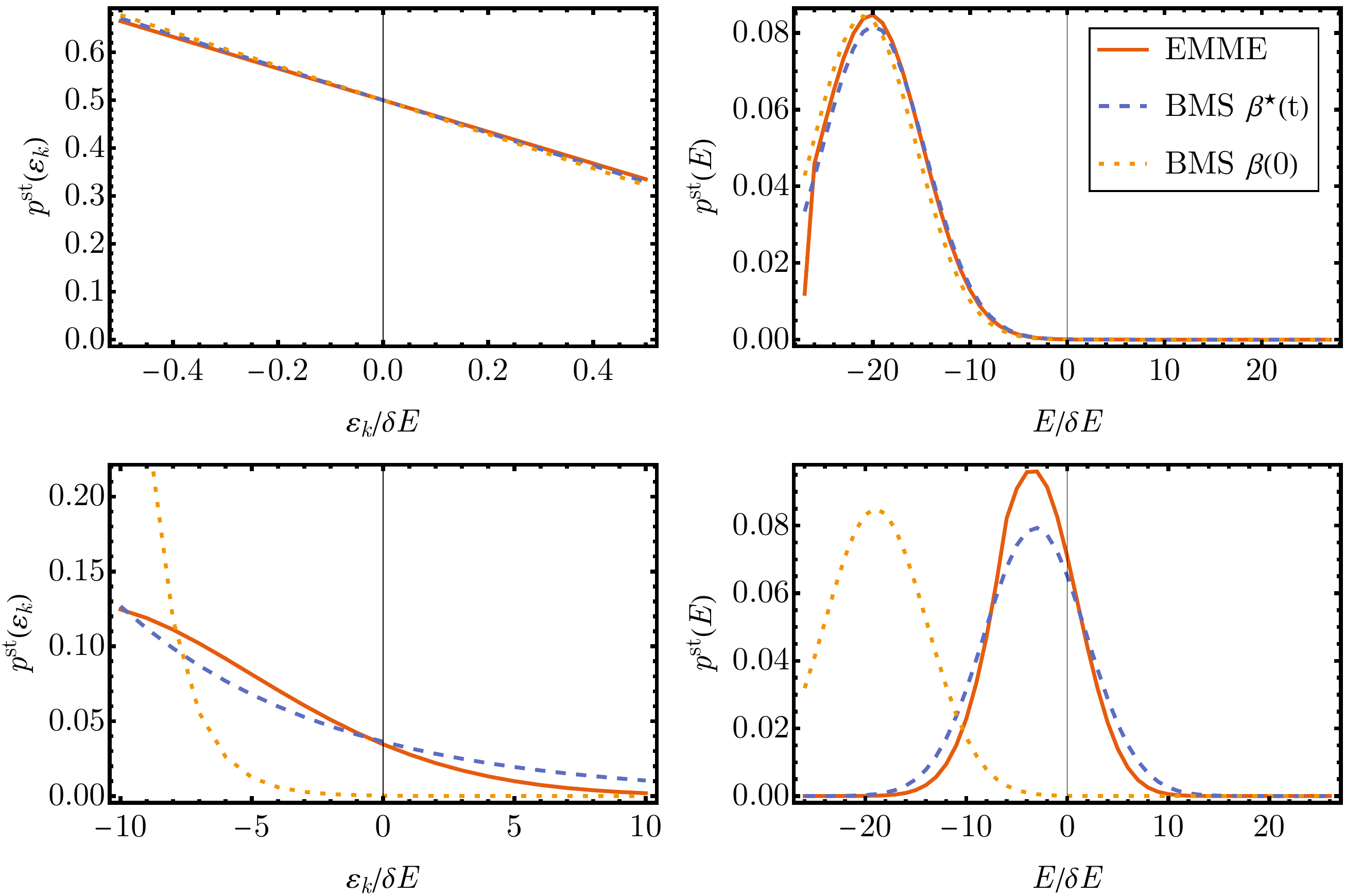}}\\
\caption{(left-column) Comparison between the stationary distributions of the system $p^\text{st}(\varepsilon_k)$ and the bath $p^\text{st}(E)$ as predicted for the EMME dynamics (solid orange line), the BMS with the effective nonequilibrium temperature $\beta^\star(t)$ (blue dashed line), and the BMS at fixed temperature $\beta_0$ (yellow dotted line). (top-row) The first excited state of a spin-1/2 particle and (bottom-row) the third excited state of a spin-10 particle. We take the initial state and parameters as in Fig.~\ref{fig:evolution}. \label{fig:stationary}}
\end{figure*}

Finally, we compare the dynamics generated by the EMME and those generated by the BMS with and without the effective time-dependent temperature $\beta^\star(t)$. At this point, we have to specify the system Hamiltonian $H_\text{S}$ and the system interaction $S$. We consider a particle of spin-s with $H_\text{S} = \omega_S S^z$ and $S = 2 S^x$, being $\omega_\text{S}$ the central spin frequency, and $S^x$ and $S^z$ are the central spin operators. In the energy eigenbasis the spin operators read
\begin{align}
    &S^z = \sum_{k=0}^{2\text{s}}(k-\text{s})\ketbra{k}{k},\nonumber\\
    &2 S^x = \sum_{k=0}^{2\text{s}-1} \sqrt{(k+1)(2s-k)}\ketbra{k}{k+1} + \text{h.c.}
\end{align}
For instance, for a spin-1/2 particle, we have $H_\text{S} = \omega_S \sigma^z/2$ and $S = \sigma^x$, being  $\sigma^z$ and $\sigma^x$ the standard Pauli operators of the central spin. For simplicity, we consider only the weighting function $W_\text{I}(E|E_i)$ where the energies $E$ can only take values $E \in \{ E = m \delta E\}$ with $m\in \mathbb{Z}$; i.e, we coarse-grain the bath into non-overlapping energy windows. With this choice of weighting function, we have the relation $\gamma_2(E,E';\omega) = \delta_{E,E'} \sum_{E''} \gamma_1(E'',E;\omega)$.

It is convenient to gather the probabilities $p(\varepsilon_k,E)$ into the probability vector $\textbf{p}$ and to define the stochastic matrix $\Lambda$ with off-diagonal elements $\Lambda(\varepsilon_k,E;\varepsilon_q,E') = |\langle k|S|q\rangle|^2 \gamma_1(E,E';\omega_{qk})$, and the diagonal ones determined by the condition $\sum_{k,E}\Lambda(\varepsilon_k, E;\varepsilon_q,E')=0$, which guarantees probability conservation. Then, we can compactly write
\begin{align}
\partial_t \textbf{p} =& \Lambda \mathbf{p}.\label{eq:compact_rate_equation}
\end{align}
Using the results of Sec.~\ref{subsec:volumes}, we explicitly find the off-diagonal elements
\begin{align}
\Lambda(\varepsilon_k,E;\varepsilon_q,E') = |\langle k| S |q\rangle|^2 J(\varepsilon_{q}-\varepsilon_k) \frac{f(E,E';\varepsilon_{q}-\varepsilon_k)}{V(E')}.
\end{align}
We are now at the position to numerically solve the dynamics of the central spin system.

\floatsetup[figure]{style=plain,subcapbesideposition=top}
\begin{figure*}
{\includegraphics[width=0.66\textwidth]{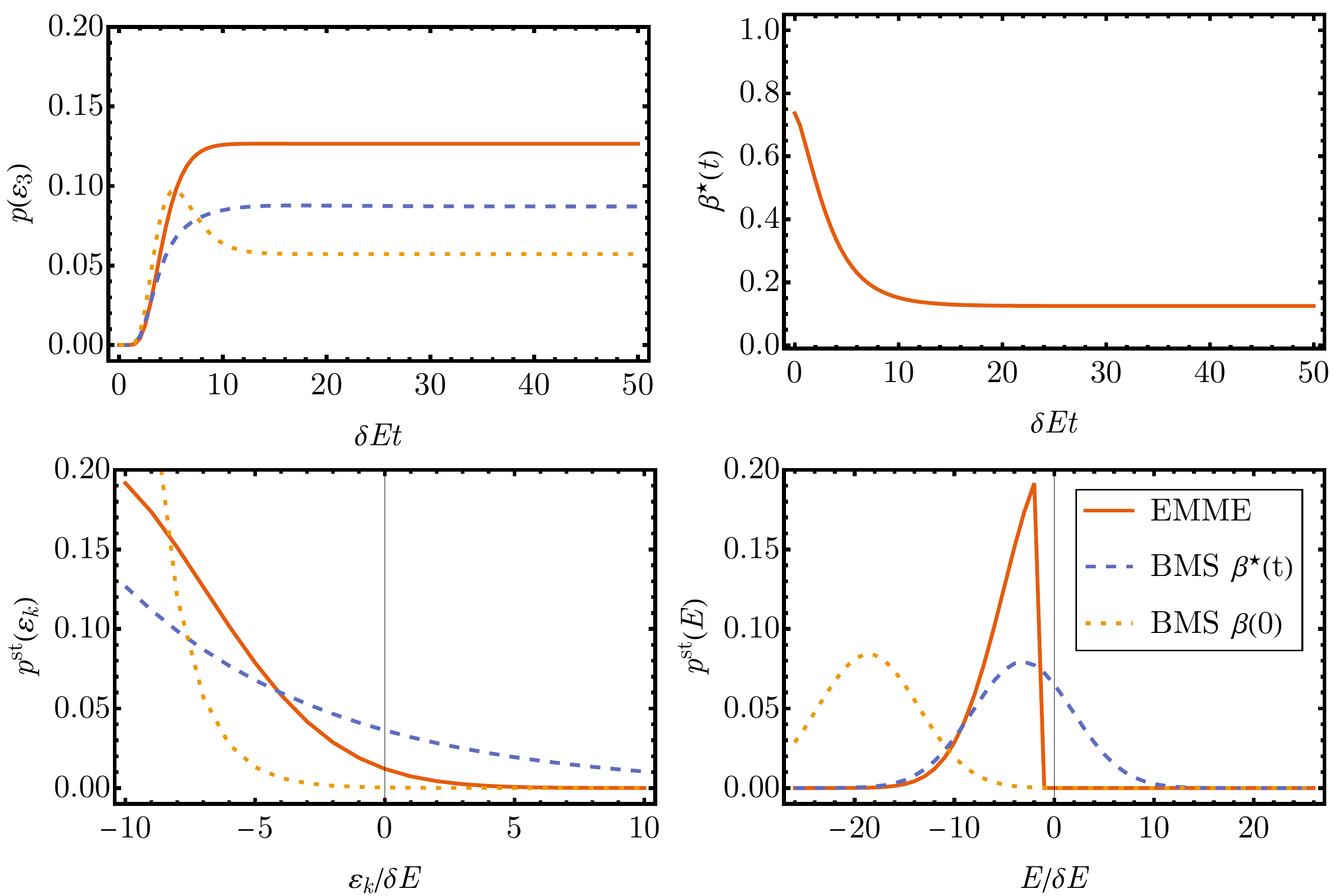}}\\
\caption{Comparison between the EMME dynamics (solid orange line), the BMS with the effective nonequilibrium temperature $\beta^\star(t)$ (blue dashed line), and the BMS at fixed temperature $\beta_0$ (yellow dotted line) for a spin-10 particle;. (Top row) (left) comparative dynamics of the third excited state and (right) corresponding evolution of the effective non-equilibrium temperature (solid orange line). (Bottom row) stationary value of the (left) system and (right) bath energy distributions. We take the initial bath state to be $\omega_\text{B}(E = -18\delta E)$ and the rest of parameters as in Fig.~\ref{fig:evolution}. \label{fig:evo_stat_mic}}
\end{figure*}

In the first column of Fig.~\ref{fig:evolution} we show a comparison between the EMME dynamics, the BMS with the nonequilibrium temperature $\beta^\star(t)$, and  the BMS at a fixed inverse temperature $\beta_0$. In the first row, we consider an initially excited spin-1/2 particle in contact with a bath of $N=100$ spin-1/2 particles; that is, we are in the limit when the relative size of the system is much smaller than the size of the bath. In this limit, the dynamical prediction of these methods is approximately the same. This is to be expected, since the BMS master equation is derived assuming an infinite bath. In the top-right panel of Fig.~\ref{fig:evolution}, we show the evolution of the effective inverse temperature $\beta^\star(t)$, which is approximately constant consistently with the fact that the infinite bath approximation holds. A different behavior is shown in the second row of Fig.~\ref{fig:evolution}, where we consider the central spin to be a spin-10 particle, still with a spin bath of $N=100$ spin-1/2 particles. In this case, the infinite bath assumption is no longer correct, and the three methods lead to different dynamical predictions. Nonetheless, as it can be shown in the bottom-left panel of Fig.~\ref{fig:evolution}, the BMS with the effective time-dependent temperature $\beta^\star(t)$ approximates much better the reduced system dynamics. Accordingly, in the bottom-right panel we observe that the effective temperature $\beta^\star(t)$ can no longer be approximated by a constant, showing that the infinite bath approximation does not hold.

After a sufficiently long time, the joint distribution $p(\varepsilon_k, E)$ reaches its stationary value $p^\text{st}(\varepsilon_k,E)$. In Fig.~\ref{fig:stationary}, we show the stationary system (left column) and bath (right column) reduced energy distributions corresponding to the dynamics of Fig.~\ref{fig:evolution}, that we denote by $p^\text{st}(\varepsilon_k)$ and $p^\text{st}(E)$, respectively. Since the BMS master equation gives no information about the bath energy distribution, it is assumed to be in a canonical state at the final bath temperature $\beta^\star(t\to\infty)$ or at initial temperature $\beta_0$. In the first row of Fig.~\ref{fig:stationary}, we show the stationary distribution corresponding to a central spin with $\text{s}=1/2$ and a bath of $N = 100$ spin-1/2 particles. Again, we observe that in this limit the stationary distributions of, both the system and the bath, are in good agreement for the three methods. Instead, noticeable differences are observed for the larger central spin with $\text{s}=10$ as it is shown in the second row of Fig.~\ref{fig:stationary}. The EMME and the BMS with the time-dependent temperature lead to similar system energy distributions $p(\varepsilon_k)$, while keeping the temperature fixed in the BMS to its initial value $\beta_0$ leads to a completely different stationary state. We understand the differences as follows. When the evolution starts, the initially excited system starts dissipating energy into the bath in the form of heat. On one hand, the difference between the EMME and the BMS at constant inverse temperature $\beta_0$ comes from the fact that the energy contribution from the system changes the bath \textit{average} energy by a non-negligible amount. On the other hand, the smaller discrepancy between the EMME and the BMS with time-dependent inverse temperature $\beta^\star(t)$ arises from the difference in the higher moments of the energy distribution, since their corresponding stationary distributions share the same average energy.  

Finally, in order to highlight the contribution of the different bath energy distributions, we consider the scenario in which the bath is initially in the microcanonical state $\omega_\text{B}(E)$ in Fig.~\ref{fig:evo_stat_mic}. In particular, we compare the dynamics of the EMME with the initial microcanonical state $\omega_\text{B}(E)$ at energy $E = -18\delta E$, with the one obtained by the BMS with the nonequilibrium temperature $\beta^\star(t)$ and the initial condition $\beta^\star(0) \approx 0.73\delta E$, which is the corresponding effective nonequilibrium temperature to the average energy $E=-18\delta E$. Even in this case, where the bath energy distributions $p(E)$ are very different, the BMS with the nonequilibrium temperature $\beta^\star(t)$ reproduces better the dynamics of the EMME. We expect more significant differences between the two approaches in scenarios where the density of states of the bath $g(e)$ changes rapidly as compared to the energy scale of the system $\omega_\text{S}$.

In App.~\ref{app:further_results}, we provide additional results regarding the central spin model. In particular, we show that under reasonable approximations the function $\kappa(E,\omega)$ becomes a linear function of the energy $E$, which is important for the effective nonequilibrium temperature $\beta^\star(t)$ to be optimal in the sense discussed in Sec.~\ref{subsec:hierarchy}. We also observe such linear behavior of $\kappa(E,\omega)$ numerically. Moreover, we show that the matrix $\Lambda$ displays a block structure in agreement with the theoretical discussion of Ref.~\cite{Riera-Campeny2021}.

\subsection{System-bath correlations}

\floatsetup[figure]{style=plain,subcapbesideposition=top}
\begin{figure}
{\includegraphics[width=0.95\textwidth]{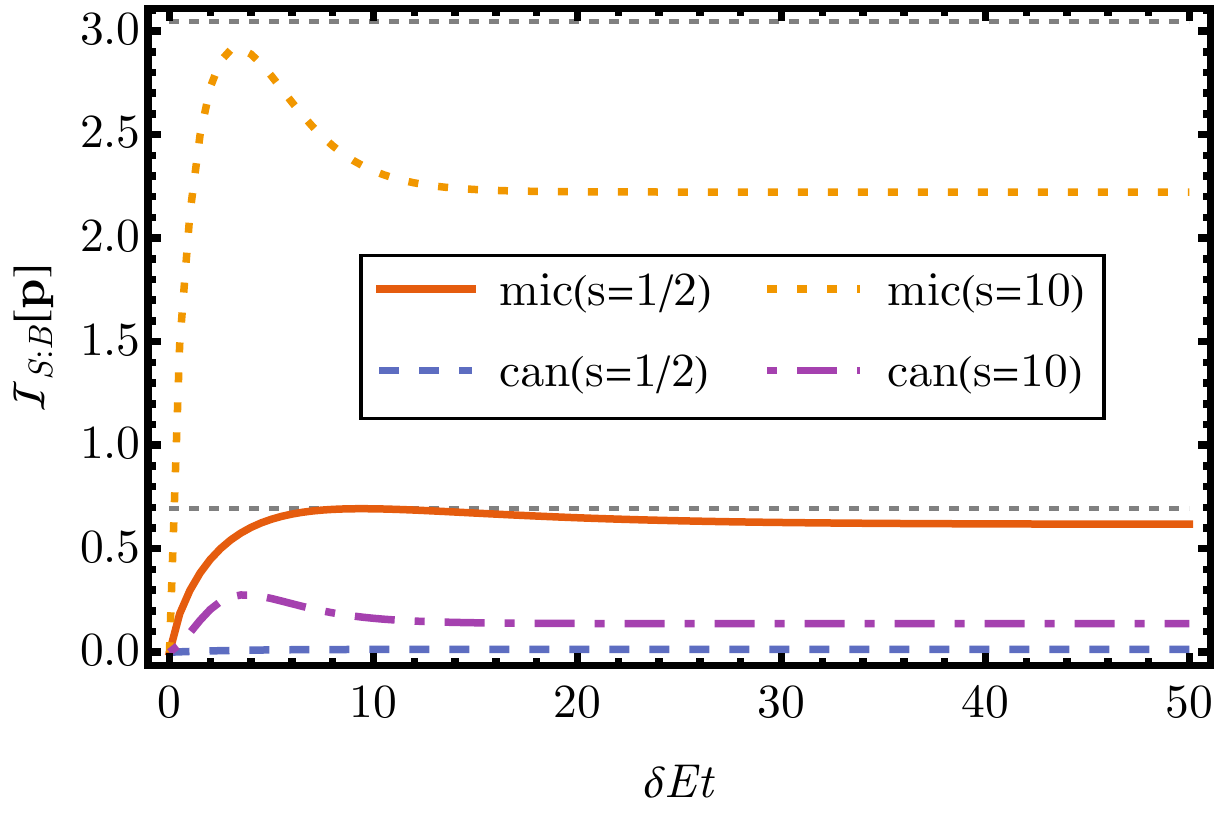}}\\
\caption{Mutual information~\eqref{eq:mutual_information} for a state of the bath in a microcanonical state $\omega_\text{B}(E)$ with $E= -18\delta E$ and corresponding Boltzmann temperature $\beta(E)\approx 0.73\delta E^{-1}$ (orange solid and yellow dotted lines); or a canonical state at initial inverse temperature $\beta_0 = 0.75\delta E^{-1}$ (blue dashed and purple dot-dashed lines). The central particle has a spin of $\text{s}=1/2$ (solid orange and blue dashed lines) or $\text{s}=10$ (yellow dotted and purple dot-dashed lines). The horizontal gray dashed lines correspond to the maximal value of the mutual information $\log(2\text{s}+1)$ for $\text{s}=1/2$ and $\text{s}=10$. The rest of the parameters are set to ${N}=100$, $\omega_\text{S}=\delta E$, $\Omega_0=\delta E$, $\sigma_\Omega = 0.2\delta E$, $c_r=1$, and $\lambda = 0.01\delta E$. \label{fig:correlations}}
\end{figure}

One of the special features of the EMME is that it is capable to describe the evolution of part of the system-bath correlations. Those correlations can be present either in the initial system-bath state $\rho(0)$ or can develope during the system-bath evolution. It is sometimes assumed that, in the weak-coupling limit, system-bath correlations are negligible. Here, we briefly investigate whether this is the case for the central spin model. To this end, we introduce the always positive mutual information
\begin{align}
    \mathcal{I}_{\text{S}:\text{B}}(\mathbf{p}) \coloneqq \sum_{k,E} p(\varepsilon_k,E) \log\left[\frac{p (\varepsilon_k,E)}{p(\varepsilon_k)p(E)}\right] \geq 0.\label{eq:mutual_information}
\end{align}
corresponding to the non-overlapping energy windows with $\{E = m\delta E\}$ and $m\in\mathbb{Z}$. The mutual information is zero if and only if the system state is uncorrelated from the coarse-grained bath energy, and it is upper bounded by $\log d_\text{S}$, where $d_\text{S}$ is the system dimension. 

In Fig.~\ref{fig:correlations}, we show the evolution of $\mathcal{I}^{\text{S}:\text{B}}(\mathbf{p})$ as a function of time. We observe that if the initial state of the bath is canonical $\pi_\text{B}(\beta)$, system-bath correlations remain small throughout the evolution. However, if the bath starts in a microcanonical state $\omega_\text{B}(E)$, system-bath correlations grow close to their maximum possible value. Therefore, system-bath correlations can grow close to their maximum value even in the weak-coupling limit due to energy conservation, in agreement with the discussion in~\cite{Riera-Campeny2021}.

\begin{figure*}
{\includegraphics[width=0.99\textwidth]{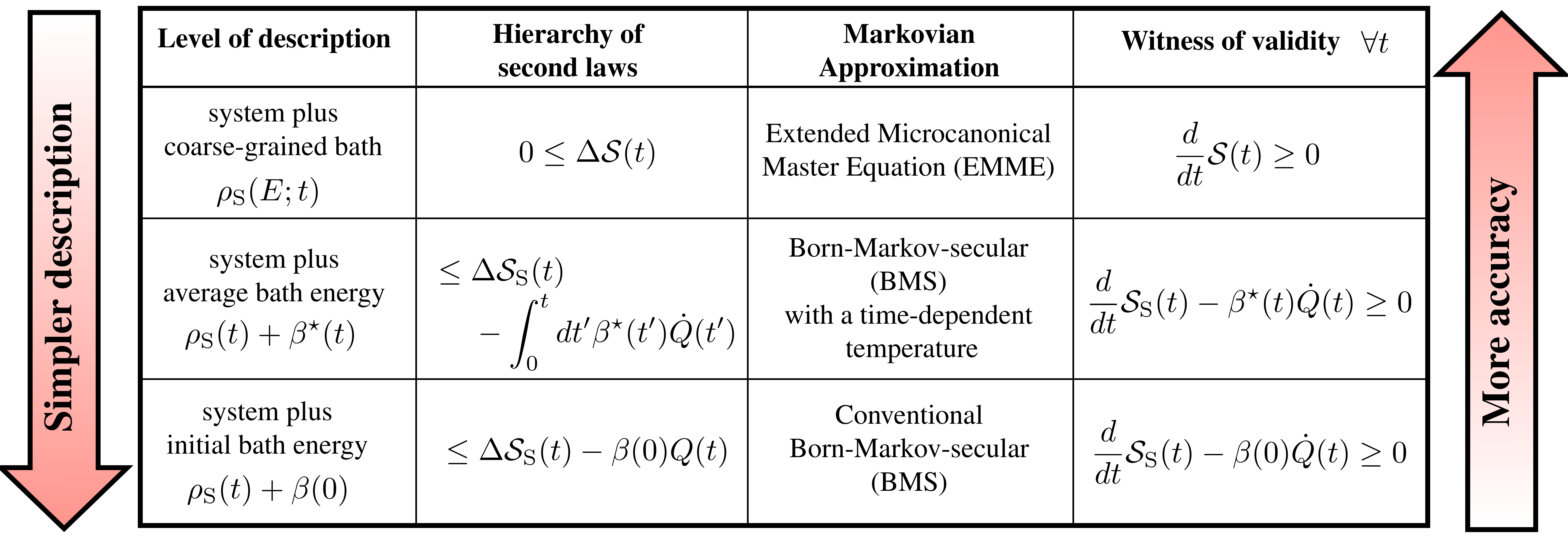}}\\
\caption{Table showing the parallelism between the different levels of description, with the corresponding second laws, and Markovian approximations with their witness of validity. The notation $\Delta \mathcal{S}(t) \coloneqq \mathcal{S}(t)-\mathcal{S}(0)$ represents the change in thermodynamic entropy at time $t$, with $\Delta\mathcal{S}_\text{S}$ being the corresponding change of only the system entropy. \label{fig:big_conclusions}}
\end{figure*}

\section{Conclusions}\label{sec:conclusions}

Together with our previous work~\cite{Riera-Campeny2021, Strasberg2021, Strasberg2021b}, we have uncovered a remarkable structural parallelism between the information one has about a system and its surrounding bath, the second law, and the corresponding Markovian master equation description. As displayed in Fig~\ref{fig:big_conclusions}, there is a hierarchy where the upper level contains the information to compute the lower level. This is obvious for the first column, the information used to describe the dynamics. However, it was only recently noted that this implies a corresponding hierarchy of second laws~\cite{Strasberg2021, Strasberg2021b}. If the bath is initially decorrelated and in a canonical state, the second laws can be ordered with the upper one bounding the lower one in the second column of Fig~\ref{fig:big_conclusions}. Moreover, at weak coupling, we can derive at each level a corresponding Markovian master equation (third column). As shown here, also these master equations form a hierarchy in terms of their accuracy.

In particular for the second level of description, we have observed that the best adapted inverse bath temperature $\beta^\star(t)$ is the same as the one appearing in Clausius inequality $\Delta \mathcal{S}_\text{S}(t) - \int_0^t dt' \beta^\star(t') \dot{Q}(t') \geq 0$ and it is in one to one correspondence with the bath average energy. Interestingly, the same inverse temperature $\beta^\star(t)$ has been proposed as a definition of non-equilibrium temperature in phenomenological nonequilibrium thermodynamics \cite{Muschik1977, Muschik1977b}, and it appears in recent microscopic derivations of the Clausius inequality~\cite{Riera-Campeny2021, Strasberg2021b, Strasberg2021}. It should be emphasized that the description in terms of $\beta^\star(t)$ remains valid for a bath, which is \emph{not} in a canonical state. However, if the bath can be approximately described by a time-dependent canonical state, our BMS equation with time dependent temperature reduces to master equations already studied, e.g., in Refs. \cite{Kolar2012,Nietner2014,Gallego-Marcos2014,Grenier2014,Schaller2014_b,Sekera2016,Grenier2016}.

Which of the master equations describes the situation most conveniently cannot be answered \emph{a priori}. However, an important witness for the validity of each master equation is a positive rate of the corresponding entropy production (fourth column, the increase of entropy for the EMME is proven in \cite{Riera-Campeny2021}). If the entropy production rate is negative at any time, we can exclude the corresponding master equation as an accurate method as it indicates a failure of the Markov approximation \cite{Strasberg2019_b}.

To conclude, we believe that the formalism presented in this work can find applications to describe the open quantum system dynamics when the environment evolves dynamically~\cite{Brantut2012, Brantut2013,Fernandez-Acebal2018}, but also in calorimetry experiments where the calorimeter has a finite heat capacity~\cite{Muller2015, Pekola2016, Halbertal2016, Muller2019, Karimi2020, Hausler2021,Pekola2015,Suomela2016,Donvil2021}, quantum thermometry~\cite{Mehboudi2019}, or even to understand the prethermalization regime of isolated quantum systems~\cite{Lazarides2014,Mori2016,Abanin2017,Mallayya2019}. 

\begin{acknowledgements}
We thank Javier Cerrillo for stimulating discussions on this and related topics. We acknowledge financial support from the Spanish Agencia Estatal de Investigación, projects  PID2019-107609GB-I00 and IJC2019-040883-I, Spanish MINECO  FIS2016-80681-P (AEI/FEDER, UE); Generalitat de Catalunya CIRIT 2017-SGR-1127, Secretaria d'Universitats i Recerca del Departament d'Empresa i Coneixement de la Generalitat de Catalunya, co-funded by the European Union Regional Development Fund within the ERDF Operational Program of Catalunya (project QuantumCat, ref. 001-P-001644), co-financed by the European Regional Development Fund (FEDER). PS also received financial support of a fellowship from “la Caixa” Foundation (ID 100010434, fellowship code LCF/BQ/PR21/11840014).
\end{acknowledgements}

\appendix
\onecolumngrid 

\section{Arbitrary bath coupling operators}\label{app:offdiagonal_b}

In the main text, we have assumed for the sake of the discussion that the coupling operator $B$ had only off-diagonal elements. In general, we would have to decompose the bath coupling operator $B$ into diagonal $B_d$ and off-diagonal $B_o$ terms in the eigenbasis $\{E_i\}$ of $H_\text{B}$, such that $B=B_d+B_o$. Then, we break the interaction Hamiltonian into
\begin{align}
H_\text{int} = \lambda S\otimes B_d + \lambda S\otimes B_o \eqqcolon \delta H + V. 
\end{align}
In the \textit{non-standard} interaction picture with respect to $H_0+ \delta H$, the evolution equation of the state $\rho(t)$ is
\begin{align}
\partial_t \hat{\rho}(t) = -i [\hat{V}(t),\rho(0)]-\int_0^t dt' \left[\hat{V}(t),[\hat{V}(t'),\hat{\rho}(t)]\right] + \mathcal{O}(\lambda^3),
\end{align}
where the hat is used to denote an operator in the interaction picture with respect to $H_0+\delta H$. Back to the Schr\"odinger picture, the evolution equation yields
\begin{align}
\partial_t \rho(t) =-i[H_0+\delta H,\rho]-i[V,\hat{\rho}(-t)] -\int_0^t dt' \left[V,[\hat{V}(t'-t),\rho]\right] + \mathcal{O}(\lambda^3).
\end{align}
We now multiply by $P(E)$ and take the trace with respect to the bath degrees of freedom to obtain
\begin{align}
\partial_t \rho_\text{S}(E) =-i[H_\text{S},\rho_\text{S}(E)]-i\text{tr}_\text{B}\{P(E)[\delta H,\rho]\} -i\text{tr}_\text{B}\{P(E)[V,\hat{\rho}(-t)]\} -\int_0^t dt' \text{tr}_\text{B}\left\{P(E) \left[V,[\tilde{V}(t'-t),\rho]\right]\right\}+ \mathcal{O}(\lambda^3).
\end{align}
where we have used $\hat{V}(t) = \tilde{V}(t)+\mathcal{O}(\lambda)$ and the tilde is used to denote an operator in the interaction picture with respect to $H_0$.
Then, following Sec.~\ref{sec:preliminaries}, we make the approximation~\eqref{eq:born_approx} which leads to second order in $\lambda$ to 
\begin{align}
\partial_t \rho_\text{S}(E) =-i[H_\text{S},\rho_\text{S}(E)]-i\int dE' [\delta H(E,E'),\rho_\text{S}(E')]\} -\int dE' \int_0^t dt' \text{tr}_\text{B}\left\{P(E) \left[V,[\tilde{V}(t'-t),\rho_\text{S}(E')\otimes \omega_\text{B}(E')]\right]\right\},
\end{align}
where we have used $\qav{E_i|V|E_i} = 0$ and have defined
\begin{align}
\delta H(E,E') =\lambda \text{tr}_\text{B}[P(E) \omega_\text{B}(E') B_d]S.
\end{align}
One could now proceed analogously to Sec.~\ref{sec:preliminaries} but explicitly taking into account the contribution of $\delta H(E,E')$ and obtain the corresponding master equation in this case. However, we note that since $\delta H(E,E')$ enters only as a commutator, its contribution can be absorbed as an order $\lambda$ modification of the Lamb-shift term; see Eq.~\eqref{eq:definitions}. Thus, after redefining $ H'_\text{LS}(E,E') = H_\text{LS}(E,E')+\delta H(E,E')$, the treatment of the Sec.~\ref{sec:preliminaries} to Sec.~\ref{sec:models} remains valid even for $B_d \neq 0$. 

\section{Conservation of the average energy}\label{app:energy_conservation}

The aim of this section is to show that Eq.~\eqref{eq:emme_povm} preserves the total energy if the weighting function is unbiased $\int dE E W(E|E_i) =E_i$. Using Eq.~\eqref{eq:emme_povm}, the change of total energy $\partial_t {U} = \int dE \text{tr}[(E\mathbf{1}_\text{S} + H_\text{S})\partial_t\rho_\text{S}(E)]$ yields
\begin{align}
\partial_t U = \iint dE dE'  \sum_\omega\left\{ E [\gamma_1(E,E';\omega) - \gamma_2(E,E';\omega)]- \omega \gamma_1(E,E';\omega)\right\} \text{tr}[S^\dagger_\omega S_\omega \rho_\text{S}(E')],
\end{align}
where we have used that $[H_\text{S},S_\omega] = -\omega S_\omega$. Next, we explicitly write down the dissipation rates
\begin{align}
\gamma_1(E,E';\omega) = 2\pi \lambda^2 \sum_{ij} |\langle E_i|B|E_j \rangle|^2 W(E|E_i) \frac{W(E'|E_j)}{V(E')} \delta( E_i -E_j-\omega),\nonumber\\
\gamma_2(E,E';\omega) = 2\pi \lambda^2 \sum_{ij} |\langle E_i|B|E_j \rangle|^2 W(E|E_j) \frac{W(E'|E_j)}{V(E')} \delta( E_i -E_j-\omega).\label{eq:dissipation_rates}
\end{align}
Using the fact that $W(E|E_i)$ is unbiased to compute the integral over $E'$, we immediately see that $\partial_t U = 0$. Note that the same result holds also if all projectors are biased by a constant amount, which is consistent with the fact that only energy differences are relevant in thermodynamics. 

\section{Equivalence of the Nakajima-Zwanzig and the finite-time Redfield equation}\label{app:projection_techniques}

In this section, we show the non-trivial correspondence between the second order Nakajima-Zwanzig \cite{Nakajima1958,Zwanzig1960,Breuer2002} equation and the finite-time Redfield equation \cite{Redfield1957,Breuer2002} in the case where we do \textit{not} assume $\mathcal{P}$ to be a projector; that is, $\mathcal{P}^2 \neq \mathcal{P}$. We start introducing $\mathcal{P}$ the super-operator that extracts the relevant degrees of freedom, together with its complementary super-operator $\mathcal{Q}\coloneqq \mathcal{I}-\mathcal{P}$. Then, the Liouville-von Neumann Eq.~\eqref{eq:vonNeumann}, can be used to derive
\begin{align}
\partial_t \mathcal{P}\tilde{\rho}(t) =& \mathcal{P}\mathcal{L}(t) \mathcal{P}\tilde{\rho}(t) + \mathcal{P}\mathcal{L}(t) \mathcal{Q}\tilde{\rho}(t),\nonumber\\
\partial_t \mathcal{Q}\tilde{\rho}(t) =& \mathcal{Q}\mathcal{L}(t) \mathcal{P}\tilde{\rho}(t) +\mathcal{Q}\mathcal{L}(t) \mathcal{Q}\tilde{\rho}(t),\label{eq:block_vonNeumann}
\end{align}
where we have used $\mathcal{P+Q=I}$. With the help of the Green function
\begin{align}
\mathcal{G}(t_2,t_1) \coloneqq \exp_+\left[\int_{t_1}^{t_2} dt \mathcal{Q}\mathcal{L}(t) \right],
\end{align}
we find the formal solution of the equation of the irrelevant part
\begin{align}
\mathcal{Q}\tilde{\rho}(t) =& \mathcal{G}(t,0)\mathcal{Q}\tilde{\rho}(0) +\int_0^t dt'  \mathcal{G}(t,t')\mathcal{Q}\mathcal{L}(t') \mathcal{P}\tilde{\rho}(t').\label{eq:irrelevant_solution}
\end{align}
Using Eq.~\eqref{eq:irrelevant_solution} into the first line of Eq.~\eqref{eq:block_vonNeumann} we arrive at the Nakajima-Zwanzig equation
\begin{align}
\partial_t \mathcal{P}\tilde{\rho}(t) =& \mathcal{P}\mathcal{L}(t) \mathcal{P}\tilde{\rho}(t) + \mathcal{P}\mathcal{L}(t) \mathcal{G}(t,0)\mathcal{Q}\tilde{\rho}(0)  +  \mathcal{P}\mathcal{L}(t) \int_0^t dt'  \mathcal{G}(t,t')\mathcal{Q}\mathcal{L}(t') \mathcal{P}\tilde{\rho}(t').\label{eq:NZ_equation}
\end{align}
It is only left to expand Eq.~\eqref{eq:NZ_equation} in $H_\text{int}$ to second order to obtain our desired equation
\begin{align}
\partial_t \mathcal{P}\tilde{\rho}(t) =& \mathcal{P}\mathcal{L}(t) \mathcal{P}\tilde{\rho}(t) + \mathcal{P}\mathcal{L}(t)\mathcal{Q}\tilde{\rho}(0) +\mathcal{P}\mathcal{L}(t) \int_0^t dt' \mathcal{Q}\mathcal{L}(t') \mathcal{Q}\tilde{\rho}(0) +  \mathcal{P}\mathcal{L}(t) \int_0^t dt'  \mathcal{Q}\mathcal{L}(t') \mathcal{P}\tilde{\rho}(t').\label{eq:NZ_equation_2}
\end{align}
We want to manipulate Eq.~\eqref{eq:NZ_equation_2} to make contact with the alternative derivation in the main text. First, note that the formal integration of the Liouville-von Neumann equation and left multiplication by $\mathcal{P}$ gives
\begin{align}
\mathcal{P}\tilde{\rho}(t) = \mathcal{P}\tilde{\rho}(0) +\mathcal{P}\int_0^t dt' \mathcal{L}(t') \tilde{\rho}(t').
\end{align}
Substituting $\mathcal{P}\tilde{\rho}(t)$ into the first term of the right-hand-side of Eq.~\eqref{eq:NZ_equation_2} and rearranging yields
\begin{align}
\partial_t \mathcal{P}\tilde{\rho}(t) =& \mathcal{P}\mathcal{L}(t)\tilde{\rho}(0) +\mathcal{P}\mathcal{L}(t) \int_0^t dt' \mathcal{L}(t')\tilde{\rho}(t')+\mathcal{P}\mathcal{L}(t) \int_0^t dt'  \mathcal{Q}\mathcal{L}(t') [\mathcal{Q}\tilde{\rho}(0)-\mathcal{Q}\tilde{\rho}(t')],\label{eq:NZ_redfield}
\end{align}
where the rightmost term can be ignored since the difference $\mathcal{Q}\tilde{\rho}(0)-\mathcal{Q}\tilde{\rho}(t')$ is of order $H_\text{int}$. Interestingly, there is a much simpler way to arrive to Eq.~\eqref{eq:NZ_redfield}. The formal solution of the Liouville-von Neumann Eq.~\eqref{eq:vonNeumann} reads
\begin{align}
\tilde{\rho}(t) = \tilde{\rho}(0) + \int_0^t dt' \mathcal{L}(t) \tilde{\rho}(0) + \int_0^t dt'\int_0^{t'} dt''\mathcal{L}(t') \mathcal{L}(t'')\tilde{\rho}(t'').
\end{align}
After taking the derivative of the above equation we get
\begin{align}
\partial_t \tilde{\rho}(t) = \mathcal{L}(t) \tilde{\rho}(0) + \mathcal{L}(t)\int_0^{t} dt' \mathcal{L}(t')\tilde{\rho}(t'),
\end{align}
which after acting with $\mathcal{P}$ from the left is identical to Eq.~\eqref{eq:NZ_redfield} to second order in $H_\text{int}$.

In conclusion, we have proven that it is equivalent to use the second-order expansion of the Nakajima-Zwanzig equation or the finite-time Redfield equation as a starting point to derive a second order master equation for the relevant degrees of freedom $\mathcal{P}\rho$ even in the case where $\mathcal{P}^2\neq \mathcal{P}$.

\section{Correlation functions for a piecewise non-interacting bath}\label{app:correlation_functions}

Assuming a Hamiltonian of the form \eqref{eq:loc_ham} and an interaction of the form \eqref{eq:loc_int} it is possible to compute the exact correlation functions appearing in $H_\text{LS}(E)$, $\gamma_1(E,E';\omega)$, $\gamma_2(E,E';\omega)$. To this end, we explicitly compute the first type of correlation function
\begin{align}
\langle B(\tau) P(E) B \rangle_{E'} =& \sum_{\textbf{n},\textbf{m}} W(E|E_\textbf{m}) \frac{W(E'|E_\textbf{n})}{V(E')} |\langle \textbf{n}| B|\textbf{m}\rangle|^2 e^{i(E_\textbf{n}-E_\textbf{m})\tau}\nonumber\\
=& \sum_R \sum_{\bar{\textbf{n}}}\sum_{m(R)}\sum_{n(R)} W(E|E_{m(R)} + E_{\bar{\textbf{n}}}) \frac{W(E'|E_{n(R)} + E_{\bar{\textbf{n}}})}{V(E')} | \langle n(R)| B_R |m(R)\rangle |^2 e^{i(E_{n(R)}-E_{m(R)})\tau},\label{eq:corr_bpbp}
\end{align}
as well as the second 
\begin{align}
\langle P(E) B(\tau) B \rangle_{E'} =& \sum_{\textbf{n},\textbf{m}} W(E|E_\textbf{n}) \frac{W(E'|E_\textbf{n})}{V(E')} |\langle \textbf{n}| B|\textbf{m}\rangle|^2 e^{i(E_\textbf{n}-E_\textbf{m})\tau}\nonumber\\
=& \sum_R \sum_{\bar{\textbf{n}}}\sum_{m(R)}\sum_{n(R)} W(E|E_{n(R)} + E_{\bar{\textbf{n}}}) \frac{W(E'|E_{n(R)} + E_{\bar{\textbf{n}}})}{V(E')} | \langle n(R)| B_R |m(R)\rangle |^2 e^{i(E_{n(R)}-E_{m(R)})\tau}.\label{eq:corr_pbbp}
\end{align}
Note that because $B$ is purely off-diagonal, the crossed terms between different regions always vanish. Hence, we can write down the correlation functions as a sum over regions of local correlation functions. 

Summing Eq.~\eqref{eq:corr_pbbp} over $E$ and taking the time-integral, one recovers the $\langle B(\tau) B \rangle_{E'}$ that appears in the computation of $\kappa(E;\omega)$, yielding the marginalized dissipation rates
\begin{align}
\kappa(E;\omega) = 2\pi\lambda^2 \sum_R \sum_{n(R)}\sum_{m(R)} |\langle n(R)|B_R|m(R)\rangle|^2 \frac{V{\bm (}E|n(R){\bm )}}{V(E)} \delta(E_{m(R)}-E_{n(R)}+\omega),\label{eq:non_interacting_kappa}
\end{align}. Similarly, the Lamb-shift $H_\text{LS}(E,E')$ is related to the time integral of~\eqref{eq:corr_pbbp} times the sign function. From the definition of $H_\text{LS}(E,E')$ in Eq.~\eqref{eq:definitions}, using $\int_0^\infty dx \sin(a x ) = \mathbb{P}(1/a)$ where $\mathbb{P}$ is the Cauchy principal value, we arrive at
\begin{align}
H_\text{LS}(E,E') = \lambda^2 \sum_\omega \sum_R \sum_{\bar{\textbf{n}}}\sum_{m(R)}\sum_{n(R)} W(E|E_{n(R)} + E_{\bar{\textbf{n}}}) \frac{W(E'|E_{n(R)} + E_{\bar{\textbf{n}}})}{V(E')} | \langle n(R)| B_R |m(R)\rangle |^2 \mathbb{P}\left( \frac{1}{E_{n(R)}-E_{m(R)}+\omega} \right) S_\omega^\dagger S_\omega.
\end{align}
Similarly, the dissipation rates $\gamma_1(E,E';\omega)$ and $\gamma_2(E,E';\omega)$ are related to the Fourier transform of Eqs.~\eqref{eq:corr_bpbp} and~\eqref{eq:corr_pbbp}. Thus, we find
\begin{align}
&\gamma_1(E,E';\omega) =  \frac{2\pi \lambda^2}{V(E')} \sum_R \sum_{m(R)}\sum_{n(R)} \delta(E_{n(R)}-E_{m(R)}+\omega) | \langle n(R)| B_R |m(R)\rangle |^2 \sum_{\bar{\textbf{n}}} W(E|E_{n(R)}+\omega+ E_{\bar{\textbf{n}}}) W(E'|E_{n(R)} + E_{\bar{\textbf{n}}}), \nonumber\\
&\gamma_2(E,E';\omega) =  \frac{2\pi \lambda^2}{V(E')} \sum_R \sum_{m(R)}\sum_{n(R)} \delta(E_{n(R)}-E_{m(R)}+\omega) | \langle n(R)| B_R |m(R)\rangle |^2 \sum_{\bar{\textbf{n}}} W(E|E_{n(R)}+ E_{\bar{\textbf{n}}}) W(E'|E_{n(R)} + E_{\bar{\textbf{n}}}).
\end{align}


\section{Details on the Gaussian density of states}\label{app:lindeberg}

In probability theory, the Lindeberg theorem provides a sufficient condition for a set of random variables to converge to a normal distribution. We present here the details showing that Lindeberg theorem guarantees the convergence of the density of states $g(e)$ of the spin bath to a normal distribution.

The Lindeberg theorem is as follows. Let $X_{N,r}$ be a triangular array of independent (but not necessarily identically distributed) random variables where $r = 1,\cdots, N$, with $\mathbb{E}[X_{N,r}] = 0$ and $\mathbb{E}[X_{N,r}^2] = \sigma_{N,r}^2$. Define the random variable of the sum $S_{N} \coloneqq \sum_{r=1}^N X_{N,r}$ with $\mathbb{E}[S_{N}] = 0$ and $\mathbb{E}[S_{N}^2] = \sigma_N^2 \coloneqq \sum_{r=1}^N \sigma_{N,r}^2$. If the Lindeberg condition
\begin{align}
    \lim_{N\to\infty}  \frac{1}{\sigma_N^2}\sum_{r=1}^N\mathbb{E}[X_{N,r}^2 : |X_{N,r}| > \eta \sigma_N] = 0 \qquad \forall \eta>0, \label{eq:lindeberg_condition}
\end{align}
holds, then, $S_N$ is normally distributed with zero mean and standard deviation $\sigma_N$ as $N$ tends to infinity.

The application to the spin bath is as follows. Consider a spin bath of $N$ non-interacting spin-1/2 particles. A bath eigenstate $\ket{e_\textbf{n}}$ is uniquely identified by the sequence $\textbf{n} = (n_1,\cdots,n_N)$ where $n_r = \{-1,+1\}$ are independent \textit{and} identically distributed random variables with probability $p(\pm 1) = 1/2$, with associated mean $\mathbb{E}[n_r] = 0$ and variance $\mathbb{E}[n_r^2] = 1$. However, the individual contribution to the energy is scaled by a prefactor $\Omega_r/2 > 0$, so we define $X_{N,r} = n_r \Omega_r/2$. We note that
\begin{align}
    \sum_{r=1}^N \mathbb{E}[X_{N,r}^2:|X_{N,r}|> \eta \sigma_N] &= \sum_{r=1}^N \frac{\Omega_r^2}{4} \mathbb{E}[n_r^2:|n_r|> \Omega_r^{-1} 2\eta \sigma_N]\nonumber\\
    &\leq \left(\sum_{r=1}^N \frac{\Omega_r^2}{4}\right) \mathbb{E}[n_1^2:|n_1|> \{\max_r \Omega_r\}^{-1} 2\eta \sigma_N],
\end{align}
where we have used the independent and identically distribution for the $n_r$'s. Hence, the Lindeberg condition holds provided that 
\begin{align}
    \lim_{N\to\infty} \frac{\max_r \Omega_r/2 }{\sqrt{\sum_r \Omega_r^2/4}} = 0, 
\end{align}
which is true for our spin bath. Therefore, we obtain a Gaussian density of states $g(e) = 2^N\mathcal{N}(e,\sigma_N)$.



\section{Further numerical results for the central spin model}\label{app:further_results}


\floatsetup[figure]{style=plain,subcapbesideposition=top}
\begin{figure}
{\includegraphics[width=\textwidth]{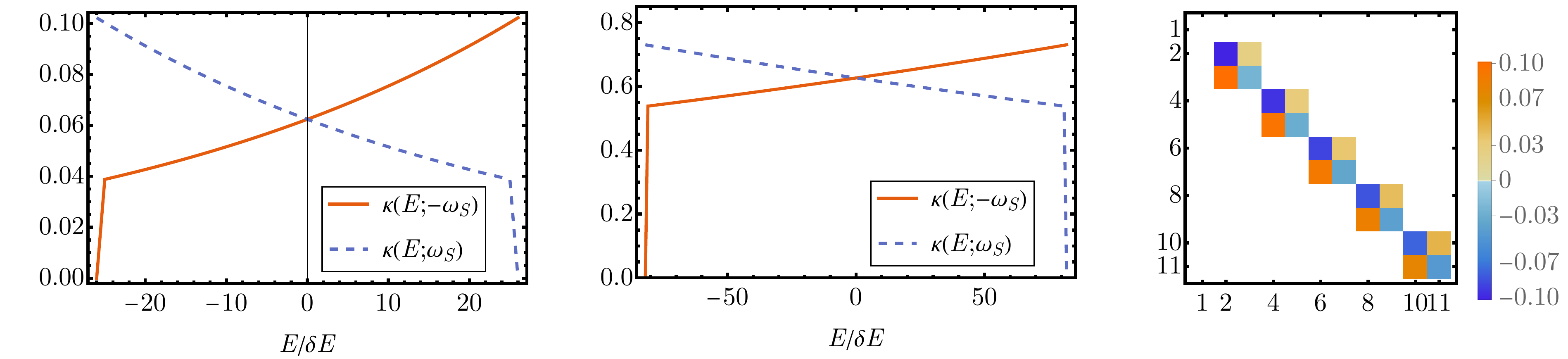}}
  \caption{
Function $\kappa(E;-\omega_\text{S})$ (solid orange line) and $\kappa(E;-\omega_\text{S})$ (blue dashed line) as a function of the energy $E$ for $N=100$ (first column) and $N=1000$ (second column) for a small central spin with $\text{s}=1/2$. The Zeeman frequencies of the bath are distributed according to $\mathcal{N}(\Omega-\Omega_0,\sigma_\Omega)$, and the central spin frequency is $\omega_\text{S} = \delta E$. In the third column, we show the block diagonal structure of the evolution matrix $\Lambda(\varepsilon_k,E;\varepsilon_q,E')$ for $N=100$ in the basis $\{\cdots, p(-\omega_\text{S}/2,E+\delta E), p(\omega_\text{S}/2,E), \cdots \}$. The numbers in the axis indicate the position in the basis, where the energies of the bath are ordered by increasing energy. The rest of the parameters are set to $\Omega_0 = \delta E$, $\sigma_\Omega = 0.2\delta E$, $\lambda = 0.01\delta E$, and $cr = 1$.  \label{fig:block_structure}} 
\end{figure}

The aim of this appendix is to partially extend the concise numerical analysis presented in Sec.~\ref{sec:models} of the main text. The following analysis focuses on two issues. 
First, we have seen that $\beta^\star$ corresponds to the best choice of inverse temperature only if $\kappa(E;\omega)$ is well approximated by a linear function of $E$ (see Sec.~\ref{subsec:hierarchy}). How accurate is this linear approximation for the model under study? Second, we know that the total average energy $U$ introduced in Sec.~\ref{sec:preliminaries} is preserved by the dynamics. Is this constraint reflected in the structure of the rate matrix $\Lambda(\varepsilon_k,E;\varepsilon_q,E')$? We give answers to those questions in the following.

\subsection{Is $\kappa(E;\omega)$ a linear function of E?}\label{subapp:further_linear}

We numerically evaluate the function $\kappa(E;\omega)$ for the non-interacting spin bath. In the left and middle panel of Fig.~\ref{fig:block_structure} we show its behavior with energy for $N=100$ and $N=1000$ particles respectively. As it can be seen, the linear approximation holds for all range of energies provided that the energy variance $\Delta E$ of $p(E)$ is not exceedingly large. Moreover, $\kappa(E;\omega)$ is better approximated by a linear function for an increasing particle number $N$.

This result can be also understood from the following rough analysis. First, we focus on the limit $\sigma_N\gg \delta E$, which is always the case for a sufficiently large number of particles $N$. Then, we approximate
\begin{align}
    V(E) = \int dE W(E|e) g(e) \approx g(E) W(E|E) \delta E. 
\end{align}
We are interested in the case where $W(E|E_i) = W(E-E_i)$, thus $W(E|E)=W(0)$. Hence, for the central spin model we obtain 
\begin{align}
    &\mathcal{S}(E) = -\frac{E^2}{2\sigma_N^2}+ \log\left(\frac{2^N W(0)\delta E^2}{\sqrt{2\pi} \sigma_N}\right),\nonumber\\
    &\beta (E) = - \frac{E}{\sigma_N^2}.
\end{align}
Also, within this approximation, the microcanonical heat capacity becomes $\mathcal{C}(E) \coloneqq - \beta(E)^2 [\partial_E \beta(E)]^{-1} =  \beta(E)^2 \sigma_N^2$.

Second, we note that from Eq.~\eqref{eq:non_interacting_kappa}, it follows that $\kappa(E;\omega)$ depends on energy only through the ratio
\begin{align}
    \frac{V(E-E_{n(R)})}{V(E)} \approx 1+\beta(E)E_{n(R)} = 1 -\frac{E_{n(R)}}{\sigma_N^2}E.
\end{align}
where we have assumed that the volume changes slowly compared to the energy scale $E_{n(R)}$. In this limit, we expect $\kappa(E;\omega)$ to be also a linear function of the energy. Despite being a very rough analysis, it indicates that, for the central spin model, the linear approximation of $\kappa(E;\omega)$ can hold true even if the variance $\Delta E$ of the energy distribution $p(E)$ becomes large. Also, it indicates that, for a fixed energy $E$ and frequency $\omega$, the function $\kappa(E;\omega)$ tends to a constant as the number of particles $N$ goes to infinity. 

\subsection{Block structure of $\Lambda$}

In the right panel of Fig.~\ref{fig:block_structure} we show the first matrix elements $\Lambda(\varepsilon_k,E;\varepsilon_q,E')$. From the figure it can be seen that the energy conservation leads to a block structure of $\Lambda$ as it was theoretically discussed in Ref.~\cite{Riera-Campeny2021} under the name of strict total energy conservation. This implies that not only the average energy is preserved, but also its probability distribution $p_\text{tot}(E) = \sum_{k} p(\varepsilon_k, E-\varepsilon_k)$ is conserved in time.

%

\bibliographystyle{apsrev4-1}
\bibliography{Refs}

\begin{thebibliography}{63}%
\makeatletter
\providecommand \@ifxundefined [1]{%
 \@ifx{#1\undefined}
}%
\providecommand \@ifnum [1]{%
 \ifnum #1\expandafter \@firstoftwo
 \else \expandafter \@secondoftwo
 \fi
}%
\providecommand \@ifx [1]{%
 \ifx #1\expandafter \@firstoftwo
 \else \expandafter \@secondoftwo
 \fi
}%
\providecommand \natexlab [1]{#1}%
\providecommand \enquote  [1]{``#1''}%
\providecommand \bibnamefont  [1]{#1}%
\providecommand \bibfnamefont [1]{#1}%
\providecommand \citenamefont [1]{#1}%
\providecommand \href@noop [0]{\@secondoftwo}%
\providecommand \href [0]{\begingroup \@sanitize@url \@href}%
\providecommand \@href[1]{\@@startlink{#1}\@@href}%
\providecommand \@@href[1]{\endgroup#1\@@endlink}%
\providecommand \@sanitize@url [0]{\catcode `\\12\catcode `\$12\catcode
  `\&12\catcode `\#12\catcode `\^12\catcode `\_12\catcode `\%12\relax}%
\providecommand \@@startlink[1]{}%
\providecommand \@@endlink[0]{}%
\providecommand \url  [0]{\begingroup\@sanitize@url \@url }%
\providecommand \@url [1]{\endgroup\@href {#1}{\urlprefix }}%
\providecommand \urlprefix  [0]{URL }%
\providecommand \Eprint [0]{\href }%
\providecommand \doibase [0]{http://dx.doi.org/}%
\providecommand \selectlanguage [0]{\@gobble}%
\providecommand \bibinfo  [0]{\@secondoftwo}%
\providecommand \bibfield  [0]{\@secondoftwo}%
\providecommand \translation [1]{[#1]}%
\providecommand \BibitemOpen [0]{}%
\providecommand \bibitemStop [0]{}%
\providecommand \bibitemNoStop [0]{.\EOS\space}%
\providecommand \EOS [0]{\spacefactor3000\relax}%
\providecommand \BibitemShut  [1]{\csname bibitem#1\endcsname}%
\let\auto@bib@innerbib\@empty
\bibitem [{\citenamefont {Brantut}\ \emph {et~al.}(2012)\citenamefont
  {Brantut}, \citenamefont {Meineke}, \citenamefont {Stadler}, \citenamefont
  {Krinner},\ and\ \citenamefont {Esslinger}}]{Brantut2012}%
  \BibitemOpen
  \bibfield  {author} {\bibinfo {author} {\bibfnamefont {J.-P.}\ \bibnamefont
  {Brantut}}, \bibinfo {author} {\bibfnamefont {J.}~\bibnamefont {Meineke}},
  \bibinfo {author} {\bibfnamefont {D.}~\bibnamefont {Stadler}}, \bibinfo
  {author} {\bibfnamefont {S.}~\bibnamefont {Krinner}}, \ and\ \bibinfo
  {author} {\bibfnamefont {T.}~\bibnamefont {Esslinger}},\ }\href {\doibase
  10.1126/science.1223175} {\bibfield  {journal} {\bibinfo  {journal}
  {Science}\ }\textbf {\bibinfo {volume} {337}},\ \bibinfo {pages} {1069}
  (\bibinfo {year} {2012})}\BibitemShut {NoStop}%
\bibitem [{\citenamefont {Brantut}\ \emph {et~al.}(2013)\citenamefont
  {Brantut}, \citenamefont {Grenier}, \citenamefont {Meineke}, \citenamefont
  {Stadler}, \citenamefont {Krinner}, \citenamefont {Kollath}, \citenamefont
  {Esslinger},\ and\ \citenamefont {Georges}}]{Brantut2013}%
  \BibitemOpen
  \bibfield  {author} {\bibinfo {author} {\bibfnamefont {J.-P.}\ \bibnamefont
  {Brantut}}, \bibinfo {author} {\bibfnamefont {C.}~\bibnamefont {Grenier}},
  \bibinfo {author} {\bibfnamefont {J.}~\bibnamefont {Meineke}}, \bibinfo
  {author} {\bibfnamefont {D.}~\bibnamefont {Stadler}}, \bibinfo {author}
  {\bibfnamefont {S.}~\bibnamefont {Krinner}}, \bibinfo {author} {\bibfnamefont
  {C.}~\bibnamefont {Kollath}}, \bibinfo {author} {\bibfnamefont
  {T.}~\bibnamefont {Esslinger}}, \ and\ \bibinfo {author} {\bibfnamefont
  {A.}~\bibnamefont {Georges}},\ }\href {\doibase 10.1126/science.1242308}
  {\bibfield  {journal} {\bibinfo  {journal} {Science}\ }\textbf {\bibinfo
  {volume} {342}},\ \bibinfo {pages} {713} (\bibinfo {year}
  {2013})}\BibitemShut {NoStop}%
\bibitem [{\citenamefont {M\"uller}\ \emph {et~al.}(2015)\citenamefont
  {M\"uller}, \citenamefont {Lisenfeld}, \citenamefont {Shnirman},\ and\
  \citenamefont {Poletto}}]{Muller2015}%
  \BibitemOpen
  \bibfield  {author} {\bibinfo {author} {\bibfnamefont {C.}~\bibnamefont
  {M\"uller}}, \bibinfo {author} {\bibfnamefont {J.}~\bibnamefont {Lisenfeld}},
  \bibinfo {author} {\bibfnamefont {A.}~\bibnamefont {Shnirman}}, \ and\
  \bibinfo {author} {\bibfnamefont {S.}~\bibnamefont {Poletto}},\ }\href
  {\doibase 10.1103/PhysRevB.92.035442} {\bibfield  {journal} {\bibinfo
  {journal} {Phys. Rev. B}\ }\textbf {\bibinfo {volume} {92}},\ \bibinfo
  {pages} {035442} (\bibinfo {year} {2015})}\BibitemShut {NoStop}%
\bibitem [{\citenamefont {Pekola}\ \emph {et~al.}(2016)\citenamefont {Pekola},
  \citenamefont {Suomela},\ and\ \citenamefont {Galperin}}]{Pekola2016}%
  \BibitemOpen
  \bibfield  {author} {\bibinfo {author} {\bibfnamefont {J.~P.}\ \bibnamefont
  {Pekola}}, \bibinfo {author} {\bibfnamefont {S.}~\bibnamefont {Suomela}}, \
  and\ \bibinfo {author} {\bibfnamefont {Y.~M.}\ \bibnamefont {Galperin}},\
  }\href {\doibase 10.1007/s10909-016-1618-5} {\bibfield  {journal} {\bibinfo
  {journal} {J. Low Temp. Phys.}\ }\textbf {\bibinfo {volume} {184}},\ \bibinfo
  {pages} {1015} (\bibinfo {year} {2016})}\BibitemShut {NoStop}%
\bibitem [{\citenamefont {Halbertal}\ \emph {et~al.}(2016)\citenamefont
  {Halbertal}, \citenamefont {Cuppens}, \citenamefont {Shalom}, \citenamefont
  {Embon}, \citenamefont {Shadmi}, \citenamefont {Anahory}, \citenamefont
  {Naren}, \citenamefont {Sarkar}, \citenamefont {Uri}, \citenamefont {Ronen},
  \citenamefont {Myasoedov}, \citenamefont {Levitov}, \citenamefont
  {Joselevich}, \citenamefont {Geim},\ and\ \citenamefont
  {Zeldov}}]{Halbertal2016}%
  \BibitemOpen
  \bibfield  {author} {\bibinfo {author} {\bibfnamefont {D.}~\bibnamefont
  {Halbertal}}, \bibinfo {author} {\bibfnamefont {J.}~\bibnamefont {Cuppens}},
  \bibinfo {author} {\bibfnamefont {M.~B.}\ \bibnamefont {Shalom}}, \bibinfo
  {author} {\bibfnamefont {L.}~\bibnamefont {Embon}}, \bibinfo {author}
  {\bibfnamefont {N.}~\bibnamefont {Shadmi}}, \bibinfo {author} {\bibfnamefont
  {Y.}~\bibnamefont {Anahory}}, \bibinfo {author} {\bibfnamefont {H.~R.}\
  \bibnamefont {Naren}}, \bibinfo {author} {\bibfnamefont {J.}~\bibnamefont
  {Sarkar}}, \bibinfo {author} {\bibfnamefont {A.}~\bibnamefont {Uri}},
  \bibinfo {author} {\bibfnamefont {Y.}~\bibnamefont {Ronen}}, \bibinfo
  {author} {\bibfnamefont {Y.}~\bibnamefont {Myasoedov}}, \bibinfo {author}
  {\bibfnamefont {L.~S.}\ \bibnamefont {Levitov}}, \bibinfo {author}
  {\bibfnamefont {E.}~\bibnamefont {Joselevich}}, \bibinfo {author}
  {\bibfnamefont {A.~K.}\ \bibnamefont {Geim}}, \ and\ \bibinfo {author}
  {\bibfnamefont {E.}~\bibnamefont {Zeldov}},\ }\href {\doibase
  10.1038/nature19843} {\bibfield  {journal} {\bibinfo  {journal} {Nature}\
  }\textbf {\bibinfo {volume} {539}},\ \bibinfo {pages} {407} (\bibinfo {year}
  {2016})}\BibitemShut {NoStop}%
\bibitem [{\citenamefont {Müller}\ \emph {et~al.}(2019)\citenamefont
  {Müller}, \citenamefont {Cole},\ and\ \citenamefont
  {Lisenfeld}}]{Muller2019}%
  \BibitemOpen
  \bibfield  {author} {\bibinfo {author} {\bibfnamefont {C.}~\bibnamefont
  {Müller}}, \bibinfo {author} {\bibfnamefont {J.~H.}\ \bibnamefont {Cole}}, \
  and\ \bibinfo {author} {\bibfnamefont {J.}~\bibnamefont {Lisenfeld}},\ }\href
  {\doibase 10.1088/1361-6633/ab3a7e} {\bibfield  {journal} {\bibinfo
  {journal} {Rep. Prog. Phys.}\ }\textbf {\bibinfo {volume} {82}},\ \bibinfo
  {pages} {124501} (\bibinfo {year} {2019})}\BibitemShut {NoStop}%
\bibitem [{\citenamefont {Karimi}\ \emph {et~al.}(2020)\citenamefont {Karimi},
  \citenamefont {Brange}, \citenamefont {Samuelsson},\ and\ \citenamefont
  {Pekola}}]{Karimi2020}%
  \BibitemOpen
  \bibfield  {author} {\bibinfo {author} {\bibfnamefont {B.}~\bibnamefont
  {Karimi}}, \bibinfo {author} {\bibfnamefont {F.}~\bibnamefont {Brange}},
  \bibinfo {author} {\bibfnamefont {P.}~\bibnamefont {Samuelsson}}, \ and\
  \bibinfo {author} {\bibfnamefont {J.~P.}\ \bibnamefont {Pekola}},\ }\href
  {\doibase 10.1038/s41467-019-14247-2} {\bibfield  {journal} {\bibinfo
  {journal} {Nat. Commun.}\ }\textbf {\bibinfo {volume} {11}},\ \bibinfo
  {pages} {367} (\bibinfo {year} {2020})}\BibitemShut {NoStop}%
\bibitem [{\citenamefont {H\"ausler}\ \emph {et~al.}(2021)\citenamefont
  {H\"ausler}, \citenamefont {Fabritius}, \citenamefont {Mohan}, \citenamefont
  {Lebrat}, \citenamefont {Corman},\ and\ \citenamefont
  {Esslinger}}]{Hausler2021}%
  \BibitemOpen
  \bibfield  {author} {\bibinfo {author} {\bibfnamefont {S.}~\bibnamefont
  {H\"ausler}}, \bibinfo {author} {\bibfnamefont {P.}~\bibnamefont
  {Fabritius}}, \bibinfo {author} {\bibfnamefont {J.}~\bibnamefont {Mohan}},
  \bibinfo {author} {\bibfnamefont {M.}~\bibnamefont {Lebrat}}, \bibinfo
  {author} {\bibfnamefont {L.}~\bibnamefont {Corman}}, \ and\ \bibinfo {author}
  {\bibfnamefont {T.}~\bibnamefont {Esslinger}},\ }\href {\doibase
  10.1103/PhysRevX.11.021034} {\bibfield  {journal} {\bibinfo  {journal} {Phys.
  Rev. X}\ }\textbf {\bibinfo {volume} {11}},\ \bibinfo {pages} {021034}
  (\bibinfo {year} {2021})}\BibitemShut {NoStop}%
\bibitem [{\citenamefont {de~Vega}\ and\ \citenamefont
  {Alonso}(2017)}]{deVega2017}%
  \BibitemOpen
  \bibfield  {author} {\bibinfo {author} {\bibfnamefont {I.}~\bibnamefont
  {de~Vega}}\ and\ \bibinfo {author} {\bibfnamefont {D.}~\bibnamefont
  {Alonso}},\ }\href {\doibase 10.1103/RevModPhys.89.015001} {\bibfield
  {journal} {\bibinfo  {journal} {Rev. Mod. Phys.}\ }\textbf {\bibinfo {volume}
  {89}},\ \bibinfo {pages} {015001} (\bibinfo {year} {2017})}\BibitemShut
  {NoStop}%
\bibitem [{\citenamefont {Weimer}\ \emph {et~al.}(2021)\citenamefont {Weimer},
  \citenamefont {Kshetrimayum},\ and\ \citenamefont {Or\'us}}]{Weimer2021}%
  \BibitemOpen
  \bibfield  {author} {\bibinfo {author} {\bibfnamefont {H.}~\bibnamefont
  {Weimer}}, \bibinfo {author} {\bibfnamefont {A.}~\bibnamefont
  {Kshetrimayum}}, \ and\ \bibinfo {author} {\bibfnamefont {R.}~\bibnamefont
  {Or\'us}},\ }\href {\doibase 10.1103/RevModPhys.93.015008} {\bibfield
  {journal} {\bibinfo  {journal} {Rev. Mod. Phys.}\ }\textbf {\bibinfo {volume}
  {93}},\ \bibinfo {pages} {015008} (\bibinfo {year} {2021})}\BibitemShut
  {NoStop}%
\bibitem [{\citenamefont {Gardiner}\ and\ \citenamefont
  {Zoller}(2000)}]{Gardiner2000}%
  \BibitemOpen
  \bibfield  {author} {\bibinfo {author} {\bibfnamefont {C.~W.}\ \bibnamefont
  {Gardiner}}\ and\ \bibinfo {author} {\bibfnamefont {P.}~\bibnamefont
  {Zoller}},\ }\href {\doibase 10.1007/978-3-662-04103-1} {\emph {\bibinfo
  {title} {Quantum Noise}}}\ (\bibinfo  {publisher} {Springer Berlin
  Heidelberg},\ \bibinfo {year} {2000})\BibitemShut {NoStop}%
\bibitem [{\citenamefont {Breuer}\ and\ \citenamefont
  {Petruccione}(2002)}]{Breuer2002}%
  \BibitemOpen
  \bibfield  {author} {\bibinfo {author} {\bibfnamefont {H.-P.}\ \bibnamefont
  {Breuer}}\ and\ \bibinfo {author} {\bibfnamefont {F.}~\bibnamefont
  {Petruccione}},\ }\href {\doibase 10.1093/acprof:oso/9780199213900.001.0001}
  {\emph {\bibinfo {title} {The theory of open quantum systems}}}\ (\bibinfo
  {publisher} {Oxford University Press},\ \bibinfo {year} {2002})\BibitemShut
  {NoStop}%
\bibitem [{\citenamefont {Schaller}(2014)}]{Schaller2014}%
  \BibitemOpen
  \bibfield  {author} {\bibinfo {author} {\bibfnamefont {G.}~\bibnamefont
  {Schaller}},\ }\href {\doibase 10.1007/978-3-319-03877-3} {\emph {\bibinfo
  {title} {Open quantum systems far from equilibrium}}},\ Vol.\ \bibinfo
  {volume} {881}\ (\bibinfo  {publisher} {Springer},\ \bibinfo {year}
  {2014})\BibitemShut {NoStop}%
\bibitem [{\citenamefont {Esposito}\ and\ \citenamefont
  {Gaspard}(2003)}]{Esposito2003a}%
  \BibitemOpen
  \bibfield  {author} {\bibinfo {author} {\bibfnamefont {M.}~\bibnamefont
  {Esposito}}\ and\ \bibinfo {author} {\bibfnamefont {P.}~\bibnamefont
  {Gaspard}},\ }\href {\doibase 10.1103/PhysRevE.68.066112} {\bibfield
  {journal} {\bibinfo  {journal} {Phys. Rev. E}\ }\textbf {\bibinfo {volume}
  {68}},\ \bibinfo {pages} {066112} (\bibinfo {year} {2003})}\BibitemShut
  {NoStop}%
\bibitem [{\citenamefont {Esposito}\ and\ \citenamefont
  {Gaspard}(2007)}]{Esposito2007}%
  \BibitemOpen
  \bibfield  {author} {\bibinfo {author} {\bibfnamefont {M.}~\bibnamefont
  {Esposito}}\ and\ \bibinfo {author} {\bibfnamefont {P.}~\bibnamefont
  {Gaspard}},\ }\href {\doibase 10.1103/PhysRevE.76.041134} {\bibfield
  {journal} {\bibinfo  {journal} {Phys. Rev. E}\ }\textbf {\bibinfo {volume}
  {76}},\ \bibinfo {pages} {041134} (\bibinfo {year} {2007})}\BibitemShut
  {NoStop}%
\bibitem [{\citenamefont {Breuer}\ \emph {et~al.}(2006)\citenamefont {Breuer},
  \citenamefont {Gemmer},\ and\ \citenamefont {Michel}}]{Breuer2006}%
  \BibitemOpen
  \bibfield  {author} {\bibinfo {author} {\bibfnamefont {H.-P.}\ \bibnamefont
  {Breuer}}, \bibinfo {author} {\bibfnamefont {J.}~\bibnamefont {Gemmer}}, \
  and\ \bibinfo {author} {\bibfnamefont {M.}~\bibnamefont {Michel}},\ }\href
  {\doibase 10.1103/PhysRevE.73.016139} {\bibfield  {journal} {\bibinfo
  {journal} {Phys. Rev. E}\ }\textbf {\bibinfo {volume} {73}},\ \bibinfo
  {pages} {016139} (\bibinfo {year} {2006})}\BibitemShut {NoStop}%
\bibitem [{\citenamefont {Breuer}(2007)}]{Breuer2007}%
  \BibitemOpen
  \bibfield  {author} {\bibinfo {author} {\bibfnamefont {H.-P.}\ \bibnamefont
  {Breuer}},\ }\href {\doibase 10.1103/PhysRevA.75.022103} {\bibfield
  {journal} {\bibinfo  {journal} {Phys. Rev. A}\ }\textbf {\bibinfo {volume}
  {75}},\ \bibinfo {pages} {022103} (\bibinfo {year} {2007})}\BibitemShut
  {NoStop}%
\bibitem [{\citenamefont {Riera-Campeny}\ \emph {et~al.}(2021)\citenamefont
  {Riera-Campeny}, \citenamefont {Sanpera},\ and\ \citenamefont
  {Strasberg}}]{Riera-Campeny2021}%
  \BibitemOpen
  \bibfield  {author} {\bibinfo {author} {\bibfnamefont {A.}~\bibnamefont
  {Riera-Campeny}}, \bibinfo {author} {\bibfnamefont {A.}~\bibnamefont
  {Sanpera}}, \ and\ \bibinfo {author} {\bibfnamefont {P.}~\bibnamefont
  {Strasberg}},\ }\href {\doibase 10.1103/PRXQuantum.2.010340} {\bibfield
  {journal} {\bibinfo  {journal} {PRX Quantum}\ }\textbf {\bibinfo {volume}
  {2}},\ \bibinfo {pages} {010340} (\bibinfo {year} {2021})}\BibitemShut
  {NoStop}%
\bibitem [{\citenamefont {Donvil}\ and\ \citenamefont
  {Ankerhold}(2021)}]{Donvil2021}%
  \BibitemOpen
  \bibfield  {author} {\bibinfo {author} {\bibfnamefont {B.}~\bibnamefont
  {Donvil}}\ and\ \bibinfo {author} {\bibfnamefont {J.}~\bibnamefont
  {Ankerhold}},\ }\href {https://arxiv.org/pdf/2104.14894.pdf} {\bibfield
  {journal} {\bibinfo  {journal} {arXiv:2104.14894}\ } (\bibinfo {year}
  {2021})}\BibitemShut {NoStop}%
\bibitem [{\citenamefont {Ronzani}\ \emph {et~al.}(2018)\citenamefont
  {Ronzani}, \citenamefont {Karimi}, \citenamefont {Senior}, \citenamefont
  {Chang}, \citenamefont {Peltonen}, \citenamefont {Chen},\ and\ \citenamefont
  {Pekola}}]{Ronzani2018}%
  \BibitemOpen
  \bibfield  {author} {\bibinfo {author} {\bibfnamefont {A.}~\bibnamefont
  {Ronzani}}, \bibinfo {author} {\bibfnamefont {B.}~\bibnamefont {Karimi}},
  \bibinfo {author} {\bibfnamefont {J.}~\bibnamefont {Senior}}, \bibinfo
  {author} {\bibfnamefont {Y.-C.}\ \bibnamefont {Chang}}, \bibinfo {author}
  {\bibfnamefont {J.~T.}\ \bibnamefont {Peltonen}}, \bibinfo {author}
  {\bibfnamefont {C.}~\bibnamefont {Chen}}, \ and\ \bibinfo {author}
  {\bibfnamefont {J.~P.}\ \bibnamefont {Pekola}},\ }\href {\doibase
  10.1038/s41567-018-0199-4} {\bibfield  {journal} {\bibinfo  {journal} {Nat.
  Phys.}\ }\textbf {\bibinfo {volume} {14}},\ \bibinfo {pages} {991} (\bibinfo
  {year} {2018})}\BibitemShut {NoStop}%
\bibitem [{\citenamefont {Kokkoniemi}\ \emph {et~al.}(2019)\citenamefont
  {Kokkoniemi}, \citenamefont {Govenius}, \citenamefont {Vesterinen},
  \citenamefont {Lake}, \citenamefont {Gunyhó}, \citenamefont {Tan},
  \citenamefont {Simbierowicz}, \citenamefont {Grönberg}, \citenamefont
  {Lehtinen}, \citenamefont {Prunnila}, \citenamefont {Hassel}, \citenamefont
  {Lamminen}, \citenamefont {Saira},\ and\ \citenamefont
  {Möttönen}}]{Kokkoniemi2019}%
  \BibitemOpen
  \bibfield  {author} {\bibinfo {author} {\bibfnamefont {R.}~\bibnamefont
  {Kokkoniemi}}, \bibinfo {author} {\bibfnamefont {J.}~\bibnamefont
  {Govenius}}, \bibinfo {author} {\bibfnamefont {V.}~\bibnamefont
  {Vesterinen}}, \bibinfo {author} {\bibfnamefont {R.~E.}\ \bibnamefont
  {Lake}}, \bibinfo {author} {\bibfnamefont {A.~M.}\ \bibnamefont {Gunyhó}},
  \bibinfo {author} {\bibfnamefont {K.~Y.}\ \bibnamefont {Tan}}, \bibinfo
  {author} {\bibfnamefont {S.}~\bibnamefont {Simbierowicz}}, \bibinfo {author}
  {\bibfnamefont {L.}~\bibnamefont {Grönberg}}, \bibinfo {author}
  {\bibfnamefont {J.}~\bibnamefont {Lehtinen}}, \bibinfo {author}
  {\bibfnamefont {M.}~\bibnamefont {Prunnila}}, \bibinfo {author}
  {\bibfnamefont {J.}~\bibnamefont {Hassel}}, \bibinfo {author} {\bibfnamefont
  {A.}~\bibnamefont {Lamminen}}, \bibinfo {author} {\bibfnamefont {O.-P.}\
  \bibnamefont {Saira}}, \ and\ \bibinfo {author} {\bibfnamefont
  {M.}~\bibnamefont {Möttönen}},\ }\href {\doibase 10.1038/s42005-019-0225-6}
  {\bibfield  {journal} {\bibinfo  {journal} {Comm. Phys.}\ }\textbf {\bibinfo
  {volume} {2}},\ \bibinfo {pages} {124} (\bibinfo {year} {2019})}\BibitemShut
  {NoStop}%
\bibitem [{\citenamefont {Senior}\ \emph {et~al.}(2020)\citenamefont {Senior},
  \citenamefont {Gubaydullin}, \citenamefont {Karimi}, \citenamefont
  {Peltonen}, \citenamefont {Ankerhold},\ and\ \citenamefont
  {Pekola}}]{Senior2020}%
  \BibitemOpen
  \bibfield  {author} {\bibinfo {author} {\bibfnamefont {J.}~\bibnamefont
  {Senior}}, \bibinfo {author} {\bibfnamefont {A.}~\bibnamefont {Gubaydullin}},
  \bibinfo {author} {\bibfnamefont {B.}~\bibnamefont {Karimi}}, \bibinfo
  {author} {\bibfnamefont {J.~T.}\ \bibnamefont {Peltonen}}, \bibinfo {author}
  {\bibfnamefont {J.}~\bibnamefont {Ankerhold}}, \ and\ \bibinfo {author}
  {\bibfnamefont {J.~P.}\ \bibnamefont {Pekola}},\ }\href {\doibase
  10.1038/s42005-020-0307-5} {\bibfield  {journal} {\bibinfo  {journal} {Comm.
  Phys.}\ }\textbf {\bibinfo {volume} {3}},\ \bibinfo {pages} {40} (\bibinfo
  {year} {2020})}\BibitemShut {NoStop}%
\bibitem [{\citenamefont {Nazir}\ and\ \citenamefont
  {Schaller}(2019)}]{Nazir2018}%
  \BibitemOpen
  \bibfield  {author} {\bibinfo {author} {\bibfnamefont {A.}~\bibnamefont
  {Nazir}}\ and\ \bibinfo {author} {\bibfnamefont {G.}~\bibnamefont
  {Schaller}},\ }in\ \href {\doibase
  https://doi.org/10.1007/978-3-319-99046-0_23} {\emph {\bibinfo {booktitle}
  {Thermodynamics in the quantum regime: fundamental aspects and new
  directions}}},\ Vol.\ \bibinfo {volume} {195}\ (\bibinfo  {publisher}
  {Springer Cham},\ \bibinfo {year} {2019})\ pp.\ \bibinfo {pages}
  {551--577}\BibitemShut {NoStop}%
\bibitem [{\citenamefont {Strasberg}\ and\ \citenamefont
  {Winter}(2021)}]{Strasberg2021b}%
  \BibitemOpen
  \bibfield  {author} {\bibinfo {author} {\bibfnamefont {P.}~\bibnamefont
  {Strasberg}}\ and\ \bibinfo {author} {\bibfnamefont {A.}~\bibnamefont
  {Winter}},\ }\href {\doibase 10.1103/PRXQuantum.2.030202} {\bibfield
  {journal} {\bibinfo  {journal} {PRX Quantum}\ }\textbf {\bibinfo {volume}
  {2}},\ \bibinfo {pages} {030202} (\bibinfo {year} {2021})}\BibitemShut
  {NoStop}%
\bibitem [{\citenamefont {Nielsen}\ and\ \citenamefont
  {Chuang}(2010)}]{Nielsen2002}%
  \BibitemOpen
  \bibfield  {author} {\bibinfo {author} {\bibfnamefont {M.~A.}\ \bibnamefont
  {Nielsen}}\ and\ \bibinfo {author} {\bibfnamefont {I.}~\bibnamefont
  {Chuang}},\ }\href {\doibase https://doi.org/10.1017/CBO9780511976667} {\emph
  {\bibinfo {title} {Quantum computation and quantum information}}}\ (\bibinfo
  {publisher} {Cambridge University Press},\ \bibinfo {year}
  {2010})\BibitemShut {NoStop}%
\bibitem [{\citenamefont {Milz}\ and\ \citenamefont {Modi}(2021)}]{Milz2021}%
  \BibitemOpen
  \bibfield  {author} {\bibinfo {author} {\bibfnamefont {S.}~\bibnamefont
  {Milz}}\ and\ \bibinfo {author} {\bibfnamefont {K.}~\bibnamefont {Modi}},\
  }\href {\doibase 10.1103/PRXQuantum.2.030201} {\bibfield  {journal} {\bibinfo
   {journal} {PRX Quantum}\ }\textbf {\bibinfo {volume} {2}},\ \bibinfo {pages}
  {030201} (\bibinfo {year} {2021})}\BibitemShut {NoStop}%
\bibitem [{\citenamefont {Muschik}\ and\ \citenamefont
  {Brunk}(1977)}]{Muschik1977}%
  \BibitemOpen
  \bibfield  {author} {\bibinfo {author} {\bibfnamefont {W.}~\bibnamefont
  {Muschik}}\ and\ \bibinfo {author} {\bibfnamefont {G.}~\bibnamefont
  {Brunk}},\ }\href {\doibase https://doi.org/10.1016/0020-7225(77)90047-7}
  {\bibfield  {journal} {\bibinfo  {journal} {Int. J. Eng. Sci.}\ }\textbf
  {\bibinfo {volume} {15}},\ \bibinfo {pages} {377} (\bibinfo {year}
  {1977})}\BibitemShut {NoStop}%
\bibitem [{\citenamefont {Muschik}(1977)}]{Muschik1977b}%
  \BibitemOpen
  \bibfield  {author} {\bibinfo {author} {\bibfnamefont {W.}~\bibnamefont
  {Muschik}},\ }\href {\doibase 10.1007/BF00248902} {\bibfield  {journal}
  {\bibinfo  {journal} {Arch. Ration. Mech. Anal.}\ }\textbf {\bibinfo {volume}
  {66}},\ \bibinfo {pages} {379} (\bibinfo {year} {1977})}\BibitemShut
  {NoStop}%
\bibitem [{\citenamefont {Strasberg}\ \emph {et~al.}(2021)\citenamefont
  {Strasberg}, \citenamefont {D\'{\i}az},\ and\ \citenamefont
  {Riera-Campeny}}]{Strasberg2021}%
  \BibitemOpen
  \bibfield  {author} {\bibinfo {author} {\bibfnamefont {P.}~\bibnamefont
  {Strasberg}}, \bibinfo {author} {\bibfnamefont {M.~G.}\ \bibnamefont
  {D\'{\i}az}}, \ and\ \bibinfo {author} {\bibfnamefont {A.}~\bibnamefont
  {Riera-Campeny}},\ }\href {\doibase 10.1103/PhysRevE.104.L022103} {\bibfield
  {journal} {\bibinfo  {journal} {Phys. Rev. E}\ }\textbf {\bibinfo {volume}
  {104}},\ \bibinfo {pages} {L022103} (\bibinfo {year} {2021})}\BibitemShut
  {NoStop}%
\bibitem [{\citenamefont {Gaudin}(1976)}]{Gaudin1976}%
  \BibitemOpen
  \bibfield  {author} {\bibinfo {author} {\bibfnamefont {M.}~\bibnamefont
  {Gaudin}},\ }\href {\doibase 10.1051 / jphys: 0197600370100108700} {\bibfield
   {journal} {\bibinfo  {journal} {J. Phys. France}\ }\textbf {\bibinfo
  {volume} {37}},\ \bibinfo {pages} {1087} (\bibinfo {year}
  {1976})}\BibitemShut {NoStop}%
\bibitem [{\citenamefont {Caldeira}\ and\ \citenamefont
  {Leggett}(1983)}]{Caldeira1983}%
  \BibitemOpen
  \bibfield  {author} {\bibinfo {author} {\bibfnamefont {A.}~\bibnamefont
  {Caldeira}}\ and\ \bibinfo {author} {\bibfnamefont {A.}~\bibnamefont
  {Leggett}},\ }\href {\doibase https://doi.org/10.1016/0378-4371(83)90013-4}
  {\bibfield  {journal} {\bibinfo  {journal} {Physica A}\ }\textbf {\bibinfo
  {volume} {121}},\ \bibinfo {pages} {587} (\bibinfo {year}
  {1983})}\BibitemShut {NoStop}%
\bibitem [{\citenamefont {Landau}\ and\ \citenamefont
  {Lifshitz}(1980)}]{Landau2013}%
  \BibitemOpen
  \bibfield  {author} {\bibinfo {author} {\bibfnamefont {L.}~\bibnamefont
  {Landau}}\ and\ \bibinfo {author} {\bibfnamefont {E.}~\bibnamefont
  {Lifshitz}},\ }\href {\doibase https://doi.org/10.1016/C2009-0-24487-4}
  {\emph {\bibinfo {title} {Statistical Physics: Volume 5}}}\ (\bibinfo
  {publisher} {Butterworth-Heinemann},\ \bibinfo {year} {1980})\BibitemShut
  {NoStop}%
\bibitem [{\citenamefont {Goldstein}\ \emph {et~al.}(2006)\citenamefont
  {Goldstein}, \citenamefont {Lebowitz}, \citenamefont {Tumulka},\ and\
  \citenamefont {Zangh\`{\i}}}]{Goldstein2006}%
  \BibitemOpen
  \bibfield  {author} {\bibinfo {author} {\bibfnamefont {S.}~\bibnamefont
  {Goldstein}}, \bibinfo {author} {\bibfnamefont {J.~L.}\ \bibnamefont
  {Lebowitz}}, \bibinfo {author} {\bibfnamefont {R.}~\bibnamefont {Tumulka}}, \
  and\ \bibinfo {author} {\bibfnamefont {N.}~\bibnamefont {Zangh\`{\i}}},\
  }\href {\doibase 10.1103/PhysRevLett.96.050403} {\bibfield  {journal}
  {\bibinfo  {journal} {Phys. Rev. Lett.}\ }\textbf {\bibinfo {volume} {96}},\
  \bibinfo {pages} {050403} (\bibinfo {year} {2006})}\BibitemShut {NoStop}%
\bibitem [{\citenamefont {Popescu}\ \emph {et~al.}(2006)\citenamefont
  {Popescu}, \citenamefont {Short},\ and\ \citenamefont
  {Winter}}]{Popescu2006}%
  \BibitemOpen
  \bibfield  {author} {\bibinfo {author} {\bibfnamefont {S.}~\bibnamefont
  {Popescu}}, \bibinfo {author} {\bibfnamefont {A.~J.}\ \bibnamefont {Short}},
  \ and\ \bibinfo {author} {\bibfnamefont {A.}~\bibnamefont {Winter}},\ }\href
  {\doibase 10.1038/nphys444} {\bibfield  {journal} {\bibinfo  {journal} {Nat.
  Phys.}\ }\textbf {\bibinfo {volume} {2}},\ \bibinfo {pages} {754} (\bibinfo
  {year} {2006})}\BibitemShut {NoStop}%
\bibitem [{\citenamefont {Kubo}(1957)}]{Kubo1957}%
  \BibitemOpen
  \bibfield  {author} {\bibinfo {author} {\bibfnamefont {R.}~\bibnamefont
  {Kubo}},\ }\href {\doibase 10.1143/JPSJ.12.570} {\bibfield  {journal}
  {\bibinfo  {journal} {J. Phys. Soc. Japan}\ }\textbf {\bibinfo {volume}
  {12}},\ \bibinfo {pages} {570} (\bibinfo {year} {1957})}\BibitemShut
  {NoStop}%
\bibitem [{\citenamefont {Martin}\ and\ \citenamefont
  {Schwinger}(1959)}]{Martin1959}%
  \BibitemOpen
  \bibfield  {author} {\bibinfo {author} {\bibfnamefont {P.~C.}\ \bibnamefont
  {Martin}}\ and\ \bibinfo {author} {\bibfnamefont {J.}~\bibnamefont
  {Schwinger}},\ }\href {\doibase 10.1103/PhysRev.115.1342} {\bibfield
  {journal} {\bibinfo  {journal} {Phys. Rev.}\ }\textbf {\bibinfo {volume}
  {115}},\ \bibinfo {pages} {1342} (\bibinfo {year} {1959})}\BibitemShut
  {NoStop}%
\bibitem [{\citenamefont {Prokof{\textquotesingle}ev}\ and\ \citenamefont
  {Stamp}(2000)}]{Prokofev2000}%
  \BibitemOpen
  \bibfield  {author} {\bibinfo {author} {\bibfnamefont {N.~V.}\ \bibnamefont
  {Prokof{\textquotesingle}ev}}\ and\ \bibinfo {author} {\bibfnamefont
  {P.~C.~E.}\ \bibnamefont {Stamp}},\ }\href {\doibase
  10.1088/0034-4885/63/4/204} {\bibfield  {journal} {\bibinfo  {journal} {Rep.
  Prog. Phys.}\ }\textbf {\bibinfo {volume} {63}},\ \bibinfo {pages} {669}
  (\bibinfo {year} {2000})}\BibitemShut {NoStop}%
\bibitem [{\citenamefont {London}\ \emph {et~al.}(2013)\citenamefont {London},
  \citenamefont {Scheuer}, \citenamefont {Cai}, \citenamefont {Schwarz},
  \citenamefont {Retzker}, \citenamefont {Plenio}, \citenamefont {Katagiri},
  \citenamefont {Teraji}, \citenamefont {Koizumi}, \citenamefont {Isoya},
  \citenamefont {Fischer}, \citenamefont {McGuinness}, \citenamefont
  {Naydenov},\ and\ \citenamefont {Jelezko}}]{London2013}%
  \BibitemOpen
  \bibfield  {author} {\bibinfo {author} {\bibfnamefont {P.}~\bibnamefont
  {London}}, \bibinfo {author} {\bibfnamefont {J.}~\bibnamefont {Scheuer}},
  \bibinfo {author} {\bibfnamefont {J.-M.}\ \bibnamefont {Cai}}, \bibinfo
  {author} {\bibfnamefont {I.}~\bibnamefont {Schwarz}}, \bibinfo {author}
  {\bibfnamefont {A.}~\bibnamefont {Retzker}}, \bibinfo {author} {\bibfnamefont
  {M.~B.}\ \bibnamefont {Plenio}}, \bibinfo {author} {\bibfnamefont
  {M.}~\bibnamefont {Katagiri}}, \bibinfo {author} {\bibfnamefont
  {T.}~\bibnamefont {Teraji}}, \bibinfo {author} {\bibfnamefont
  {S.}~\bibnamefont {Koizumi}}, \bibinfo {author} {\bibfnamefont
  {J.}~\bibnamefont {Isoya}}, \bibinfo {author} {\bibfnamefont
  {R.}~\bibnamefont {Fischer}}, \bibinfo {author} {\bibfnamefont {L.~P.}\
  \bibnamefont {McGuinness}}, \bibinfo {author} {\bibfnamefont
  {B.}~\bibnamefont {Naydenov}}, \ and\ \bibinfo {author} {\bibfnamefont
  {F.}~\bibnamefont {Jelezko}},\ }\href {\doibase
  10.1103/PhysRevLett.111.067601} {\bibfield  {journal} {\bibinfo  {journal}
  {Phys. Rev. Lett.}\ }\textbf {\bibinfo {volume} {111}},\ \bibinfo {pages}
  {067601} (\bibinfo {year} {2013})}\BibitemShut {NoStop}%
\bibitem [{\citenamefont {Sushkov}\ \emph {et~al.}(2014)\citenamefont
  {Sushkov}, \citenamefont {Lovchinsky}, \citenamefont {Chisholm},
  \citenamefont {Walsworth}, \citenamefont {Park},\ and\ \citenamefont
  {Lukin}}]{Sushkov2014}%
  \BibitemOpen
  \bibfield  {author} {\bibinfo {author} {\bibfnamefont {A.~O.}\ \bibnamefont
  {Sushkov}}, \bibinfo {author} {\bibfnamefont {I.}~\bibnamefont {Lovchinsky}},
  \bibinfo {author} {\bibfnamefont {N.}~\bibnamefont {Chisholm}}, \bibinfo
  {author} {\bibfnamefont {R.~L.}\ \bibnamefont {Walsworth}}, \bibinfo {author}
  {\bibfnamefont {H.}~\bibnamefont {Park}}, \ and\ \bibinfo {author}
  {\bibfnamefont {M.~D.}\ \bibnamefont {Lukin}},\ }\href {\doibase
  10.1103/PhysRevLett.113.197601} {\bibfield  {journal} {\bibinfo  {journal}
  {Phys. Rev. Lett.}\ }\textbf {\bibinfo {volume} {113}},\ \bibinfo {pages}
  {197601} (\bibinfo {year} {2014})}\BibitemShut {NoStop}%
\bibitem [{\citenamefont {Schwartz}\ \emph {et~al.}(2018)\citenamefont
  {Schwartz}, \citenamefont {Scheuer}, \citenamefont {Tratzmiller},
  \citenamefont {Müller}, \citenamefont {Chen}, \citenamefont {Dhand},
  \citenamefont {Wang}, \citenamefont {Müller}, \citenamefont {Naydenov},
  \citenamefont {Jelezko},\ and\ \citenamefont {Plenio}}]{Scwartz2018}%
  \BibitemOpen
  \bibfield  {author} {\bibinfo {author} {\bibfnamefont {I.}~\bibnamefont
  {Schwartz}}, \bibinfo {author} {\bibfnamefont {J.}~\bibnamefont {Scheuer}},
  \bibinfo {author} {\bibfnamefont {B.}~\bibnamefont {Tratzmiller}}, \bibinfo
  {author} {\bibfnamefont {S.}~\bibnamefont {Müller}}, \bibinfo {author}
  {\bibfnamefont {Q.}~\bibnamefont {Chen}}, \bibinfo {author} {\bibfnamefont
  {I.}~\bibnamefont {Dhand}}, \bibinfo {author} {\bibfnamefont {Z.-Y.}\
  \bibnamefont {Wang}}, \bibinfo {author} {\bibfnamefont {C.}~\bibnamefont
  {Müller}}, \bibinfo {author} {\bibfnamefont {B.}~\bibnamefont {Naydenov}},
  \bibinfo {author} {\bibfnamefont {F.}~\bibnamefont {Jelezko}}, \ and\
  \bibinfo {author} {\bibfnamefont {M.~B.}\ \bibnamefont {Plenio}},\ }\href
  {\doibase 10.1126/sciadv.aat8978} {\bibfield  {journal} {\bibinfo  {journal}
  {Sci. Adv.}\ }\textbf {\bibinfo {volume} {4}},\ \bibinfo {pages} {eaat8978}
  (\bibinfo {year} {2018})}\BibitemShut {NoStop}%
\bibitem [{\citenamefont {Hanson}\ \emph {et~al.}(2007)\citenamefont {Hanson},
  \citenamefont {Kouwenhoven}, \citenamefont {Petta}, \citenamefont {Tarucha},\
  and\ \citenamefont {Vandersypen}}]{Hanson2007}%
  \BibitemOpen
  \bibfield  {author} {\bibinfo {author} {\bibfnamefont {R.}~\bibnamefont
  {Hanson}}, \bibinfo {author} {\bibfnamefont {L.~P.}\ \bibnamefont
  {Kouwenhoven}}, \bibinfo {author} {\bibfnamefont {J.~R.}\ \bibnamefont
  {Petta}}, \bibinfo {author} {\bibfnamefont {S.}~\bibnamefont {Tarucha}}, \
  and\ \bibinfo {author} {\bibfnamefont {L.~M.~K.}\ \bibnamefont
  {Vandersypen}},\ }\href {\doibase 10.1103/RevModPhys.79.1217} {\bibfield
  {journal} {\bibinfo  {journal} {Rev. Mod. Phys.}\ }\textbf {\bibinfo {volume}
  {79}},\ \bibinfo {pages} {1217} (\bibinfo {year} {2007})}\BibitemShut
  {NoStop}%
\bibitem [{\citenamefont {Urbaszek}\ \emph {et~al.}(2013)\citenamefont
  {Urbaszek}, \citenamefont {Marie}, \citenamefont {Amand}, \citenamefont
  {Krebs}, \citenamefont {Voisin}, \citenamefont {Maletinsky}, \citenamefont
  {H\"ogele},\ and\ \citenamefont {Imamoglu}}]{Urbaszek2013}%
  \BibitemOpen
  \bibfield  {author} {\bibinfo {author} {\bibfnamefont {B.}~\bibnamefont
  {Urbaszek}}, \bibinfo {author} {\bibfnamefont {X.}~\bibnamefont {Marie}},
  \bibinfo {author} {\bibfnamefont {T.}~\bibnamefont {Amand}}, \bibinfo
  {author} {\bibfnamefont {O.}~\bibnamefont {Krebs}}, \bibinfo {author}
  {\bibfnamefont {P.}~\bibnamefont {Voisin}}, \bibinfo {author} {\bibfnamefont
  {P.}~\bibnamefont {Maletinsky}}, \bibinfo {author} {\bibfnamefont
  {A.}~\bibnamefont {H\"ogele}}, \ and\ \bibinfo {author} {\bibfnamefont
  {A.}~\bibnamefont {Imamoglu}},\ }\href {\doibase 10.1103/RevModPhys.85.79}
  {\bibfield  {journal} {\bibinfo  {journal} {Rev. Mod. Phys.}\ }\textbf
  {\bibinfo {volume} {85}},\ \bibinfo {pages} {79} (\bibinfo {year}
  {2013})}\BibitemShut {NoStop}%
\bibitem [{\citenamefont {Niknam}\ \emph {et~al.}(2020)\citenamefont {Niknam},
  \citenamefont {Santos},\ and\ \citenamefont {Cory}}]{Niknam2020}%
  \BibitemOpen
  \bibfield  {author} {\bibinfo {author} {\bibfnamefont {M.}~\bibnamefont
  {Niknam}}, \bibinfo {author} {\bibfnamefont {L.~F.}\ \bibnamefont {Santos}},
  \ and\ \bibinfo {author} {\bibfnamefont {D.~G.}\ \bibnamefont {Cory}},\
  }\href {\doibase 10.1103/PhysRevResearch.2.013200} {\bibfield  {journal}
  {\bibinfo  {journal} {Phys. Rev. Research}\ }\textbf {\bibinfo {volume}
  {2}},\ \bibinfo {pages} {013200} (\bibinfo {year} {2020})}\BibitemShut
  {NoStop}%
\bibitem [{\citenamefont {Niknam}\ \emph {et~al.}(2021)\citenamefont {Niknam},
  \citenamefont {Santos},\ and\ \citenamefont {Cory}}]{Niknam2021}%
  \BibitemOpen
  \bibfield  {author} {\bibinfo {author} {\bibfnamefont {M.}~\bibnamefont
  {Niknam}}, \bibinfo {author} {\bibfnamefont {L.~F.}\ \bibnamefont {Santos}},
  \ and\ \bibinfo {author} {\bibfnamefont {D.~G.}\ \bibnamefont {Cory}},\
  }\href {\doibase 10.1103/PhysRevLett.127.080401} {\bibfield  {journal}
  {\bibinfo  {journal} {Phys. Rev. Lett.}\ }\textbf {\bibinfo {volume} {127}},\
  \bibinfo {pages} {080401} (\bibinfo {year} {2021})}\BibitemShut {NoStop}%
\bibitem [{\citenamefont {Kol\'a\ifmmode~\check{r}\else \v{r}\fi{}}\ \emph
  {et~al.}(2012)\citenamefont {Kol\'a\ifmmode~\check{r}\else \v{r}\fi{}},
  \citenamefont {Gelbwaser-Klimovsky}, \citenamefont {Alicki},\ and\
  \citenamefont {Kurizki}}]{Kolar2012}%
  \BibitemOpen
  \bibfield  {author} {\bibinfo {author} {\bibfnamefont {M.}~\bibnamefont
  {Kol\'a\ifmmode~\check{r}\else \v{r}\fi{}}}, \bibinfo {author} {\bibfnamefont
  {D.}~\bibnamefont {Gelbwaser-Klimovsky}}, \bibinfo {author} {\bibfnamefont
  {R.}~\bibnamefont {Alicki}}, \ and\ \bibinfo {author} {\bibfnamefont
  {G.}~\bibnamefont {Kurizki}},\ }\href {\doibase
  10.1103/PhysRevLett.109.090601} {\bibfield  {journal} {\bibinfo  {journal}
  {Phys. Rev. Lett.}\ }\textbf {\bibinfo {volume} {109}},\ \bibinfo {pages}
  {090601} (\bibinfo {year} {2012})}\BibitemShut {NoStop}%
\bibitem [{\citenamefont {Nietner}\ \emph {et~al.}(2014)\citenamefont
  {Nietner}, \citenamefont {Schaller},\ and\ \citenamefont
  {Brandes}}]{Nietner2014}%
  \BibitemOpen
  \bibfield  {author} {\bibinfo {author} {\bibfnamefont {C.}~\bibnamefont
  {Nietner}}, \bibinfo {author} {\bibfnamefont {G.}~\bibnamefont {Schaller}}, \
  and\ \bibinfo {author} {\bibfnamefont {T.}~\bibnamefont {Brandes}},\ }\href
  {\doibase 10.1103/PhysRevA.89.013605} {\bibfield  {journal} {\bibinfo
  {journal} {Phys. Rev. A}\ }\textbf {\bibinfo {volume} {89}},\ \bibinfo
  {pages} {013605} (\bibinfo {year} {2014})}\BibitemShut {NoStop}%
\bibitem [{\citenamefont {Gallego-Marcos}\ \emph {et~al.}(2014)\citenamefont
  {Gallego-Marcos}, \citenamefont {Platero}, \citenamefont {Nietner},
  \citenamefont {Schaller},\ and\ \citenamefont
  {Brandes}}]{Gallego-Marcos2014}%
  \BibitemOpen
  \bibfield  {author} {\bibinfo {author} {\bibfnamefont {F.}~\bibnamefont
  {Gallego-Marcos}}, \bibinfo {author} {\bibfnamefont {G.}~\bibnamefont
  {Platero}}, \bibinfo {author} {\bibfnamefont {C.}~\bibnamefont {Nietner}},
  \bibinfo {author} {\bibfnamefont {G.}~\bibnamefont {Schaller}}, \ and\
  \bibinfo {author} {\bibfnamefont {T.}~\bibnamefont {Brandes}},\ }\href
  {\doibase 10.1103/PhysRevA.90.033614} {\bibfield  {journal} {\bibinfo
  {journal} {Phys. Rev. A}\ }\textbf {\bibinfo {volume} {90}},\ \bibinfo
  {pages} {033614} (\bibinfo {year} {2014})}\BibitemShut {NoStop}%
\bibitem [{\citenamefont {Grenier}\ \emph {et~al.}(2014)\citenamefont
  {Grenier}, \citenamefont {Georges},\ and\ \citenamefont
  {Kollath}}]{Grenier2014}%
  \BibitemOpen
  \bibfield  {author} {\bibinfo {author} {\bibfnamefont {C.}~\bibnamefont
  {Grenier}}, \bibinfo {author} {\bibfnamefont {A.}~\bibnamefont {Georges}}, \
  and\ \bibinfo {author} {\bibfnamefont {C.}~\bibnamefont {Kollath}},\ }\href
  {\doibase 10.1103/PhysRevLett.113.200601} {\bibfield  {journal} {\bibinfo
  {journal} {Phys. Rev. Lett.}\ }\textbf {\bibinfo {volume} {113}},\ \bibinfo
  {pages} {200601} (\bibinfo {year} {2014})}\BibitemShut {NoStop}%
\bibitem [{\citenamefont {Schaller}\ \emph {et~al.}(2014)\citenamefont
  {Schaller}, \citenamefont {Nietner},\ and\ \citenamefont
  {Brandes}}]{Schaller2014_b}%
  \BibitemOpen
  \bibfield  {author} {\bibinfo {author} {\bibfnamefont {G.}~\bibnamefont
  {Schaller}}, \bibinfo {author} {\bibfnamefont {C.}~\bibnamefont {Nietner}}, \
  and\ \bibinfo {author} {\bibfnamefont {T.}~\bibnamefont {Brandes}},\ }\href
  {\doibase 10.1088/1367-2630/16/12/125011} {\bibfield  {journal} {\bibinfo
  {journal} {New J. Phys.}\ }\textbf {\bibinfo {volume} {16}},\ \bibinfo
  {pages} {125011} (\bibinfo {year} {2014})}\BibitemShut {NoStop}%
\bibitem [{\citenamefont {Sekera}\ \emph {et~al.}(2016)\citenamefont {Sekera},
  \citenamefont {Bruder},\ and\ \citenamefont {Belzig}}]{Sekera2016}%
  \BibitemOpen
  \bibfield  {author} {\bibinfo {author} {\bibfnamefont {T.}~\bibnamefont
  {Sekera}}, \bibinfo {author} {\bibfnamefont {C.}~\bibnamefont {Bruder}}, \
  and\ \bibinfo {author} {\bibfnamefont {W.}~\bibnamefont {Belzig}},\ }\href
  {\doibase 10.1103/PhysRevA.94.033618} {\bibfield  {journal} {\bibinfo
  {journal} {Phys. Rev. A}\ }\textbf {\bibinfo {volume} {94}},\ \bibinfo
  {pages} {033618} (\bibinfo {year} {2016})}\BibitemShut {NoStop}%
\bibitem [{\citenamefont {Grenier}\ \emph {et~al.}(2016)\citenamefont
  {Grenier}, \citenamefont {Kollath},\ and\ \citenamefont
  {Georges}}]{Grenier2016}%
  \BibitemOpen
  \bibfield  {author} {\bibinfo {author} {\bibfnamefont {C.}~\bibnamefont
  {Grenier}}, \bibinfo {author} {\bibfnamefont {C.}~\bibnamefont {Kollath}}, \
  and\ \bibinfo {author} {\bibfnamefont {A.}~\bibnamefont {Georges}},\ }\href
  {\doibase https://doi.org/10.1016/j.crhy.2016.08.013} {\bibfield  {journal}
  {\bibinfo  {journal} {C. R. Phys}\ }\textbf {\bibinfo {volume} {17}},\
  \bibinfo {pages} {1161} (\bibinfo {year} {2016})},\ \bibinfo {note}
  {mesoscopic thermoelectric phenomena / Phénomènes thermoélectriques
  mésoscopiques}\BibitemShut {NoStop}%
\bibitem [{\citenamefont {Strasberg}\ and\ \citenamefont
  {Esposito}(2019)}]{Strasberg2019_b}%
  \BibitemOpen
  \bibfield  {author} {\bibinfo {author} {\bibfnamefont {P.}~\bibnamefont
  {Strasberg}}\ and\ \bibinfo {author} {\bibfnamefont {M.}~\bibnamefont
  {Esposito}},\ }\href {\doibase 10.1103/PhysRevE.99.012120} {\bibfield
  {journal} {\bibinfo  {journal} {Phys. Rev. E}\ }\textbf {\bibinfo {volume}
  {99}},\ \bibinfo {pages} {012120} (\bibinfo {year} {2019})}\BibitemShut
  {NoStop}%
\bibitem [{\citenamefont {Fernández-Acebal}\ \emph {et~al.}(2018)\citenamefont
  {Fernández-Acebal}, \citenamefont {Rosolio}, \citenamefont {Scheuer},
  \citenamefont {Müller}, \citenamefont {Müller}, \citenamefont {Schmitt},
  \citenamefont {McGuinness}, \citenamefont {Schwarz}, \citenamefont {Chen},
  \citenamefont {Retzker}, \citenamefont {Naydenov}, \citenamefont {Jelezko},\
  and\ \citenamefont {Plenio}}]{Fernandez-Acebal2018}%
  \BibitemOpen
  \bibfield  {author} {\bibinfo {author} {\bibfnamefont {P.}~\bibnamefont
  {Fernández-Acebal}}, \bibinfo {author} {\bibfnamefont {O.}~\bibnamefont
  {Rosolio}}, \bibinfo {author} {\bibfnamefont {J.}~\bibnamefont {Scheuer}},
  \bibinfo {author} {\bibfnamefont {C.}~\bibnamefont {Müller}}, \bibinfo
  {author} {\bibfnamefont {S.}~\bibnamefont {Müller}}, \bibinfo {author}
  {\bibfnamefont {S.}~\bibnamefont {Schmitt}}, \bibinfo {author} {\bibfnamefont
  {L.}~\bibnamefont {McGuinness}}, \bibinfo {author} {\bibfnamefont
  {I.}~\bibnamefont {Schwarz}}, \bibinfo {author} {\bibfnamefont
  {Q.}~\bibnamefont {Chen}}, \bibinfo {author} {\bibfnamefont {A.}~\bibnamefont
  {Retzker}}, \bibinfo {author} {\bibfnamefont {B.}~\bibnamefont {Naydenov}},
  \bibinfo {author} {\bibfnamefont {F.}~\bibnamefont {Jelezko}}, \ and\
  \bibinfo {author} {\bibfnamefont {M.}~\bibnamefont {Plenio}},\ }\href
  {\doibase 10.1021/acs.nanolett.7b05175} {\bibfield  {journal} {\bibinfo
  {journal} {Nano Lett.}\ }\textbf {\bibinfo {volume} {18}},\ \bibinfo {pages}
  {1882} (\bibinfo {year} {2018})}\BibitemShut {NoStop}%
\bibitem [{\citenamefont {Pekola}(2015)}]{Pekola2015}%
  \BibitemOpen
  \bibfield  {author} {\bibinfo {author} {\bibfnamefont {J.~P.}\ \bibnamefont
  {Pekola}},\ }\href {\doibase 10.1038/nphys3169} {\bibfield  {journal}
  {\bibinfo  {journal} {Nat. Phys.}\ }\textbf {\bibinfo {volume} {11}},\
  \bibinfo {pages} {118} (\bibinfo {year} {2015})}\BibitemShut {NoStop}%
\bibitem [{\citenamefont {Suomela}\ \emph {et~al.}(2016)\citenamefont
  {Suomela}, \citenamefont {Kutvonen},\ and\ \citenamefont
  {Ala-Nissila}}]{Suomela2016}%
  \BibitemOpen
  \bibfield  {author} {\bibinfo {author} {\bibfnamefont {S.}~\bibnamefont
  {Suomela}}, \bibinfo {author} {\bibfnamefont {A.}~\bibnamefont {Kutvonen}}, \
  and\ \bibinfo {author} {\bibfnamefont {T.}~\bibnamefont {Ala-Nissila}},\
  }\href {\doibase 10.1103/PhysRevE.93.062106} {\bibfield  {journal} {\bibinfo
  {journal} {Phys. Rev. E}\ }\textbf {\bibinfo {volume} {93}},\ \bibinfo
  {pages} {062106} (\bibinfo {year} {2016})}\BibitemShut {NoStop}%
\bibitem [{\citenamefont {Mehboudi}\ \emph {et~al.}(2019)\citenamefont
  {Mehboudi}, \citenamefont {Sanpera},\ and\ \citenamefont
  {Correa}}]{Mehboudi2019}%
  \BibitemOpen
  \bibfield  {author} {\bibinfo {author} {\bibfnamefont {M.}~\bibnamefont
  {Mehboudi}}, \bibinfo {author} {\bibfnamefont {A.}~\bibnamefont {Sanpera}}, \
  and\ \bibinfo {author} {\bibfnamefont {L.~A.}\ \bibnamefont {Correa}},\
  }\href {\doibase 10.1088/1751-8121/ab2828} {\bibfield  {journal} {\bibinfo
  {journal} {J. Phys. A: Math. Theor.}\ }\textbf {\bibinfo {volume} {52}},\
  \bibinfo {pages} {303001} (\bibinfo {year} {2019})}\BibitemShut {NoStop}%
\bibitem [{\citenamefont {Lazarides}\ \emph {et~al.}(2014)\citenamefont
  {Lazarides}, \citenamefont {Das},\ and\ \citenamefont
  {Moessner}}]{Lazarides2014}%
  \BibitemOpen
  \bibfield  {author} {\bibinfo {author} {\bibfnamefont {A.}~\bibnamefont
  {Lazarides}}, \bibinfo {author} {\bibfnamefont {A.}~\bibnamefont {Das}}, \
  and\ \bibinfo {author} {\bibfnamefont {R.}~\bibnamefont {Moessner}},\ }\href
  {\doibase 10.1103/PhysRevE.90.012110} {\bibfield  {journal} {\bibinfo
  {journal} {Phys. Rev. E}\ }\textbf {\bibinfo {volume} {90}},\ \bibinfo
  {pages} {012110} (\bibinfo {year} {2014})}\BibitemShut {NoStop}%
\bibitem [{\citenamefont {Mori}\ \emph {et~al.}(2016)\citenamefont {Mori},
  \citenamefont {Kuwahara},\ and\ \citenamefont {Saito}}]{Mori2016}%
  \BibitemOpen
  \bibfield  {author} {\bibinfo {author} {\bibfnamefont {T.}~\bibnamefont
  {Mori}}, \bibinfo {author} {\bibfnamefont {T.}~\bibnamefont {Kuwahara}}, \
  and\ \bibinfo {author} {\bibfnamefont {K.}~\bibnamefont {Saito}},\ }\href
  {\doibase 10.1103/PhysRevLett.116.120401} {\bibfield  {journal} {\bibinfo
  {journal} {Phys. Rev. Lett.}\ }\textbf {\bibinfo {volume} {116}},\ \bibinfo
  {pages} {120401} (\bibinfo {year} {2016})}\BibitemShut {NoStop}%
\bibitem [{\citenamefont {Abanin}\ \emph {et~al.}(2017)\citenamefont {Abanin},
  \citenamefont {De~Roeck}, \citenamefont {Ho},\ and\ \citenamefont
  {Huveneers}}]{Abanin2017}%
  \BibitemOpen
  \bibfield  {author} {\bibinfo {author} {\bibfnamefont {D.}~\bibnamefont
  {Abanin}}, \bibinfo {author} {\bibfnamefont {W.}~\bibnamefont {De~Roeck}},
  \bibinfo {author} {\bibfnamefont {W.~W.}\ \bibnamefont {Ho}}, \ and\ \bibinfo
  {author} {\bibfnamefont {F.}~\bibnamefont {Huveneers}},\ }\href {\doibase
  10.1007/s00220-017-2930-x} {\bibfield  {journal} {\bibinfo  {journal} {Comm.
  Math. Phys.}\ }\textbf {\bibinfo {volume} {354}},\ \bibinfo {pages} {809}
  (\bibinfo {year} {2017})}\BibitemShut {NoStop}%
\bibitem [{\citenamefont {Mallayya}\ \emph {et~al.}(2019)\citenamefont
  {Mallayya}, \citenamefont {Rigol},\ and\ \citenamefont
  {De~Roeck}}]{Mallayya2019}%
  \BibitemOpen
  \bibfield  {author} {\bibinfo {author} {\bibfnamefont {K.}~\bibnamefont
  {Mallayya}}, \bibinfo {author} {\bibfnamefont {M.}~\bibnamefont {Rigol}}, \
  and\ \bibinfo {author} {\bibfnamefont {W.}~\bibnamefont {De~Roeck}},\ }\href
  {\doibase 10.1103/PhysRevX.9.021027} {\bibfield  {journal} {\bibinfo
  {journal} {Phys. Rev. X}\ }\textbf {\bibinfo {volume} {9}},\ \bibinfo {pages}
  {021027} (\bibinfo {year} {2019})}\BibitemShut {NoStop}%
\bibitem [{\citenamefont {Nakajima}(1958)}]{Nakajima1958}%
  \BibitemOpen
  \bibfield  {author} {\bibinfo {author} {\bibfnamefont {S.}~\bibnamefont
  {Nakajima}},\ }\href {\doibase https://doi.org/10.1143/PTP.20.948} {\bibfield
   {journal} {\bibinfo  {journal} {Prog. Theor. Phys.}\ }\textbf {\bibinfo
  {volume} {20}},\ \bibinfo {pages} {948} (\bibinfo {year} {1958})}\BibitemShut
  {NoStop}%
\bibitem [{\citenamefont {Zwanzig}(1960)}]{Zwanzig1960}%
  \BibitemOpen
  \bibfield  {author} {\bibinfo {author} {\bibfnamefont {R.}~\bibnamefont
  {Zwanzig}},\ }\href {\doibase https://doi.org/10.1063/1.1731409} {\bibfield
  {journal} {\bibinfo  {journal} {{J}. {C}hem. {P}hys.}\ }\textbf {\bibinfo
  {volume} {33}},\ \bibinfo {pages} {1338} (\bibinfo {year}
  {1960})}\BibitemShut {NoStop}%
\bibitem [{\citenamefont {Redfield}(1957)}]{Redfield1957}%
  \BibitemOpen
  \bibfield  {author} {\bibinfo {author} {\bibfnamefont {A.~G.}\ \bibnamefont
  {Redfield}},\ }\href {\doibase 10.1147/rd.11.0019} {\bibfield  {journal}
  {\bibinfo  {journal} {IBM J. Res. Dev.}\ }\textbf {\bibinfo {volume} {1}},\
  \bibinfo {pages} {19} (\bibinfo {year} {1957})}\BibitemShut {NoStop}%
\end{thebibliography}%

\end{document}